\documentclass[aps,prb,twocolumn,superscriptaddress,floatfix,showpacs]{revtex4}
\pdfoutput=1
\usepackage{latexsym,amsmath,amssymb,amsfonts,mathbbol,graphicx,bm}
\usepackage{dcolumn} % Align table columns on decimal point
\usepackage[T1]{fontenc}
\usepackage[latin1]{inputenc}

\begin{document}

\title{Multiband {\bf $s$}-wave topological superconductors: \\ 
role of dimensionality and magnetic field response}

\author{Shusa Deng} \affiliation{\mbox{Department of Physics and
Astronomy, Dartmouth College, Hanover, New Hampshire 03755, USA}}

\author{Gerardo Ortiz}
\affiliation{\mbox{Department of Physics,  Indiana University,
Bloomington, Indiana 47405, USA}}

\author{Lorenza Viola} \affiliation{\mbox{Department of Physics and
Astronomy, Dartmouth College, Hanover, New Hampshire 03755, USA}}

\begin{abstract}
We further investigate a class of time-reversal-invariant two-band 
$s$-wave topological superconductors introduced in Phys. Rev. Lett. {\bf 108},
036803 (2012).
Provided that a {\em sign reversal} between the two superconducting pairing 
gaps is realized, the  topological phase diagram can be determined exactly 
(within mean field) in one and two dimensions, as well as in three dimensions upon 
restricting to the excitation spectrum of time-reversal invariant momentum modes.  
We show how, in the presence of time-reversal symmetry, ${\mathbb Z}_2$ invariants 
that distinguish between trivial and non-trivial quantum phases can be 
constructed by considering {\em only one} of the Kramers' sectors in which the 
Hamiltonian decouples into.  We find that the main features identified in our original 
two-dimensional setting remain qualitatively unchanged, with non-trivial topological 
superconducting phases supporting an {\em odd number of Kramers' pairs} of helical 
Majorana modes on each boundary, as long as the required $\pi$ phase difference 
between gaps is maintained.  We also analyze the consequences of 
time-reversal symmetry-breaking either due to the presence of an applied or impurity 
magnetic field or to a deviation from the intended phase matching between the 
superconducting gaps.   We demonstrate how the relevant notion of 
topological invariance must be modified when time-reversal symmetry is broken, and 
how both the persistence of gapless Majorana modes and their robustness properties 
depend in general upon the way in which the original Hamiltonian is perturbed.  
Interestingly, a topological quantum phase transition between helical and chiral 
superconducting phases can be induced by suitably tuning a Zeeman field in 
conjunction with a phase mismatch between the gaps.  Recent experiments in doped 
semiconducting crystals, of potential relevance to the proposed model, and possible 
candidate material realizations in superconductors with 
$s_\pm$ pairing symmetry are discussed.
\end{abstract}

\maketitle

\section{Introduction}

Obtaining a complete understanding of topological quantum matter has a
fundamental significance across condensed-matter physics as well as
potential practical implications within quantum science. On the one
hand, characterizing ``non-local'' topological order and unveiling
connections with the emergence of ``topologically protected'' edge
states are prerequisites for developing a unified classification of
matter beyond Landau's paradigm of symmetry breaking \cite{GBook}.  On
the other hand, taking full advantage of the distinctive robustness
features that the degenerate ground-state manifold enjoys may offer
new pathways to fault-tolerance in topological quantum memory and
quantum computation \cite{Kitaev03,Nayak}.

Building on the paradigmatic example of quantum Hall liquids and
the recent discovery of topological insulators
\cite{Kane05,Nayak,Kanereview} (TIs), {\em topological superconductors} 
\cite{Green,Ivanov,Kitaev01,Qi} (TSs) are attracting a growing
theoretical and experimental interest in this context.  TSs are gapped
phases of fermionic quantum matter whose ``zero-energy'' edge states are
naturally associated to {\em Majorana quasi-particles}, that is,
fermions which are their own antiparticle, as originally suggested by
Ettore Majorana back in 1937 \cite{Ettore}. Remarkably, 
%%% LV: Changed
braiding of Majorana fermions leads to non-Abelian exchange statistics
\cite{Ivanov,Moore,Frank}, making Majorana states 
%%% 
uniquely suited in principle to 
both fundamental quantum studies and topological qubit
implementations.  As a result, a variety of proposals have been put
forward in recent years to engineer Majorana fermions in different  
condensed-matter platforms.

A number of proposed TS realizations involve explicit breaking
of time-reversal (TR) symmetry -- notably, the seminal ``Kitaev's
wire'' in one dimension (1D) \cite{Kitaev01} and the chiral
superconductors with $p+ip$ pairing symmetry \cite{Green}, as well as
subsequent 1D hybrid semiconductor-superconductor nano-wires as well as 
heterostructures in 2D, see
Refs. \onlinecite{Lutchyn,Oreg,Potter10,2dhetero,Shen11,Brouwer,Chung} 
for representative contributions.  Our interest, however, is in {\em TR-invariant 
topological superconductivity}.  Existing proposals for TR-invariant TSs have
thus far largely relied on the proximity effect between a 3D TI and a
conventional ($s$-wave) superconductor \cite{Kane,Sarma} (see also Ref. 
\onlinecite{Volkov} for early contributions), or on access to unconventional
($p+ip$ and/or spin-triplet) superconducting order parameter 
\cite{Qi09,SatoOdd,Fu10}.

Our motivation is to explore whether alternative routes to
TR-invariant TSs exist based on conventional $s$-wave bulk pairing
symmetry.  Our key physical insight is to take advantage of {\em
multiband} superconductivity, directly in the spirit of the original
proposal by Suhl-Matthias-Walker (SMW) for two-band $s$-wave
superconductors \cite{SMW}.  Following the experimental discovery of
MgB$_2$ in 2001 \cite{Naga}, signatures
of multi-band superconductivity have been reported by now for a
variety of materials \cite{materials}, including newly discovered
iron-based superconductors \cite{MazinNat}. A model Hamiltonian for a
2D TR-invariant centro-symmetric  two-band $s$-wave TS supporting 
Majorana edge states was proposed in Ref. \onlinecite{Deng}, under 
the condition that a {\em sign reversal} be enforced between the two 
pairing gaps.  From a mathematical viewpoint, such a 
%%% LV: 
% 2D 
two-band model can equally describe a bi-layer system where the band 
index is replaced by a layer index \cite{Deng,Nakosai,Babak,FuKaneNote}.  
%%% LV: See new footnote
In this work, we aim to continue our exploration of TR-invariant 
multi-band TS, with the goal of (i) obtaining a more complete 
characterization of the non-trivial topological features that emerge for 
this class of Hamiltonians; and (ii) gaining a deeper insight on basic
aspects of TS and their edge states in general.  The content 
is organized as follows.

We begin in Sec. II by introducing the relevant class of two-band
TR-invariant model Hamiltonians for systems of different spatial
dimension.  While it is not possible to analytically determine the
excitation spectrum for general parameter values, we show how an {\em
exact} solution leading to non-trivial topological behavior may be
obtained under the assumption that the two {\em pairing gaps are
$\pi$-shifted} from one another, including in 3D as long as we focus on 
the excitation spectrum of  TR-invariant modes. We argue that, as long 
as two decoupled ``Kramers' sectors'' are identifiable,  
restricting to a {\em single} such sectors allows to naturally construct
${\mathbb Z}_2$ topological invariants applicable when TR symmetry is
preserved.  In particular, we demonstrate how topological numbers
originally defined in low dimension (such as, notably, Berry phases)
can be extended beyond 1D upon restriction to 
low-dimensional manifolds in parameter space.

Section III is devoted to characterizing the topological features that
emerge in our model as the spatial dimension increases from 1D to 3D.
This is done by first examining the bulk topological response, and
then by establishing the extent to which bulk properties relate to the
existence of Majorana edge states via a bulk-boundary
correspondence. In all dimensions, we find that a ${\mathbb Z}_2$
invariant distinguishes topologically trivial from non-trivial TS phases.  
The latter are found to support an {\em odd number of Kramers' pairs of
counter-propagating (helical) Majorana modes on each boundary}.  We
explicitly show that gapless Majorana modes existing in topologically 
trivial phases are generally not robust against perturbations, even if 
TR-symmetry is preserved.

In Sec. IV, we explore the consequences of explicitly breaking TR
according to different mechanisms, with emphasis on
determining the conditions under which gapless Majorana modes may persist,
and characterizing their degree of robustness compared to the
TR-invariant case. We first analyze the effect of a static magnetic
field in various physical scenarios, allowing for the field to act in
different directions on the bulk or solely on the boundary,
respectively.  In particular, we show how the {\em full} excitation
spectrum can still be exactly determined in the presence of a longitudinal
Zeeman field in 2D, and how  topological invariants must
be appropriately redefined as a consequence of TR symmetry being
broken. In general, we find that gapless Majorana modes may exist
in the presence of applied or impurity magnetic fields.  However, {\em their 
degree of robustness against subsequent perturbations
depends in general on both the symmetry and the details of the
latter}. Similar conclusions apply if TR symmetry is explicitly broken
due to a phase mismatch between the pairing gaps relative to the
intended value of $\pi$.  Remarkably, the application of a suitable
Zeeman field may then be used to {\em restore gapless Majorana modes} that 
would be gapped in under a phase mismatch alone, and to induce a topological
quantum phase transition (QPT) between helical and chiral phases,
accompanied by a vanishing bulk gap.

In Sec. V, we present some considerations on the practical feasibility
of our proposal in real materials.
%% LV: changed 
In particular, we qualitatively discuss implications in the context of 
the search for topological superconductivity in 3D doped TI
materials \cite{Sasaki11,Sasaki12,Stroscio}, 
and follow up on our original  
suggestion \cite{Deng} of realizations in iron-based 
superconductors exhibiting $s_{\pm}$ {\em pairing symmetry}.  We 
conclude in Sec. VI with a summary of our main results and an outlook 
to open problems.  Additional technical 
details are included in the Appendices. Specifically, Appendix \ref{appA} 
presents a complete symmetry analysis of our TS model Hamiltonian, thereby 
defining its symmetry class. Appendix \ref{self} briefly discusses the role of the 
self-consistency constraint in determining the physical accessibility and stability 
of different phases, whereas Appendix \ref{appB} shows the explicit form of 
the Hamiltonian matrix when arbitrary static magnetic fields perturb the TS system. 
Finally, Appendix \ref{appC} addresses the important problem of a consistent, 
numerically gauge-invariant evaluation of topological invariants, by discussing 
Berry phase and Chern number computations.

\section{Time-reversal invariant topological superconductors}
\label{dimension}

\subsection{Model Hamiltonian}
\label{model}

Our starting point is a class of TR-invariant two-band Hamiltonians of
the form introduced in Ref.~\onlinecite{Deng}.  Consider a regular
lattice in $D$ spatial dimensions, with orthonormal vectors $\{\hat{e}_\nu \,|\, 
\nu \in u_D \} \equiv \{\hat{x},\hat{y},\hat{z},\ldots\} $, where the relevant set of 
indexes $u_D$ depends upon dimensionality and $N_{\nu}$ is the number of 
lattice sites in the $\nu$th direction (for example, ${u}_2 = \{ {x}, {y} \}$ with a total
number of sites $N=N_x N_y $ in 2D, and so on).  Let $c$ and
$d$ label two orbitals (bands) and for each lattice site $j$ introduce
$$\psi_j \equiv
(c_{j,\uparrow},c_{j,\downarrow},d_{j,\uparrow},d_{j,\downarrow})^T,$$
\noindent 
in terms of fermionic annihilation operators $c_{j, \sigma}, d_{j\sigma}$,
with $\sigma= \uparrow, \downarrow$ denoting the spin quantum number.
Let us also introduce Pauli matrices $\tau_{\nu}$ and $\sigma_{\nu}$
($\nu=x,y,z$) that act on the orbital and spin part, respectively.  Then the
relevant Hamiltonian may be written as
\begin{equation}
H_D = H_{\sf cd}+H_{\sf so}+H_{\sf sw}+ \text{H.c.},
\label{Ham0}
\end{equation}
where the different terms have the following expression:
\begin{eqnarray}
H_{\sf cd}&=&\frac{1}{2}\sum_j(u_{cd} \psi_j^\dag \tau_x
\psi_j^{\;}-\mu \psi_j^\dag \psi_j^{\;}) -
t\hspace*{-0.5mm}\sum_{\langle i,j\rangle} \psi_i^\dag \tau_x
\psi_j^{\;}, \nonumber \\ H_{\sf so}&=&i \lambda
\hspace*{-1.5mm}\sum_{j, \nu \in {u}_D} \psi_j^\dag \tau_z
\sigma_\nu \psi_{j+\hat{e}_\nu}, \nonumber \\ H_{\sf sw}&=& \sum_j (\Delta_c
c_{j,\uparrow}^\dag c_{j,\downarrow}^\dag+ \Delta_d
d_{j,\uparrow}^\dag d_{j,\downarrow}^\dag).
\label{Ham}
\end{eqnarray}
Physically, $H_{\sf cd}$, $H_{\sf so}$, and $H_{\sf sw}$ represents
the two-band internal dynamics, the spin-orbit (SO) interaction, and
the $s$-wave intra-band superconducting fluctuations, respectively.
The parameter $\mu$ is the chemical potential, $u_{cd}$ is an on-site
spin-independent hybridization term between the two bands, whereas $t$
quantifies the strength of interband hopping, with $\langle i,j\rangle$ 
representing nearest-neighbor sites $i$ and $j$.  Note that, for
simplicity, we have assumed here that no inter-band SO coupling is
present and, without loss of generality, we have taken the strength of
the SO coupling to be the same, equal to $\lambda$, for each band.
The parameters $\Delta_c$ and $\Delta_d$ represent the $s$-wave
pairing gaps of the two bands within the mean-field approximation,
that is, in momentum space, 
\begin{equation}
\Delta_c=-V_{cc} \langle c_{{\bf k},\uparrow}
c_{-{\bf k},\downarrow}\rangle, \;\;\;\Delta_d=-V_{dd}\langle
d_{{\bf k},\uparrow}d_{-{\bf k},\downarrow}\rangle,
\label{gaps}
\end{equation}
with $V_{cc}>0$, $V_{dd} >0$ being the effective attraction strength
for fermions in each band \cite{SMW}.

Notice that in the limit where $\mu=0$ and no superconducting fluctuations are
present, $H_{\sf sw}=0$, the Hamiltonian $H_D$ in
Eq. (\ref{Ham0}) reduces to the so-called Dimmock model for a TI
\cite{Franz,Isaev}. From the point of view of the symmetry classification
introduced by Altland and Zirnbauer \cite{Altland}, one may explicitly
verify (see Appendix A) that $H_D$ exhibits manifest invariance under
both TR and particle-hole (PH) transformations, 
in addition to exhibiting inversion symmetry, 
indicating that the model can be taken to belong to DIII symmetry class. 
In particular, TR symmetry constrains each of the mean-field pairing gaps 
to be real, with a phase equal to $0$ or $\pi$.

For general parameter values and periodic boundary conditions (PBC),
$H_D$ can be block-diagonalized by a Fourier transformation in all
spatial directions. That is, we may rewrite
\begin{equation}
H_D=\frac{1}{2}\sum_{\bf{k}} (\hat{A}_{\bf{k}}^{\dag}
\hat{H}_{\bf{k}}\hat{A}_{\bf{k}}^{\;}-4\mu),
\label{fourier}
\end{equation}
where 
$$\hat{A}_{\bf{k}}^\dag=(c_{{\bf{k}},\uparrow}^\dag, c_{{\bf{k}},
\downarrow}^\dag,d_{{\bf{k}},\uparrow}^\dag,d_{{\bf{k}},
\downarrow}^\dag,c_{-{\bf{k}},\uparrow}^{\;},c_{-{\bf{k}},\downarrow}^{\;},
d_{-{\bf{k}},\uparrow}^{\;},d_{-{\bf{k}},\downarrow}^{\;}),$$ and
$\hat{H}_{\bf{k}}$ is a $8 \times 8$ matrix in general:
\begin{eqnarray}
\hat{H}_{\bf{k}}\hspace{-1mm}=\hspace{-1mm}
\hspace{-0.5mm}\left (\hspace{-1mm}
\begin{array}{cccc}
-\hspace{-0.5mm}\mu+\hspace{-0.5mm}{\vec{\lambda}_{\bf k}}
\hspace{-0.5mm}\cdot\hspace{-0.5mm}
\vec{\sigma}\hspace{-0.5mm} &\hspace{-0.5mm}m_{\bf k}\hspace{-0.5mm}
& \hspace{-0.5mm}i\Delta_c\sigma_y\hspace{-0.5mm} &
\hspace{0.5mm}0\hspace{-0.5mm} \\
\hspace{-0.5mm}m_{\bf k}\hspace{-0.5mm} &
\hspace{-0.5mm}-\mu\hspace{-0.5mm}-\hspace{-0.5mm}{\vec{\lambda}_{\bf
k}}\hspace{-0.5mm}\cdot\hspace{-0.5mm} \vec{\sigma} &
\hspace{-0.5mm}0\hspace{-0.5mm} & i\Delta_d\sigma_y\hspace{-0.5mm}\\
\hspace{-0.5mm}-i\Delta_c\sigma_y\hspace{-0.5mm} &
\hspace{-0.5mm}0\hspace{-0.5mm} &
\hspace{-0.5mm}\mu\hspace{-0.5mm}+\hspace{-0.5mm}{\vec{\lambda}_{\bf
k}}\hspace{-0.5mm}\cdot\hspace{-0.5mm} \vec{\sigma}^* & -m_{\bf k} \\
\hspace{-0.5mm}0\hspace{-0.5mm} &
\hspace{-0.5mm}-i\Delta_d\sigma_y\hspace{-0.5mm} &
\hspace{-0.5mm}-m_{\bf k} \hspace{-0.5mm} &
\hspace{-0.5mm}\mu\hspace{-0.5mm}-\hspace{-0.5mm}{\vec{\lambda}_{\bf
k}}\hspace{-0.5mm}\cdot\hspace{-0.5mm} \vec{\sigma}^*\hspace{-1mm}
\end{array}  \hspace{-0.5mm}\right)\hspace{-0.5mm}.
\label{8by8}
\end{eqnarray}
Here, ${}^*$ denotes complex conjugation and we have introduced the
compact notations 
\begin{eqnarray*}
\left \{ \begin{array}{c} \displaystyle
{\vec{\lambda}_{\bf k}} \equiv -2 \lambda \sum_{\nu
\in u_D} \sin{k_\nu} \, \hat{e}_{\nu} , \\ \displaystyle
\,\,m_{\bf{k}} \equiv u_{cd}-2 t\sum_{\nu \in u_D} \cos{k_\nu}, \\ \displaystyle
\hspace*{-10mm}\vec{\sigma} \equiv \sum_{\nu \in u_D} \sigma_\nu \, \hat{e}_{\nu}.
\end{array} \right.
\end{eqnarray*}

Remarkably, an exact analytical solution exists in both 1D and 2D in
the limit where the pairing gaps are $\pi$-shifted, that is, 
\begin{equation}
\Delta_c = -\Delta_d\equiv \Delta, 
\label{pi}
\end{equation}
in which case $\hat{H}_{\bf{k}}$ in Eq. (\ref{8by8}) decouples into
two $4 \times 4$ matrices.  Specifically, upon introducing new
canonical fermionic operators,
\begin{eqnarray}
\left \{ \begin{array}{c}
a_{{\bf{k}},\sigma}=\frac{1}{\sqrt{2}}(c_{{\bf{k}},\sigma} +
d_{{\bf{k}},\sigma} ), \\ b_{{\bf{k}},\sigma}=\frac{1}{\sqrt{2}}
(c_{{\bf{k}},\sigma} - d_{{\bf{k}},\sigma}) ,
\end{array} \right.
\label{canonicalf}
\end{eqnarray}
we may rewrite
$$H_D=\frac{1}{2}\sum_{\bf{k}} (\hat{B}_{\bf{k}}^{\dag}
\hat{H}'_{\bf{k}} \hat{B}_{\bf{k}}^{\;}-4\mu),$$ with
$$\hat{B}^\dag_{\bf k} =(a_{{\bf{k}},\uparrow}^\dag, b_{{\bf{k}},
\downarrow}^\dag,a_{{-\bf{k}},\uparrow}^{\;},b_{{-\bf{k}},\downarrow}^{\;},
a_{{-\bf{k}},\downarrow}^\dag,
b_{{-\bf{k}},\uparrow}^\dag,a_{{\bf{k}},
\downarrow}^{\;},b_{{\bf{k}},\uparrow}^{\;}),$$ 
and
$\hat{H}'_{\bf{k}}=\hat{H}'_{+,\bf{k}} \oplus \hat{H}'_{-,\bf{k}}$,
where
\begin{eqnarray}
\hat{H}'_{+,\bf{k}}\hspace{-1mm}=\hspace{-1mm}\left (\begin{array}{cc}
\hspace{-1mm}m_{\bf{k}}\sigma_z-\mu+{\vec{\lambda}_{\bf k}}\cdot
\vec{\sigma} & i\Delta\sigma_y\hspace{-1mm} \\
\hspace{-1mm}- i\Delta\sigma_y &
-m_{\bf{k}}\sigma_z+\mu+{\vec{\lambda}_{\bf k}}\cdot
\vec{\sigma}^*\hspace{-1mm}
\end{array} \hspace{-0.5mm}\right)\hspace{-0.5mm}, \label{4by4p} \\
\hat{H}'_{-,\bf{k}}\hspace{-1mm}=\hspace{-1mm} \left
(\begin{array}{cc}
\hspace{-1mm}m_{\bf{k}}\sigma_z-\mu-{\vec{\lambda}_{\bf k}}\cdot
\vec{\sigma}^* & -i\Delta\sigma_y\hspace{-1mm} \\
\hspace{-1mm}i\Delta\sigma_y &
-m_{\bf{k}}\sigma_z+\mu-{\vec{\lambda}_{\bf k}}\cdot
\vec{\sigma}\hspace{-1mm}
\end{array} \hspace{-0.5mm}\right)\hspace{-0.5mm}.
\label{4by4m}
\end{eqnarray}
Notice that the $4 \times 4$ matrices $\hat{H}'_{+,\bf{k}}$ and $\hat{H}'_{-,\bf{k}} $ may be 
regarded as TR of one another, in the following sense: by partitioning the fermonic operators  
$\hat{B}_{\bf{k}} \equiv \hat{B}_{+, \bf{k}} \oplus \hat{B}_{-, \bf{k}}$ and using the explicit 
form of the TR transformation ${\cal T}$ given in Appendix A, that is:
$$\mathcal{T} a(b)_{{\bf k},\uparrow} \mathcal{T}^{-1}=a(b)_{{-\bf
k},\downarrow}, \;\; \mathcal{T} a^\dag(b^\dag)_{{\bf k},\downarrow}
\mathcal{T}^{-1}=-a^\dag(b^\dag)_{{-\bf k},\uparrow},$$
$$\mathcal{T} a(b)_{{\bf k},\downarrow} \mathcal{T}^{-1}=-a(b)_{{-\bf
k},\uparrow}, \;\; \mathcal{T} a^\dag(b^\dag)_{{\bf k},\uparrow}
\mathcal{T}^{-1}=a^\dag(b^\dag)_{{-\bf k},\downarrow},$$ 
\noindent 
 %${\cal T}\equiv 
%(I_{4\times4} \otimes i \sigma_y) K$, where $K$ is complex conjugation,
we may write:
$$ {\cal T} \left( \hat{B}^\dag_{+, \bf{k}} \hat{H}'_{+,\bf{k}} \hat{B}_{+, \bf{k}}\right) 
{\cal T}^{-1} = \hat{B}^\dag_{-, \bf{k}} \hat{H}'_{-,\bf{k}} \hat{B}_{-, \bf{k}}.$$
\noindent 
The excitation spectrum obtained from diagonalizing either ``Kramers' sector'' 
$\hat{H}'_{+,\bf{k}}$ or $\hat{H}'_{-,\bf{k}}$ is given by 
\begin{eqnarray}
\epsilon_{n,{\bf k}}=\pm \sqrt{m_{\bf{k}}^2\hspace{-0.5mm}+
\hspace{-0.7mm}\Omega^2\hspace{-0.7mm}+
\hspace{-0.7mm}|\vec{\lambda}_{\bf{k}}|^2\hspace{-0.7mm}\pm2\sqrt{
m_{\bf{k}}^2 \Omega^2
\hspace{-0.7mm}+\hspace{-0.7mm}\mu^2|\vec{\lambda}_{\bf{k}}|^2}},
\label{spectrum2D}
\end{eqnarray}
where $\Omega^2\equiv \mu^2 + \Delta^2$ and we have assumed the energy
ordering $\epsilon_{1,{\bf k}} \leq \epsilon_{2,{\bf k}} \leq 0 \leq
\epsilon_{3,{\bf k}} \leq \epsilon_{4,{\bf k}}$. 
Clearly,  $\epsilon_{n,{\bf k}}= \epsilon_{n,-{\bf k}}$, as implied by 
inversion symmetry.  QPTs occur when the gap closes, 
that is, $\epsilon_{2,{\bf k}}=0$, for general $\Delta \ne 0$, leading to the
QPT lines determined by 
$$m_{{\bf k}_c}=\pm \,\Omega, \;\;\; \;k_{\nu,c} \in \{0, \pi \}, \;\nu \in u_D.$$
\noindent 
Note that the above condition is {\em independent} upon the SO strength 
$\lambda$, as long as $\lambda \ne 0$.

In 3D, the Hamiltonian $H_D$ can no longer be decoupled into
$4\times4$ matrices due to the SO coupling along the ${z}$
direction (see Appendix C), implying that no analytical solution of
the excitation spectrum can be obtained in general.  However, since
for the TR-invariant modes the gap closes only at the QCPs, and the SO
coupling term vanishes for the TR-invariant modes, that is, for
$k_{\nu,c} \in \{0, \pi \}$, $\nu\in u_D$, we may focus on the
excitation spectrum for $k_z = k_{z,c}$. In this case,
the decoupled structure into two TR Hamiltonians $\hat{H}'_{\pm, {\bf
k}}$ still holds, and thus we obtain exactly the same form of
excitation spectrum as in Eq.~(\ref{spectrum2D}) for the corresponding
energies $\epsilon_{n,{\bf k}=(k_x, k_y, k_{z,c})}$.  The phase
diagram of the Hamiltonian $H_D$, as the dimension changes from 1D to
3D, will be discussed in detail in Sec.~\ref{RD}, with appropriate
topological numbers being identified and labeling each phase.

Two additional remarks are in order in regard to the Hamiltonian $H_D$
in Eqs. (\ref{Ham0})-(\ref{Ham}).  First, the contribution $H_{\sf
sw}$ is a {\em special case} of the general SMW Hamiltonian for a
two-band superconductor with $s$-wave pairing symmetry \cite{SMW},
corresponding to the situation where no inter-band electron-phonon
process takes place.  In this case, the two superconducting gaps are
also associated with two distinct transition temperatures. In
principle (and in fact most likely in real materials), an inter-band
interaction Hamiltonian of the form 
\begin{eqnarray*}
H_{\sf cd}= -\sum_{\bf k, k'} V_{cd} (c_{{\bf k},\uparrow}^\dag
c_{-{\bf k},\downarrow}^\dag d_{-{\bf k'},\downarrow} d_{{\bf
k}',\uparrow} + \text{H.c.} ),
\label{hcd}
\end{eqnarray*}
may also be present, in addition to the two intra-band interactions
with strength $V_{cc}$ and $V_{dd}$ in Eqs. (\ref{Ham})-(\ref{gaps}).  
Within a mean-field description, it is easy to check that the effect
of the additional term amounts to a renormalization of the attraction
strengths for each band, 
\begin{eqnarray*}
\Delta_c \mapsto \Delta_c'\equiv V_{cc} \Delta_c + V_{cd} \Delta_d ,\\ 
\Delta_d \mapsto \Delta_d'\equiv V_{cd} \Delta_c + V_{dd} \Delta_d .
\end{eqnarray*}
Thus, our analysis can be straightforwardly extended to the general
case $V_{cd}\ne 0$.  In particular, requiring that $\Delta_c' =-
\Delta_d'$ still yields $\Delta_c =- \Delta_d$ as a unique solution if
$V_{cc}=V_{dd}$, therefore all the results obtained for $V_{cd}=0$
apply with no modification in this case.  While different attraction
strengths, as determined by band structure details, can be easily
accommodated in principle, we shall show in Sec. \ref{away} that 
the specific values of these parameters do not play a crucial role 
(cf. Fig. \ref{mismatch1}). Thus, for simplicity we shall focus on the
already rich behavior emerging for $V_{cc}=V_{dd}$ henceforth.

Second, as written in Eq. (\ref{Ham}), the superconducting term in
$H_D$ involves $s$-wave pairing {\em in each band}
separately. Consider, however, the following unitary transformation
for each mode ${\bf k}$:
\begin{eqnarray*}
U=\frac{1}{\sqrt{2}}\Big{\{}I_{2 \times 2} \otimes \Big{[}(\sigma_x+\sigma_z)\otimes I_{2\times2}\Big{]}  \Big{\}} .
\end{eqnarray*}
Then, at the symmetry point
defined by Eq. (\ref{pi}), $H_D$ transforms into:
$$\tilde{H}_D =\frac{1}{2}\sum_{\bf{k}} \,[\hat{A}_{\bf{k}}^{\dag}
(U\hat{H}_{\bf{k}}U^\dag) \hat{A}_{\bf{k}}^{\;}-4\mu],$$
\noindent
which can in turn be written in real space as follows:
\begin{eqnarray}
\tilde{H}_D&=&\frac{1}{2}\sum_j(u_{cd} \psi_j^\dag \tau_z
\psi_j^{\;}-\mu \psi_j^\dag \psi_j^{\;}) -
t\hspace*{-0.5mm}\sum_{\langle i,j\rangle} \psi_i^\dag \tau_z
\psi_j^{\;} \nonumber \\ &+& i \lambda
\hspace*{-1.5mm}\sum_{j, \nu \in u_D} \psi_j^\dag \tau_x \sigma_\nu
\psi_{j+\hat{e}_\nu} \nonumber \\ &+& \sum_j \Delta (c_{j,\uparrow}^\dag
d_{j,\downarrow}^\dag- c_{j,\downarrow}^\dag d_{j,\uparrow}^\dag) +
{\text H.c.},
\label{Hamtilde}
\end{eqnarray}
That is, the original intra-band superconductivity is transformed into
{\em inter-band spin-singlet superconductivity}, with the original
intra-band SO interaction being correspondingly transformed into
inter-band SO interaction.  Thus, all the results obtained from
investigating $H_D$ in the limit $\Delta_c=-\Delta_d$ can be directly
generalized to the class of models described by $\tilde{H}_D$.

\subsection{Topological Indicators}
\label{indicators}

Since the Hamiltonian $H_D$ preserves TR symmetry, topological
invariants that are applicable to the TR-broken case will, in general,
fail to characterize the system's topological response.  While
different approaches have been pursued \cite{Kane06,Kane,Roy,Roy09}, 
our strategy is to build on previous work \cite{Ortiz94,Ortiz96} and 
construct suitable ``partial'' topological quantum numbers 
in order to distinguish between topologically trivial and non-trivial
phases \cite{Deng}. In particular, in this work we shall focus on three 
kinds of topological indicators, which we define in what follows.  We
anticipate that despite their different form, these indicators are in
essence equivalent when applicable, as we will demonstrate for our
system in Sec. \ref{RD}.

\vspace*{1mm}

{\bf $\bullet$ Partial Berry phase sum parity.} Consider the simplest
1D case first. By taking advantage of the decoupled structure between
TR-pairs in the limit of $\pi$-shifted gaps [Eq. (\ref{pi})], we may
consider the sum of the Berry phases \cite{Berry,Ortiz94,Ortiz96} for 
the {\em two occupied negative bands of one Kramers' sector only}, say,
$|\psi_{1,k_x}\rangle$ and $|\psi_{2,k_x}\rangle$ of $\hat{H}'_{+,\bf{k}}$, 
with each Berry phase given by (see also Appendix D for more detail on the actual
numerical calculation):
\begin{eqnarray}
B_n=i \int_{-\pi}^\pi dk_x \langle \psi_{n,k_x}|\partial_{k_x}
\psi_{n,k_x}\rangle.
\label{bn}
\end{eqnarray} 
Since $B_n$ may only attain the values $0$ or $\pi$ (mod $2 \pi$) for
a system with inversion symmetry \cite{Zak}, a ${\mathbb Z}_2$
topological invariant may be naturally constructed as follows:
\begin{eqnarray}
P_B=(-1)^{{\rm mod}_{2\pi}(B_+)/\pi}, \;\; B_+ \equiv B_1+B_2.
\label{bnnp}
\end{eqnarray} 

For $D>1$, the basic idea is to use lower-dimensional 
topological numbers upon restricting to a lower-dimensional 
manifold in parameter space \cite{Ortiz96}.
In 2D, for instance, note that the superconductor in the momentum
planes $k_{y,c} \in \{0,\pi\}$ is mapped to itself under inversion and has the
topology of a 1D ring under PBC.  Thus, it is possible to define a
$\mathbb{Z}_2$ invariant by analyzing the parity of the partial Berry
phase sum {\em restricted to the planes $k_y = k_{y,c}$}. That is,
\begin{eqnarray}
P_B=(-1)^{ {\rm mod}_{2\pi}(B_+)/\pi}, \;
B_+=\hspace*{-3mm}\sum_{k_y =0, \pi }
\hspace*{-2mm}(B_{k_y,1}+B_{k_y,2}).
\label{bnnp2D}
\end{eqnarray} 
The 3D case can be treated in a similar fashion and is discussed in
more detail directly in Sec. \ref{3d}.

\vspace*{1mm}

{\bf $\bullet$ Partial Chern sum parity.} This invariant was originally 
introduced and used in Ref.~\onlinecite{Deng} for the 2D geometry.
In a similar spirit to the above, the idea is to use the Chern numbers
(CNs) of the {\em two occupied negative bands of one Kramers' sector only},
say $\hat{H}'_{+,\bf{k}}$ as before.  Call these CNs $C_1$ and $C_2$, and let
$|\psi_{n,{\bf k}}\rangle$ denote the band-$n$ eigenvector of
$\hat{H}'_{+,\bf{k}}$. Then the CNs $C_n \in {\mathbb Z}$, $n=1,2$,
are given by (see also Appendix D for its numerical implementation):
\begin{eqnarray}
C_n=\frac{1}{\pi} \int_{-\pi}^\pi dk_x\int_{-\pi}^\pi dk_y \,
\text{Im} \,\langle \partial_{k_x}\psi_{n,{\bf k}}|\partial_{k_y}
\psi_{n,{\bf k}}\rangle.
\label{cn}
\end{eqnarray}
Thus, a ${\mathbb Z}_2$ invariant may be constructed as follows:
\begin{eqnarray}
\label{invariant1}
P_C \equiv (-1)^{{\rm mod}_2(C_+)}, \;\;\; C_+ \equiv C_1+C_2.
\end{eqnarray}
Again, the extension to the 3D case will be addressed in
Sec. \ref{3d}.

\vspace*{1mm}

{\bf $\bullet$ Partial fermion number parity.} As discussed in
Ref.~\onlinecite{Deng}, there is a direct connection between the
invariant $P_C$ defined in Eq. (\ref{invariant1}) and the fermion
number parity of the TR-invariant modes. Let us focus on the
ground-state fermion number parity of the TR-invariant points 
${\bf k}_c$ in the first Brillouin zone.
Since for these modes, the two TR Hamiltonians $\hat{H}'_{+,\bf{k}_c}$
and $\hat{H}'_{-,\bf{k}_c}$ are decoupled, we need only concentrate on
the ground-state parity property of $\hat{H}'_{+,\bf{k}_c}$. Let us
introduce the new basis given by 
$$u_{{\bf k}_c}\equiv\{ a_{{\bf k}_c,\uparrow}^\dag
|\text{vac}\rangle, b_{{\bf k}_c,\downarrow}^\dag |\text{vac}\rangle,
|\text{vac}\rangle, a_{{\bf k}_c,\uparrow}^\dag b_{{\bf
k}_c,\downarrow}^\dag |\text{vac}\rangle \}.$$
\noindent 
In this basis, $\hat{H}'_{+,\bf{k}_c}$ becomes 
$$\widehat{H}_{+,{\bf k}_c}=-\mu I_{4\times4}+ [m_{{\bf k}_c}\sigma_z
\oplus ( \Delta \sigma_x + \mu \sigma_z)],$$
\noindent
with eigenvalues $-\mu \pm m_{{\bf k}_c}, -\mu\pm |\Omega|$. When
$|m_{{\bf k}_c}| > |\Omega|$, the ground state of each mode ${\bf
k}_c$ is in the sector with odd fermion parity, $P_{{\bf k}_c}=e^{i
\pi (a_{{\bf k}_c,\uparrow}^\dag a_{{\bf k}_c,\uparrow}^{\;}+b_{{\bf
k}_c,\downarrow}^\dag b_{{\bf k}_c,\downarrow}^{\;})}=-1$; conversely,
when $|m_{{\bf k}_c}| < |\Omega|$, it is in the sector with even
fermion parity $P_{{\bf k}_c}=1$. Thus, we define a ${\mathbb Z}_2$
partial fermion number parity invariant as follows:
\begin{eqnarray}
P_F=\prod_{{\bf k_c}} P_{\bf k_c}.  
\label{fermion}
\end{eqnarray}  
Computing the fermion number parity of the TR-invariant modes {\em
from one representative of each Kramers' pairs} is consistent with the
fact that only a partial CN (or Berry phase) sum can detect TS phases
in the presence of TR symmetry.

\section{Role of Dimensionality} 
\label{RD}

A relevant question regarding the existence of non-trivial TS phases
and their physical properties is the role played by spatial
dimensionality. In this section, we will explore the topological
response of our Hamiltonian $H_D$ as we move from 1D to 3D, first
through {\em bulk properties}, that is, by computing the topological
numbers from the bulk Hamiltonian; and then through the {\em
bulk-boundary correspondence}, that is, the relationship between the
nature of the bulk vacuum in the thermodynamic limit and the existence
of surface modes on the boundary.

\subsection{Bulk properties}

\subsubsection{Topological response in 1D} 

The so-called Kitaev wire \cite{Kitaev01}, which is essentially the XY
chain in a transverse magnetic field written in fermionic language,
has attracted a lot of attention recently for supporting an odd number
of Majorana modes on the boundary, as originally remarked in
Ref. \onlinecite{LSM} (see also Ref. \onlinecite{Chakravarty}). 
%% LV: Ref added... [even if I do not think it is relevant at all!!] 
Does our TR-invariant 
two-band TS model in 1D ($u_1={x}$) also exhibit non-trivial
topological phases?
\begin{figure}[thb]
\includegraphics[width=7cm]{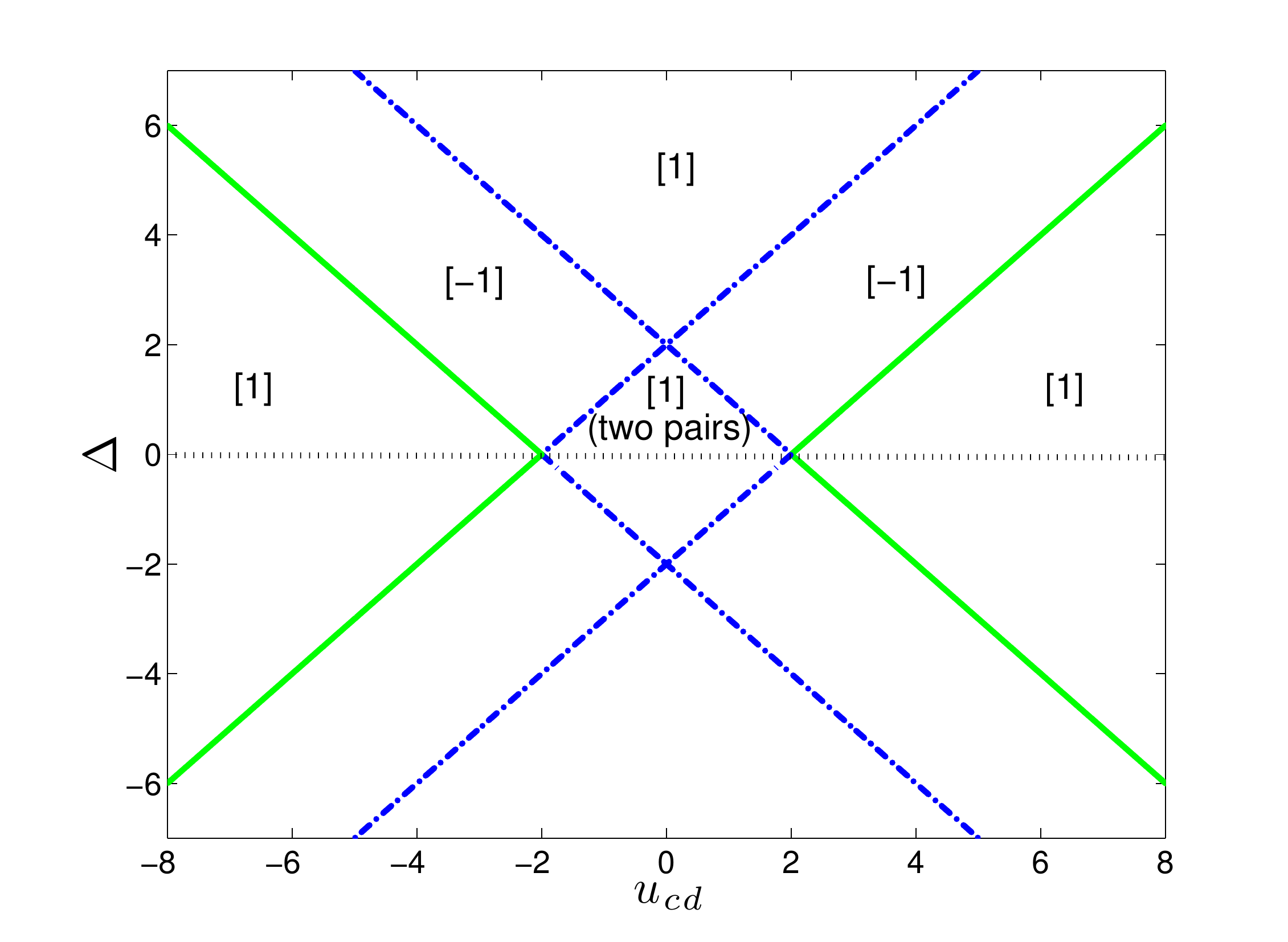}
\vspace*{-3mm} 
\caption{\label{1Dpd} (Color online) Phase diagram of the Hamiltonian
$H_D$ in 1D as a function of $u_{cd}$ and $\Delta$, with $t=1=\lambda$, 
for representative chemical potential $\mu=0$.  As noted, the phase 
diagram is independent upon $\lambda$ provided that $\lambda \ne 0$. 
The horizontal black dotted line at $\Delta=0$ represents an
insulator or metal phase, depending on the filling. The topological
response is characterized via the partial Berry phase sum parity
$P_B$, given in square bracket. Notice that the phase diagram is
symmetric under $\Delta \mapsto - \Delta$. System size: $N_x=100$. }
\end{figure}

The natural choice of topological indicator is the partial Berry phase
sum parity, instead of the partial Chern sum parity, which requires at 
least a 2D parameter space [cf. Eq.~(\ref{cn})]. 
The phase diagram  obtained (for $\mu=0$) from requiring
$\epsilon_{2,{\bf k}_c}=0$ is depicted in Fig.~\ref{1Dpd}, where the 
integer numbers in square brackets are the partial Berry phase
sum parity $P_B$, with $P_B=-1(1)$ corresponding to non-trivial
(trivial) topological phases, respectively. The phase diagram shows
the topological phases of the 1D Hamiltonian indexed by a
$\mathbb{Z}_2$ number, as expected from general 
classification arguments in this simple case 
\cite{Altland,Ryu}.

\begin{figure*}[htb]
\includegraphics[width=6.8cm]{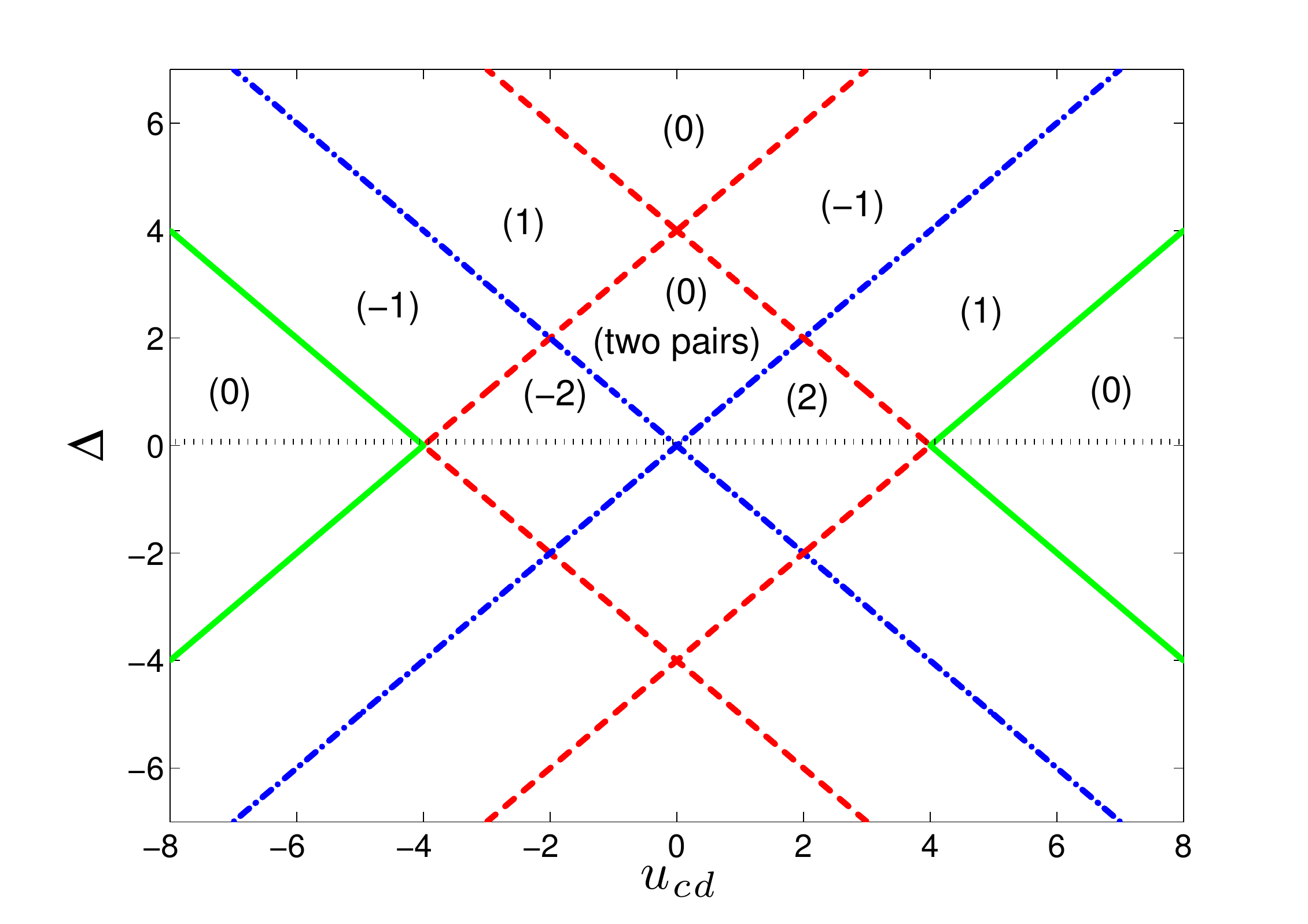}\includegraphics[width=6.8cm]{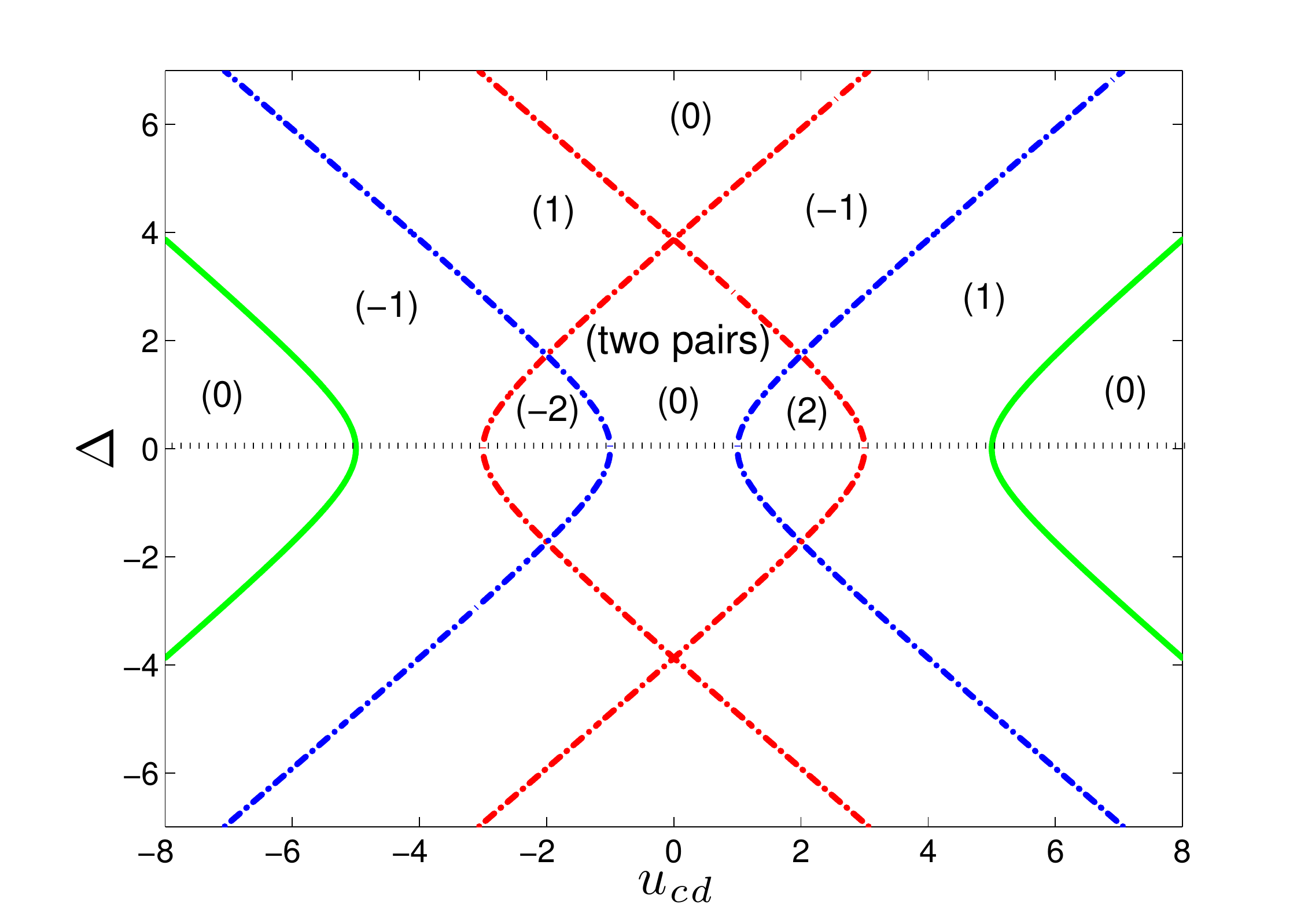}
\includegraphics[width=6.8cm]{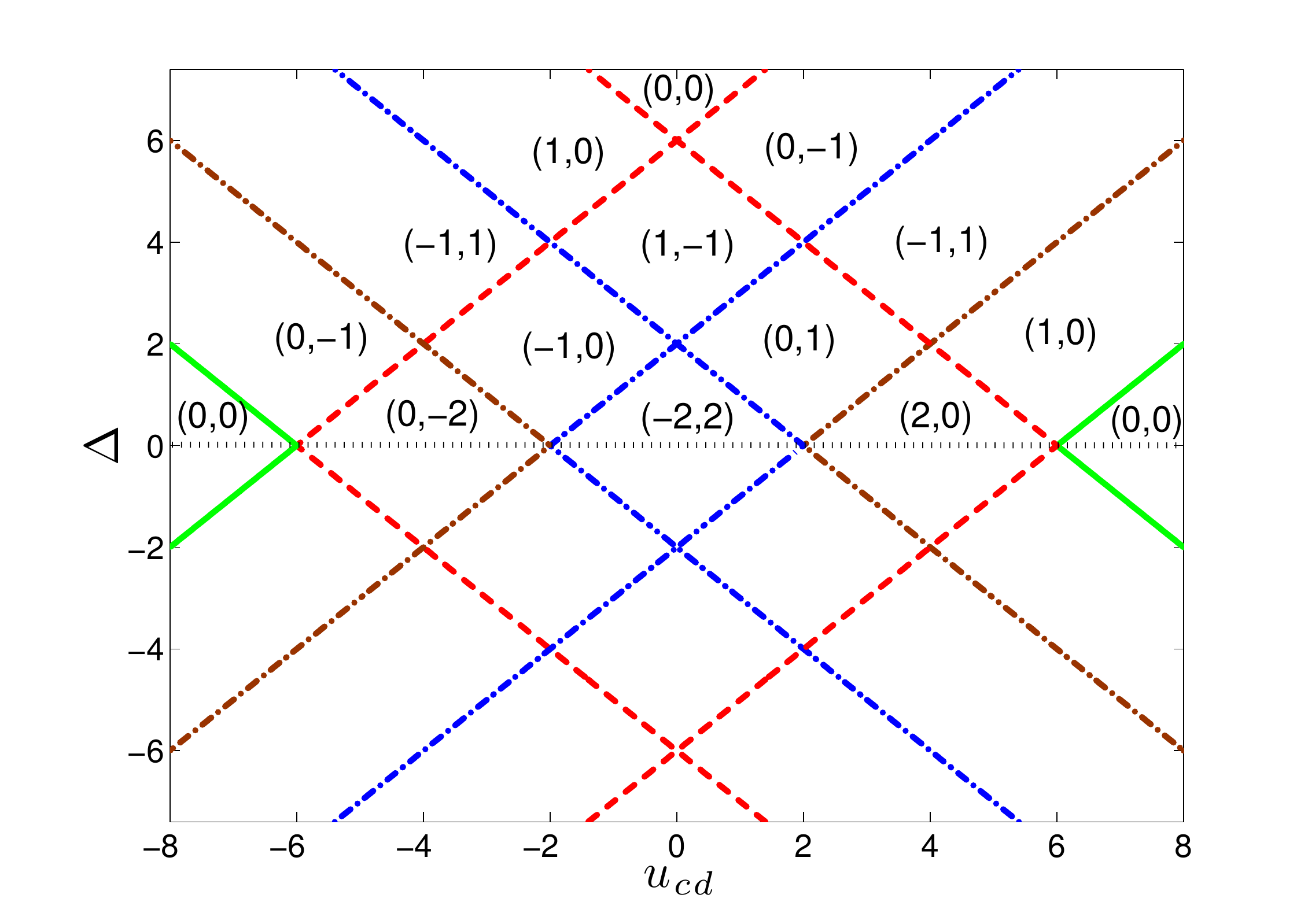}\includegraphics[width=6.8cm]{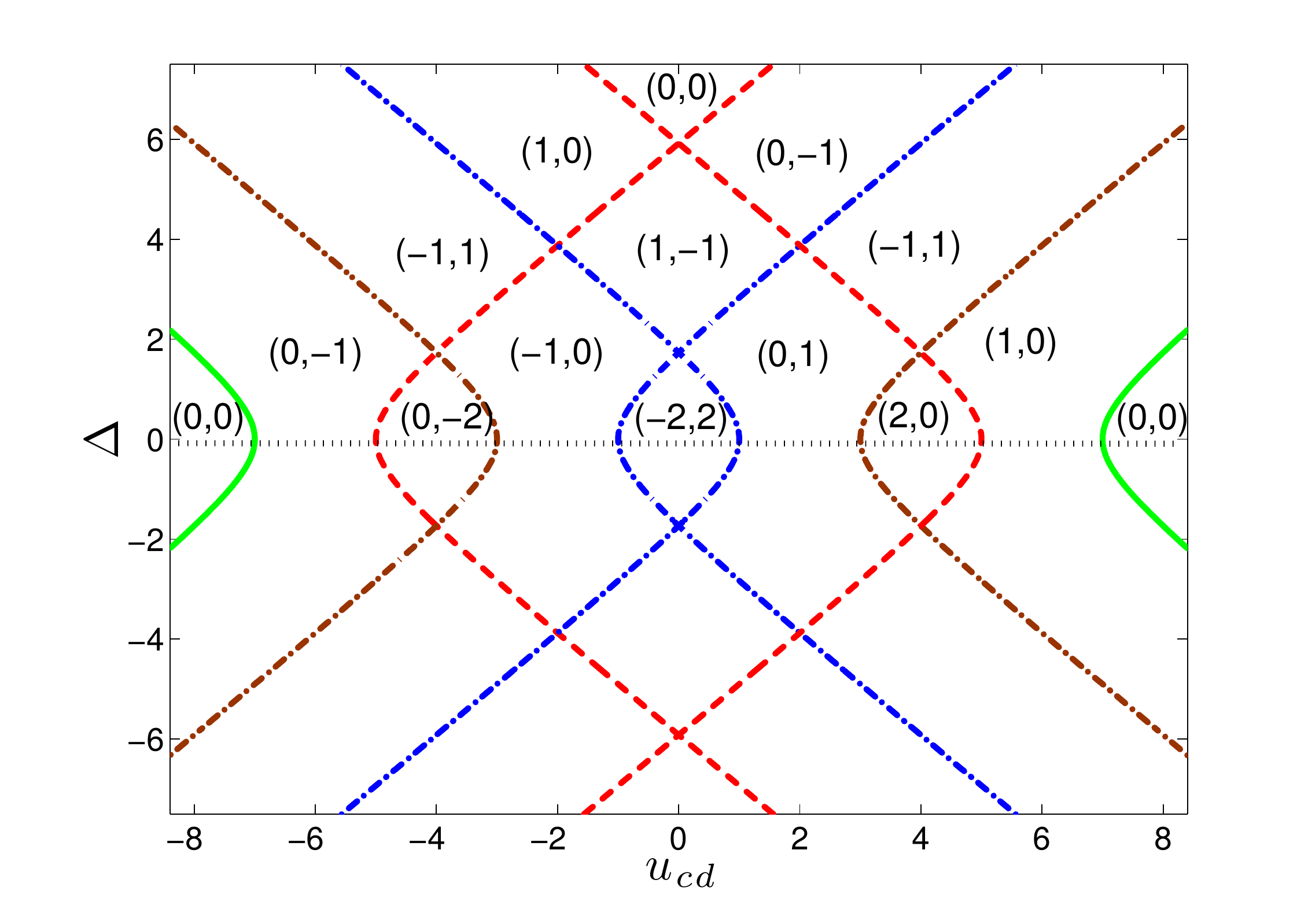}
\vspace*{-3mm} 
\caption{\label{pd} (Color online) Phase diagrams of the Hamiltonian
$H_D$ in 2D [top panels, $\mu=0$ and $\mu=-1$, respectively] and 3D
[bottom panels, $\mu=0$ and $\mu=-1$, respectively] 
 as a function of $u_{cd}$ and $\Delta$.  The topological
response is characterized in terms of the partial CN sum $C_+$ in 2D,
and in terms of the partial CN sum of the modes $k_z= k_{z,c}$
$(C^0_+,C^\pi_+)$ in 3D, the respective values being given in
parentheses. The corresponding $\mathbb{Z}_2$ invariant is the parity
of the partial CN sum, Eq. (\ref{invariant2}).  System size:
$N_x=N_y=N_z=100$.}
\end{figure*}

\subsubsection{Topological response in 2D} 
  
In 2D, we may compute both the partial CNs [$C_n$, Eq. (\ref{cn})] and
the partial Berry phase of each occupied band [$B_{k_{y,c},n}$,
Eq. (\ref{bnnp2D})] by following the procedure outlined in Appendix D.
We find that the parity of the partial Berry phase sum, $P_B$, is
consistent with the parity of partial Chern sum, $P_C$, in the whole
parameter space.

The topological responses for $\mu=0$ and $\mu=-1$ as representative
examples are shown in the two top panels of Fig.~\ref{pd}, with the
integer numbers in parenthesis giving the partial Chern sum $C_+$ 
[see also Fig. 1 in Ref. \onlinecite{Deng}].  As
it will become explicit from analyzing the bulk-boundary
correspondence (Sec. \ref{BBC}), an odd (even) value of $C_+$
corresponds to non-trivial (trivial) topological phases, respectively.
We stress that in 2D QPTs are present also between phases
carrying the {\em same} partial Chern sum parity, for instance separating
phases with $C_+=1$ and $C_+=-1$, as well as $C_+=0$ and
$C_+=\pm 2$ (blue dash-dotted lines).  Thus, a {\em
$\mathbb{Z}$ invariant is necessary in order to identify all the
topological phases} in the bulk phase diagram.  A similar
characterization was encountered for the model analyzed in
Ref. \onlinecite{Kub}, where, however, TR symmetry is explicitly
broken.  Since the partial Berry phase sum is, by definition, a
${\mathbb Z}_2$ quantity, this also makes $B_+$ inadequate to fully
characterize the phase diagram, although $P_B$ still correctly
diagnoses the presence of non-trivial topological features.

\subsubsection{Topological response in 3D}   
\label{3d}

The 3D Hamiltonian $H_D$ is especially interesting in the light of
recent discoveries of candidate TIs \cite{experiments} and 
%%% LV: softened... and added ref... 
investigations of possible TS materials 
in 3D geometries \cite{Sasaki11,Sasaki12,Stroscio}.  
Since our 3D superconductor in the momentum planes 
$k_z = k_{z,c} \in \{0,\pi\}$ is mapped to
itself under inversion and has the topology of a 2D torus under PBC,
we may define a $\mathbb{Z}_2$ invariant by analyzing the partial
Chern sum restricted to such planes.  If $C^{k_{z,c}}_+$ denotes the
corresponding partial Chern sum, then we can define the following
$\mathbb{Z}_2$ parity invariant:
\begin{eqnarray}
\label{invariant2}
P_C \equiv (-1)^{{\rm mod}_2 (C_+)},\;\; C_+ \equiv C^0_+ + C^\pi_+ .
\end{eqnarray}
Note that we could choose the partial Chern sum on the planes
$k_x = k_{x,c}$ or $k_y = k_{y,c}$. It is easy to verify that the
resulting values of $P_C$ would be the same on the planes
$k_x = k_{x,c}$, $k_y = k_{y,c}$, and $k_z = k_{z,c}$.  In fact, since only
the parity of $(C^0_+ + C^\pi_+)$ matters, the definition in
Eq. (\ref{invariant2}) is equivalent to the parity of $(C^0_+ -
C^\pi_+)$.  This is similar in spirit to the ``strong
$\mathbb{Z}_2$ invariant'' that has been invoked to distinguish strong
vs.  weak (trivial) TI phases \cite{Roy}.

The topological phase diagrams for $\mu=0$ and $\mu=-1$
are shown in the two bottom panels of Fig.~\ref{pd}, where an odd
(even) value of $C_+$ corresponds to non-trivial (trivial) topological
phases, respectively.  Notice that although we use, as in 2D, a $\mathbb{Z}$
number to map out the 3D phase diagrams, {\em the parity $P_C$} (hence
a ${\mathbb Z}_2$ invariant) {\em suffices to identify all the phases},
since all the QPT lines now separate phases with different
parity, unlike in 2D.

Similarly, we may fix $k_y = k_{y,c}, k_z =k_{z,c}$, and define the
partial Berry phase sum parity as:
\begin{eqnarray*}
P_B=(-1)^{{\rm mod}_{2\pi}(B_+)/{\pi}}, \; B_+\equiv
\hspace{-4mm}\sum_{k_y , k_z =0, \pi }
\hspace{-3mm}(B_{k_y,k_z,1}+B_{k_y,k_z,2}).
\label{bnnp3D}
\end{eqnarray*}
We find that the resulting values of $P_B$ are consistent with both $P_C$
and $P_F$ throughout the phase diagram.  We also recall that in
Ref.~\onlinecite{Ortiz94} a many-body generalization of the one-body
Berry phase was constructed in the presence of interaction, by
properly defining twisted boundary conditions.  Thus, an interesting
possibility for further exploration is whether the partial Berry phase
sum defined here might still characterize the {\em topological response of
an interacting system} with TR and inversion symmetries, provided 
that only one representative from each TR-pair is selected. 

In summary, we may conclude that the parity
of the partial Berry phase sum, of the partial Chern sum, and of the
partial fermion number are all equivalent to one another:
$P_C=P_B=P_F. $
Thus, any of them can be used to characterize the $\mathbb{Z}_2$
invariance in our TR-invariant system irrespective of dimensionality,
with odd (even) parity corresponding to non-trivial (trivial)
topological phases, respectively.  That being said, if there are
QPT lines between phases that share the same parity (such as in
the 2D phase diagram), then a $\mathbb{Z}$ invariant (the partial
Chern sum), is needed to distinguish and label all the phases.

\subsection{Bulk-boundary correspondence}
\label{BBC}

As mentioned, a bulk-boundary correspondence generally refers to the
relationship between bulk properties of the system in a given phase
and the existence and robustness of the corresponding edge states.
Specifically, for our TR-invariant Hamiltonian, we formulate the
bulk-boundary correspondence in terms of the relation between the bulk
${\mathbb Z}_2$ topological invariants and the {\em parity of the
number of TR-pairs of boundary modes} \cite{Isaev}.  That is, a bulk
phase characterized by odd (even) $P_C$ (or, equivalently, $P_B, P_F$)
is expected to correspond to an odd (even) number of TR-{\em pairs} of
edge states {\em per boundary}, which are robust against perturbations
that preserve the symmetry class of the system (DIII in our case).
One should, however, notice that there is no {\em a priori} reason to
expect that the topological numbers that characterize the phases in the
bulk phase diagram should be the same that also characterize the
bulk-boundary correspondence.

In order to characterize the bulk-boundary correspondence
in our model Hamiltonian, we investigate $H_D$ with open boundary
conditions (OBC) along one spatial direction, while keeping PBC along
the remaining direction(s). Our numerical results indicate that the
bulk property $P_C(P_B)=-1 (1) $ does correspond to an odd (even)
number of {\em Kramers' pairs of helical edge modes on each
boundary}. Since our Hamiltonian exhibits PH symmetry, the quasi-particle
annihilation operator $\gamma_{\epsilon_{n,\bf{k}}}$ for eigenvalue 
$\epsilon_{n,\bf{k}}$ obeys $\gamma^\dag_{\epsilon_{n,\bf{k}}} =
\gamma_{-\epsilon_{n, \bf{k}}}$, which identifies such zero-energy modes 
as Majorana fermions,  $\gamma^\dag_0=\gamma_0$. 

The situation is simplest for the 1D system: for instance, when
$u_{cd}=2$, $\Delta=2$ ($P_B=-1$ in the phase diagram of
Fig.~\ref{1Dpd}), one pair of Majorana edge modes exists at each end,
while when $u_{cd}=6$, $\Delta=2$ ($P_B=1$ in the phase diagram of
Fig.~\ref{1Dpd}), no edge mode is found. Direct calculation also shows
that a topologically trivial phase ($P_B=1$ with $u_{cd}=0$,
$\Delta=1$ in Fig.~\ref{1Dpd}) can likewise support two pairs of Majorana modes
per edge, corresponding to $k_x=0,k_x=\pi$, respectively. 
Since the bulk-boundary correspondence for the
2D case has already been extensively discussed in
Ref.~\onlinecite{Deng}, we focus next on addressing in detail the 3D
cubic geometry.

\begin{figure}[t]
\begin{center}
\includegraphics[width=9.1cm]{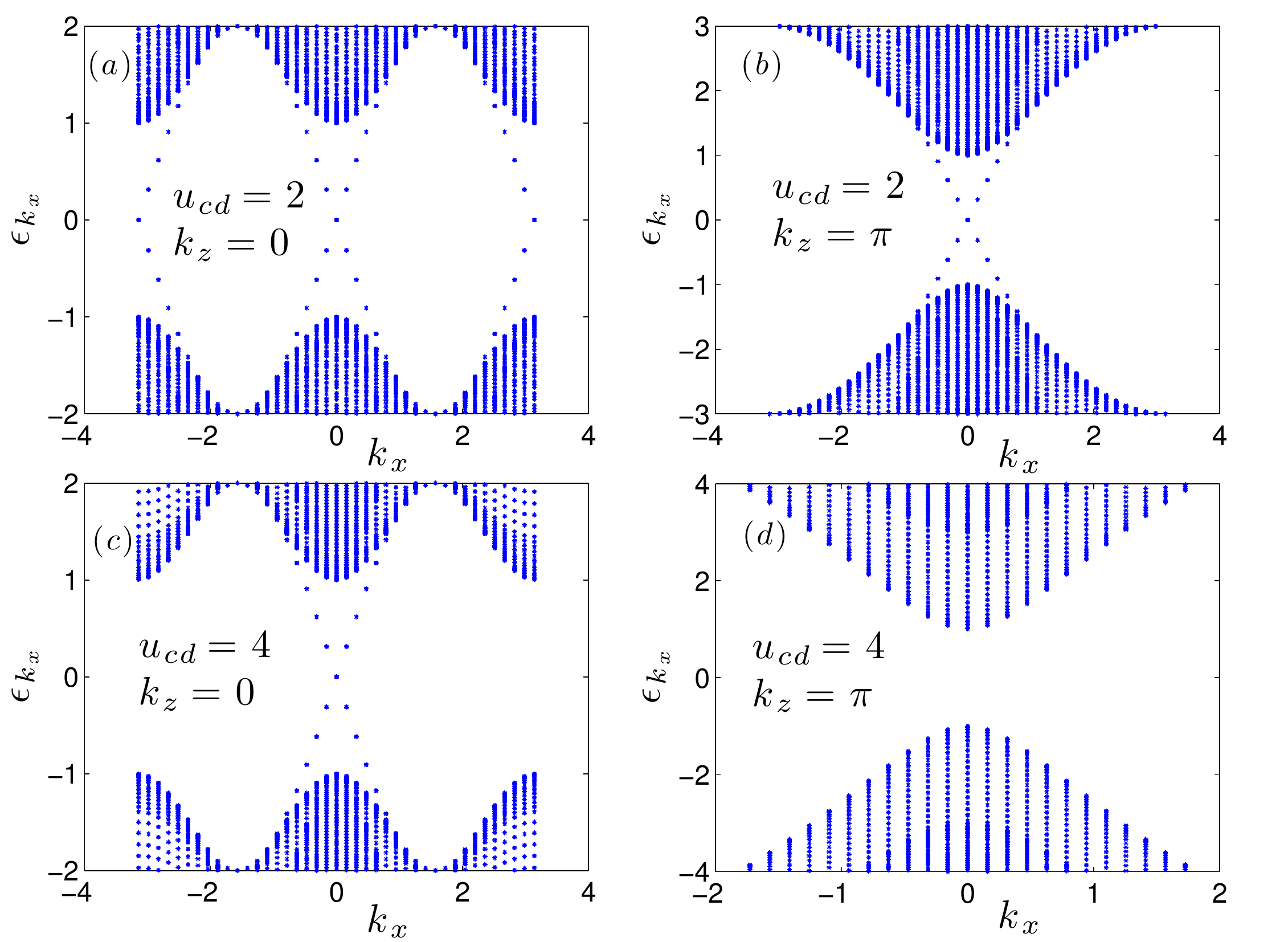}
\vspace*{-3mm}\caption{\label{tso3D} (Color online) Excitation
spectrum of the 3D Hamiltonian $H_D$ with OBC along the ${y}$
direction, for $\mu=0, t=1, \lambda=1=\Delta$.  Top panels: $u_{cd}=
2$, with $C^0_+=0$ for $k_z=0$ in panel (a), and $C^\pi_+=1$ for
$k_z=\pi$ in panel (b).  Bottom panels: $u_{cd}= 4$, with $C^0_+=2$
for $k_z=0$ in panel (c), and $C^\pi_+=0$ for $k_z=\pi$ in panel (d).
Note that the bulk gap scales as $\min(\lambda,\Delta)$, indicating 
that the edge modes are more stable for stronger SO and superconductivity.
System size: $(N_x, N_y,N_z)=(40,100,40)$. }
\end{center}
\end{figure}

In 3D, we maintain PBC along the ${x}$ and ${z}$ directions,
and use instead OBC along the ${y}$ direction (sometimes referring
to the resulting 2D boundaries as the right(left) edge).  Thus, we can
obtain the excitation spectrum $\epsilon_{n,k_x,k_{z,c}}$, by applying a
Fourier transformation in the ${x}$ and ${z}$
directions.   For simplicity, let us focus on the case
$\mu=0$.  Since the Dirac cones exist at the TR-invariant modes, we may fix
additionally $k_z = k_{z,c} \in \{0, \pi \}$. The resulting excitation spectrum is depicted in
Fig.~\ref{tso3D} for representative parameter choices.  Specifically,
panel (a) and (b) correspond to phases that support a total {\em odd}
number of Dirac cones on each edge, that is, two Dirac cones (at
$k_x=0$ and $k_x=\pi$) for $k_z=0$ and one Dirac cone (at $k_x=0$) for
$k_z=\pi$.  This is consistent with the partial Chern sum $C^0_+=0$
for $k_z=0$ and $C^\pi_+=1$ for $k_z=\pi$, hence
[Eq. (\ref{invariant2})] $P_C=-1$.  Panel (c) and (d) correspond
instead to phases supporting a total {\em even} number of Dirac cones,
that is, two Dirac cones at $k_x=0$ for $k_z=0$, with two
corresponding Kramers' pairs of Majorana modes on each boundary, and no 
Dirac cone for $k_z=\pi$. This is consistent with the partial Chern
sum $C^0_+=2$ for $k_z=0$ and $C^\pi_+=0$ for $k_z=\pi$.  
Similar to both the 1D and 2D cases, an even value of,
say, $C_+^0$ may correspond to a pair of Dirac cones [as in (a)] 
or it may indicate the absence of Dirac cones altogether [as in (d)].

\begin{figure}[t]
\begin{center}
\includegraphics[width=6.6cm]{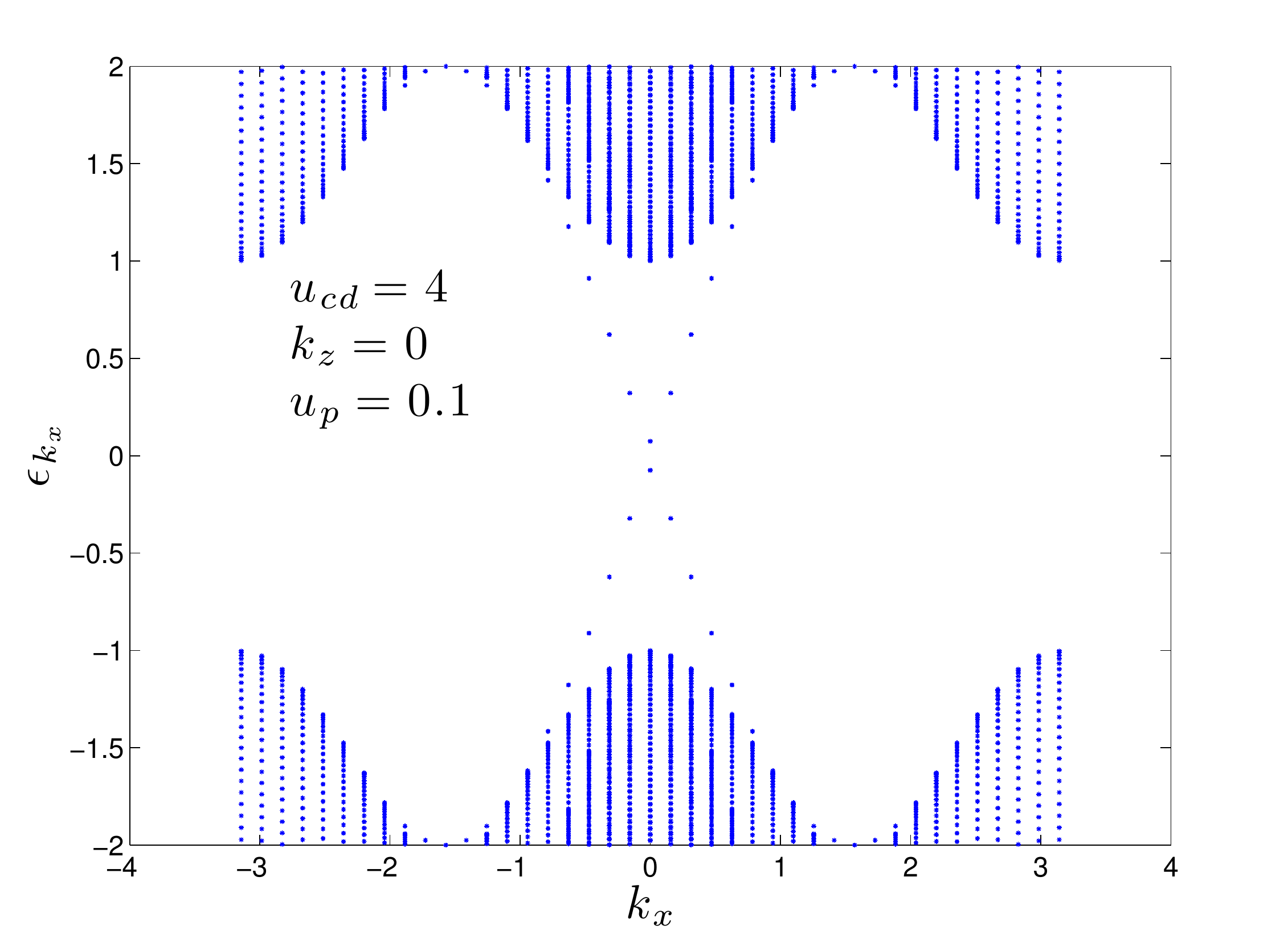}
\vspace*{-3mm}\caption{\label{3Dback} (Color online) Excitation
spectrum of the 3D Hamiltonian $H_D$ with OBC in the ${y}$
direction and in the presence of the boundary perturbation $H_p$ given
in Eq. (\ref{pert0}). The relevant parameters are: $\mu=0, t=1=
\lambda=\Delta$, and $u_{cd}= 4$, with perturbation strength
$u_p=0.1$. The gapless surface modes clearly become gapped under
$H_p$.  System size: $(N_x, N_y,N_z)=(40,100,40)$. }
\end{center}
\end{figure}

While gapless Majorana modes in non-trivial TS phases ($P_C=-1$) are
protected against boundary perturbations that respect TR
and PH symmetry (thus do not change the symmetry class), it is
interesting to explicitly verify what happens if $P_C=1$.  Consider,
in particular, the situation we discussed above with $C_+=2$ [panel
(c) in Fig. \ref{tso3D}], in which case two Kramers' pairs of helical Majorana modes, 
say, $(\gamma_1^{(i)}, \gamma_2^{(i)}),$ with $ \gamma_2^{(i)} ={\cal T}
\gamma_1^{(i)} {\cal T}^{-1}$, $i=1,2$, exist at $k_x=0,
k_z=0$. Suppose that we can express the two co-propagating modes
$\gamma_1^{(1)}$ and $\gamma_1^{(2)}$ from each Kramers' pair on a given
edge as follows:
\begin{eqnarray}
\left \{ \begin{array}{c} 
\gamma_1^{(1)}=\sum_{n=1}^{N_y}
\:\, (\alpha^{(1)}_n a_{n,\uparrow}^\dag+\beta_n^{(1)}
b_{n,\downarrow}^\dag+ \text{H.c.}), \\ \gamma_{1}^{(2)}=
\sum_{n=1}^{N_y} i (\alpha_n^{(2)} a_{n,\uparrow}^\dag+\beta_n^{(2)} 
b_{n,\downarrow}^\dag- \text{H.c.}),
\end{array} \right. 
\label{gammacn2}
\end{eqnarray}
for real coefficients $\alpha_n^{(i)}, \beta_n^{(i)}$, where we have
used the canonical fermion operators defined in Eq. (\ref{canonicalf})
along with $\gamma_1^{(i)\dag} =\gamma_1^{(i)}$, $i=1,2$.
Then consider, for example, the following perturbation that acts 
on the boundary, and preserves TR and PH symmetry, as well as 
inversion (see also Appendix A for further discussion):
\begin{eqnarray}
H_p = \hspace*{-2mm} \sum_{k_x,j,k_z,\sigma} & 
\hspace*{-4mm} u_p^{(j)} \hspace*{-2mm}
&(c_{k_x,j,k_z,\sigma}^\dag c_{-k_x,j,-k_z,\sigma} \nonumber \\ 
& \hspace*{-3mm} + \: &
d_{k_x,j,k_z,\sigma}^\dag d_{-k_x,j,-k_z,\sigma}) + \text{H.c.},
\label{pert0}
\end{eqnarray} 
where $u_p^{(j)} \equiv u_p \neq 0$ for $j=1$ or $N_y$, and $u_p^{(j)}=0$
otherwise.  By invoking degenerate perturbation theory, we can infer that the
degeneracy of the zero-energy surface modes is lifted, since explicit
calculation yields
$$\langle\Psi_{\sf gs}|\gamma_1^{(2)\dag} H_p \gamma_1^{(1)}
|\Psi_{\sf gs}\rangle \ne 0,$$ 
\noindent 
where $|\Psi_{\sf gs}\rangle$ is the many-body ground state. 
The exact excitation spectrum in the presence of $H_p$ is shown in
Fig.~\ref{3Dback} for $k_z=0$.

This explicitly illustrates how the presence of an even number of
pairs of Majorana modes, such as in the $C_+=2$ phase in 3D, makes
such modes generally non-robust against boundary perturbations, even
if the symmetry class of the system is unchanged.  
Therefore, a distinctive property of a 3D TR-invariant TS is the presence of an
{\em odd} number of Dirac cones, corresponding to an {\em odd} number
of pairs of helical Majorana surface modes, similar to the 1D and 2D
cases.  While this might seem to conflict with the robust behavior predicted
from the topological classification \cite{Ryu} for the DIII class in
3D, this apparent contradiction can be resolved by noting that such a 
classification strictly must be applied within ``irreducible'' blocks where 
{\it only} generic symmetries such as TR and PH are satisfied. While care is 
needed in interpreting the free-fermion topological classification in the presence 
%%% LV: Ref added 
of inversion symmetry [see Appendix C in Ref. \onlinecite{Ryu}] and, more 
generally, interaction effects \cite{Kitaev2010},
our model  additionally exhibits ``hidden'' discrete symmetries which, although hard to 
identify {\em a priori}, should be taken into account in principle. For instance, 
we have explicitly verified that one such  
${\mathbb Z}_2 \otimes {\mathbb Z}_2 \otimes {\mathbb Z}_2 \otimes {\mathbb Z}_2$ 
hidden symmetry exists in the limit $\mu=0$, $\lambda=t$, and is 
broken by the above perturbation $H_p$ (see Appendix A). 

In summary, topologically non-trivial (trivial) phases with odd (even)
$P_C$ ($P_B$, $P_F$) corresponds to an odd (even) number of Dirac 
cones (or points, depending on dimension). 
Thus, our model Hamiltonian $H_D$ in Eq. (\ref{Ham}) exhibits
qualitatively similar behavior regardless of dimensionality,
in the sense that both trivial and non-trivial TS phases exist in
all cases.  Thanks to the richer phase
structure, however, more possibilities arise in 3D for different
phases to exist. Interestingly, while QPT lines always separate odd-parity TS 
from even-parity trivial phases, a vanishing gap and thus QPT lines can 
also occur between regions of the phase diagram sharing similar 
topological features in 2D.

\section{Response to time-reversal symmetry breaking perturbations}
\label{break}

Since the Hamiltonian $H_D$ preserves TR symmetry, the gapless nature
of the Kramers' pair(s) of helical Majorana modes in a TS phase is
protected against perturbations that preserve TR symmetry.  It is
nevertheless interesting to explore how such robustness properties are
modified when TR symmetry is broken in different ways,
and the extent to which these changes are reflected in the bulk
topological indicators.  As a result of explicitly breaking TR
symmetry (thus bringing the system to class D, according to
Ref. \onlinecite{Altland}), we expect that different topological
invariants are needed to characterize the topological phases, as well
as a different formulation of the bulk-boundary correspondence.
Specifically, as we shall see in what follows, ${\mathbb Z}_2$
topological invariants may now be constructed from the parity of the
{\em full} (rather than partial) Chern or Berry phase sum, with the
bulk-boundary correspondence requiring that an odd (even) value of the
latter also implies an {\em odd (even) number of Majorana edge modes (as
opposed to Majorana pairs) on each boundary}.

To the best of our knowledge, most investigations aiming to explore
the robustness of topological phases under TR-symmetry breaking have
involved a TI model as their starting point, see {\em e.g.}
Refs. \onlinecite{Liu,Yang,Goldman,Dahm}.  In this section, we shall focus
on studying our TS model first in the presence of a uniform (bulk)
Zeeman field along different directions, intended as an external
control parameter, and then in the presence of different kinds of
internal (uncontrollable) magnetic impurities. Next, we will proceed
to quantitatively investigate the effect of TR-symmetry breaking by
moving away from the limit of exactly $\pi$-shifted superconducting
gaps [Eq. (\ref{pi})], and again reconsider the effect of and interplay with 
an applied Zeeman field.  While we shall primarily address a 2D geometry, 
in the light of our analysis in Sec.  \ref{RD}  and as it will become
clear through the discussion, the main features emerging for the
2D case will remain qualitatively valid with some natural modifications 
for 1D and 3D as well.

\subsection{Majorana modes under a static magnetic field}

Throughout this section, we shall continue to assume that $\Delta_c
=-\Delta_d$, and consider a total Hamiltonian of the form $H \equiv H_D +
H_M$, where $H_D$ is given in Eq. (\ref{Ham}) and 
\begin{eqnarray}
H_M= \sum_{\nu = x,y,z}   H_M^{(\nu)}, \;\;\; 
H_M^{(\nu)}=\sum_j h_\nu^{(j)} \psi_j^\dag \sigma_\nu 
\psi_j . 
\label{HM}
\end{eqnarray}
Here, $h_\nu^{(j)}$ represents the strength of the magnetic field/impurity along the
$\nu$ direction at site $j$, and the sum extends over all lattice
sites or over boundary sites only, depending on whether a bulk or
boundary field is considered.
%$\nu=z (x,y)$ corresponds to longitudinal (transverse) magnetic field, 
%and $h_\nu^{(j)}$ is the strength of the Zeeman field 
%% LV: at this stage it can be an impurity no?
%% Shusa: Yes, I added impurity.
With reference to the spin (${z}$) quantization axis, we shall
refer to the ${z}$ $({x}, {y})$ as longitudinal
(transverse) directions, respectively.

\subsubsection{Effect of a uniform longitudinal magnetic field}
\label{uniformz} 

Let us begin by considering the response of the bulk to a uniform
${z}$-magnetic field, that is, $h_\nu^{(j)}= h_z\neq 0.$
Remarkably, an analytical solution for the full spectrum still exists for PBC 
by employing the diagonalization procedure 
described in Sec.~\ref{model}, thanks to the fact that the SO coupling 
$H_{\sf so}$ has no component along ${z}$ (see also Appendix C). 
That is, the total Hamiltonian can still be rewritten as in Eq. (\ref{fourier}),
with $\hat{H}'_{\bf{k}}=\hat{H}'_{+,\bf{k}} \oplus
\hat{H}'_{-,\bf{k}}$ defined in Eqs.~(\ref{4by4p})-(\ref{4by4m}),
except that now we replace
$m_{\bf k} \mapsto 
m_{\pm ,\bf{k}}= m_{\bf{k}}\pm h_z,$
in the corresponding expression for $\hat{H}'_{\pm,\bf{k}}$.  With
this substitution, the excitation spectrum $\epsilon_{n,{\bf k},+}$
($\epsilon_{n,{\bf k},-}$) obtained from diagonalizing
$\hat{H}'_{+,\bf{k}}$ ($\hat{H}'_{-,\bf{k}}$) is formally still given
by Eq.~(\ref{spectrum2D}).
Thus, it is clear that the effect of the longitudinal Zeeman field is
to formally replace 
$u_{cd}  \mapsto u_{cd}\pm h_z$
for $\hat{H}'_{\pm,\bf{k}}$,
respectively.  QPTs occur when the excitation gap closes, that is,
when either $\epsilon_{2,{\bf k},+}=0$ or $\epsilon_{2,{\bf k},-}=0$
(for general $\Delta \ne 0$), which determines the QPT lines as
$m_{\pm,{\bf k}_c}=\pm \,\Omega$.

Since TR symmetry is broken, the full sum of the CNs over the two
occupied negative bands of {\em both} $\hat{H}'_{+,\bf{k}}$
and $\hat{H}'_{-,\bf{k}}$ need no longer be
zero. Thus, we can use the parity of this full Chern sum,
$\tilde{P}_C$, as a ${\mathbb Z}_2$ invariant.  However, in order to
make a comparison between {\em partial} Chern sums with and without
magnetic field, we still calculate, following Eq.~(\ref{cn}), the
partial Chern sums ($C_{+,\pm}$) of the two occupied negative bands of
$\hat{H}'_{\pm,\bf{k}}$ separately (say, $C_{1,\pm}$ and $C_{2,\pm}$)
and construct the parity invariant as follows:
\begin{eqnarray*}
\label{invariant3}
\tilde{P}_C \equiv (-1)^{{\rm mod}_2(C_{+,+}+C_{+,-})}, \;\;\;
C_{+,\pm} \equiv C_{1,\pm}+C_{2,\pm},
\end{eqnarray*}
in such a way that $C_{+,+} = - C_{+,-} =C_+$ [given by
Eq. (\ref{invariant1})] in the absence of magnetic field.

\begin{figure}[t]
\includegraphics[width=8.5cm]{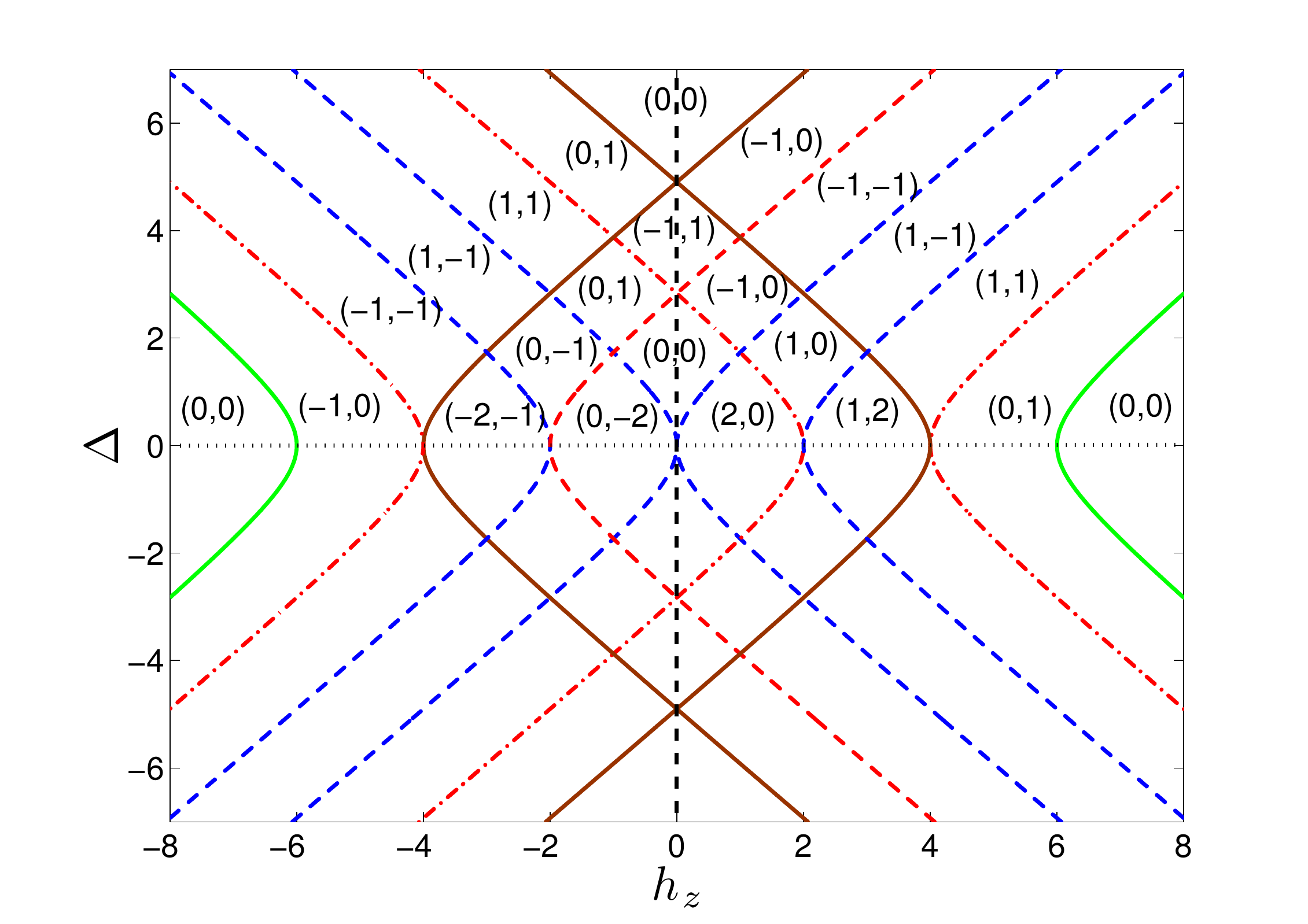}
\vspace*{-3mm} \caption{\label{phasedh} (Color online) Topological
phase diagram of the 2D Hamiltonian $H_D+H_M$, with $H_M \equiv
H_M^{(z)}$ being a longitudinal Zeeman field with strength $h_z$.  The
remaining parameters are $t=1$, $u_{cd}=1$, $\mu=-1$, for arbitrary
$\lambda \ne 0$.  The topological response is characterized via the
partial CN sum $(C_{+,+},C_{+,-})$ from the occupied bands of
$\hat{H}'_{+,\bf{k}}$ and $\hat{H}'_{-,\bf{k}}$, respectively.  The
horizontal black dotted-line has the same meaning of the zero-field
case.  The vertical black dotted-line indicates that a different
classification of phases applies along the TR-invariant line $h_z=0$,
see text. }
\end{figure}

The    resulting   topological   phase    structure   is    shown   in
Fig.~\ref{phasedh}, where the two  numbers reported in parentheses are
the  two partial  Chern sums,  $C_{+,+}$ and  $C_{+,-}$, respectively.
Two remarks  are in order. First, despite the  fact that, as  we have
verified,  {\em  no}  singular  behavior  of the  derivatives  of  the
many-body  ground-state  energy  with  respect to  $h_z$  develops  at
$h_z=0$, one should treat the phases along the TR-invariant line
$h_z=0$ as being different from  those in the TR-broken region $h_z\ne
0$. This  is because the  process of adiabatic connection  between two
regions in parameter space must  be carried out {\em without} changing
the basic symmetry class in order for these regions to be meaningfully
thought as belonging to one and the same topological phase \cite{Ryu}.
Second,  in a  real physical  system, an  excessively  strong magnetic
field  can destroy  superconductivity.  Thus,  we have  also performed
self-consistent calculations,  along the lines of Appendix \ref{self}, in
order to have at least some indication on the degree of stability of 
different phases  in Fig.~\ref{phasedh}.  Our numerical results suggest 
that the required range of $\Delta$ remains accessible in the presence of 
magnetic field for all the phases not too far from the center of the phase 
diagram, labeled with $(C_{+,+},C_{+,-})=(2,0),(0,-2),(0,0), (-1,0), (0,1), (-1,1)$. 
Specifically, in order to get a rough order-of-magnitude estimate, let us 
assume a material with narrow bandwidth, say between $10\,$meV $\sim$ $100\,$meV, 
indicating that the tight-binding coupling strength $t$ is in the range $2\,$meV $\sim$ 
$20\,$meV.  The value $h_z=0.1$ may then correspond to a field strength 
between $0.2\,$meV $\sim$ $2\,$meV (2 $\sim$ 20 T, respectively). 
Since $h_z$ can be arbitrarily close to zero in the above-mentioned phases 
however, the corresponding magnetic field strengths can safely be 
below the critical field strength in superconductors. 

Similar to the case when TR symmetry is conserved, we may still
establish a direct connection between the invariant $\tilde{P}_C$
defined above and the {\em full} fermion number parity of the
TR-invariant modes $\tilde{P}_F$.  Following the same procedure outlined in
Sec.~\ref{indicators}, we find that when $|m_{+,{\bf k}_c}| >
|\Omega|$, the ground state of each mode ${\bf k}_c$ in
$\hat{H}'_{+,\bf{k}}$ belongs to the sector with odd fermion parity,
that is, $P_{+,{\bf k}_c}=e^{i \pi (a_{{\bf k}_c,\uparrow}^\dag
a_{{\bf k}_c,\uparrow}^{\;}+b_{{\bf k}_c,\downarrow}^\dag b_{{\bf
k}_c,\downarrow}^{\;})}=-1$, otherwise it is in the sector with even
fermion parity $P_{+,{\bf k}_c}=1$.  Similar results for the fermion
parity $P_{-,{\bf k}_c}$ of the ground state of each mode ${\bf k}_c$
in $\hat{H}'_{-,\bf{k}}$ are obtained by analyzing the relation
between $|m_{-,{\bf k}_c}|$ and $|\Omega|$. As a result of breaking TR
symmetry, however, $P_{-,{\bf k}_c}$ {\em need not} be equal to $P_{+,{\bf
k}_c}$, just like $C_{+,-}$ is not necessarily the opposite to
$C_{+,+}$. Nonetheless, let us define, in analogy to Eq. (\ref{fermion}), 
$$\tilde{P}_F \equiv \prod_{ {\bf k}_c} P_{+,{\bf k}_c} P_{-,{\bf k}_c}.$$
\noindent 
Then, by analyzing the relation between $|m_{\pm,{\bf k}_c}|$ and
$|\Delta|$ for each ${\bf k}_c$, we can see that the TS (trivial)
phases with $\tilde{P}_C=-1(1)$ correspond to the ground state with $
\tilde{P}_F=-1(1)$, as anticipated.  In contrast to the TR-invariant
case, note that it is  {\em necessary} to take into account the CNs or
fermion parity of the occupied bands in both $\hat{H}'_{\pm,\bf{k}}$,
in order for this correspondence to hold.

\begin{figure}[t] 
\includegraphics[width=5.8cm]{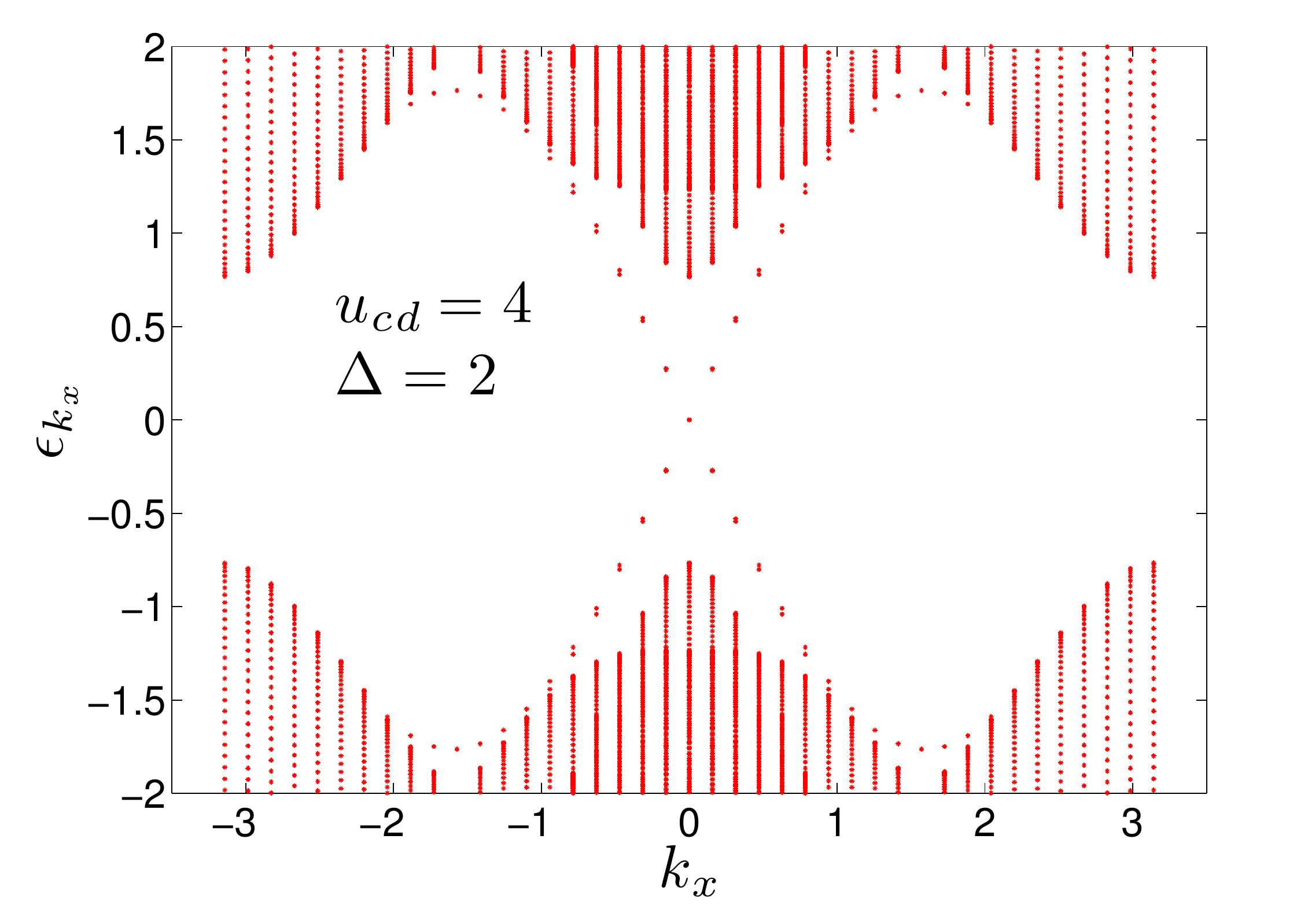}
\includegraphics[width=5.9cm]{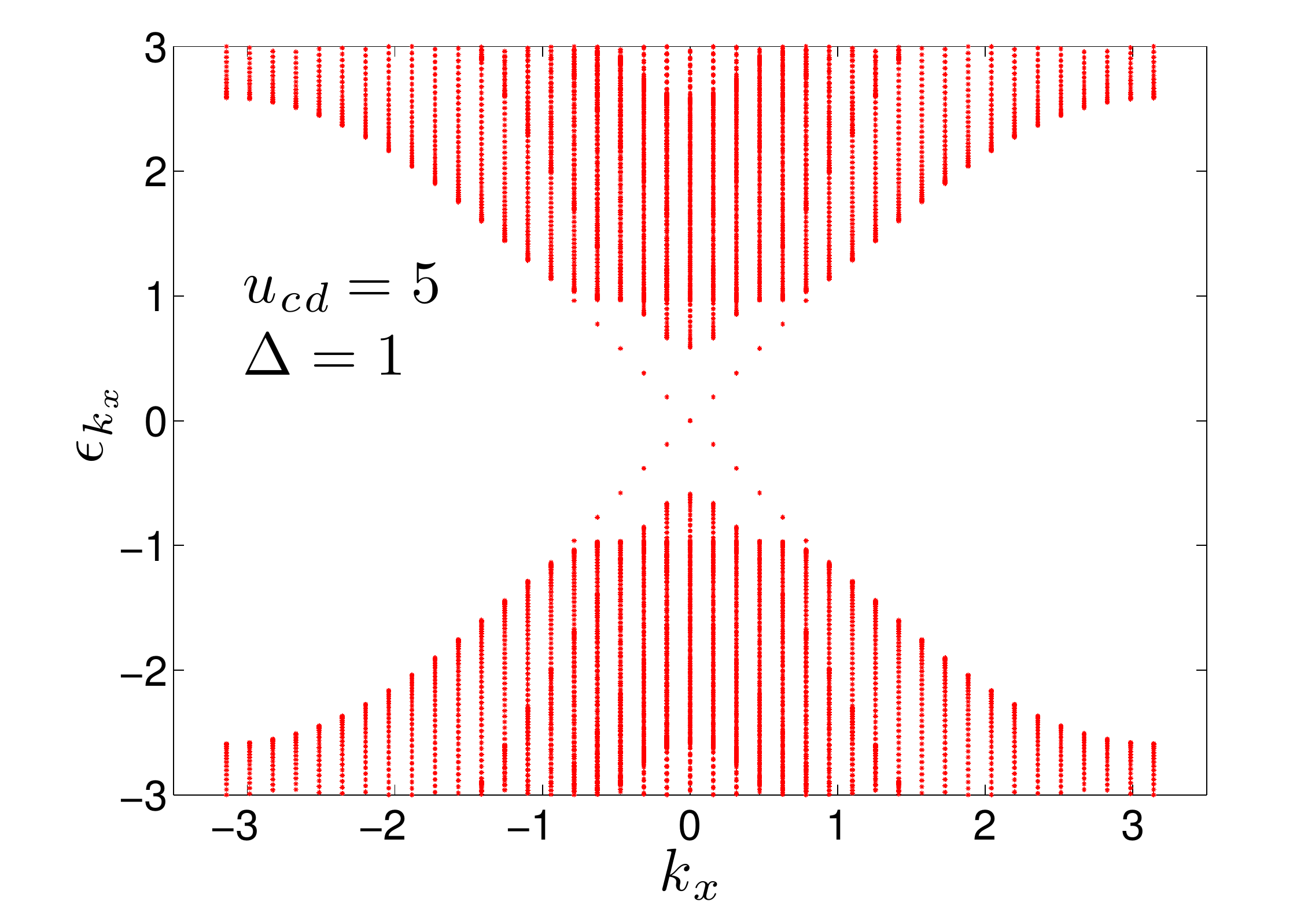}
\vspace*{-2mm}\caption{\label{tso} (Color online) Excitation spectrum
of the 2D Hamiltonian $H_D+H_M$ on a cylinder, with $H_M=H_M^{(z)}$ being
a longitudinal Zeeman field with strength $h_z=1$, and $\mu=-1, t=1=
\lambda$.  Top panel: $u_{cd}=4,\Delta=2$, $\tilde{P}_C=1$. Two
(helical) Majorana edge states exist on each boundary, corresponding
to a trivial phase. Bottom panel: $u_{cd}=5,\Delta=1$,
$\tilde{P}_C=-1$. One (chiral) Majorana edge state exist on each
boundary, corresponding to a TS phase. System size: $(N_x,
N_y)=(40,100)$. }
\end{figure}

Let us now focus on exploring the effect of the magnetic field on the
Majorana edge states.  To do so, as before we study the 2D Hamiltonian
$H=H_D+H_M$ on a cylinder (PBC along ${x}$, and OBC along ${y}$), 
with the corresponding excitation spectrum, $\epsilon_{n,k_x}$, obtained by
applying a Fourier transformation in the ${x}$-direction only.  In
order to demonstrate the bulk-boundary correspondence when TR symmetry
is broken, that is, the correspondence between a bulk with
$\tilde{P}_C=1(-1)$ and the presence of an even (odd) number of
Majorana edge modes, we show in Fig.~\ref{tso} two representative
cases with $\tilde{P}_C=1(-1)$ respectively.  Specifically, for even
$\tilde{P}_C$ [top panel, $C_{+,+}=1,C_{+,-}=-1$], two helical
Majorana edge states exist on each boundary, whereas for odd
$\tilde{P}_C$ [bottom panel, $C_{+,+}=0, C_{+,-}=-1$], only one chiral
Majorana edge state on each boundary, which is consistent with the
above bulk-boundary correspondence. Accordingly, the top (bottom)
panel of Fig.~\ref{tso} corresponds to a trivial (non-trivial)
topological phase. 

\begin{figure}[t]
\includegraphics[width=5.8cm]{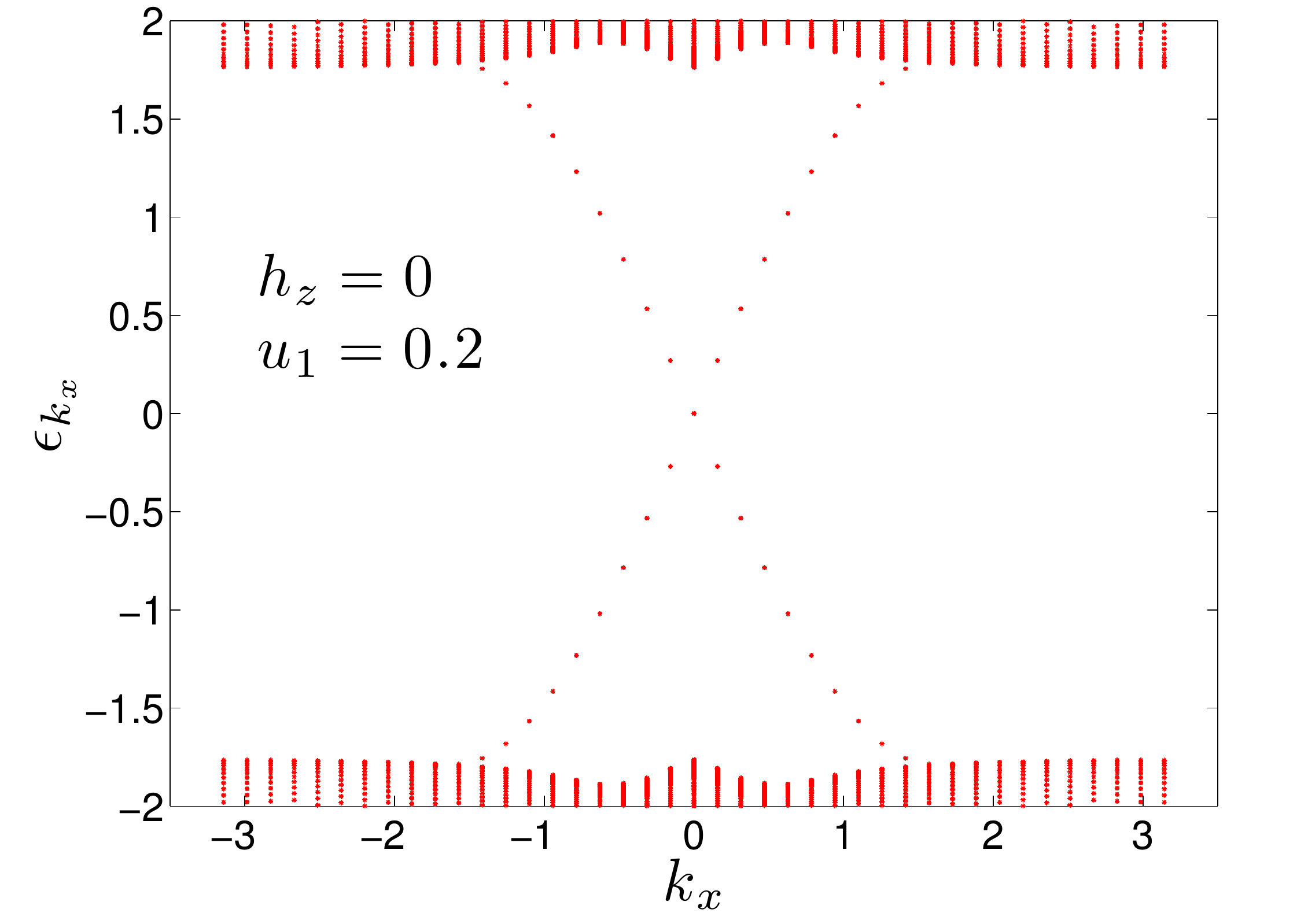}
\includegraphics[width=5.9cm]{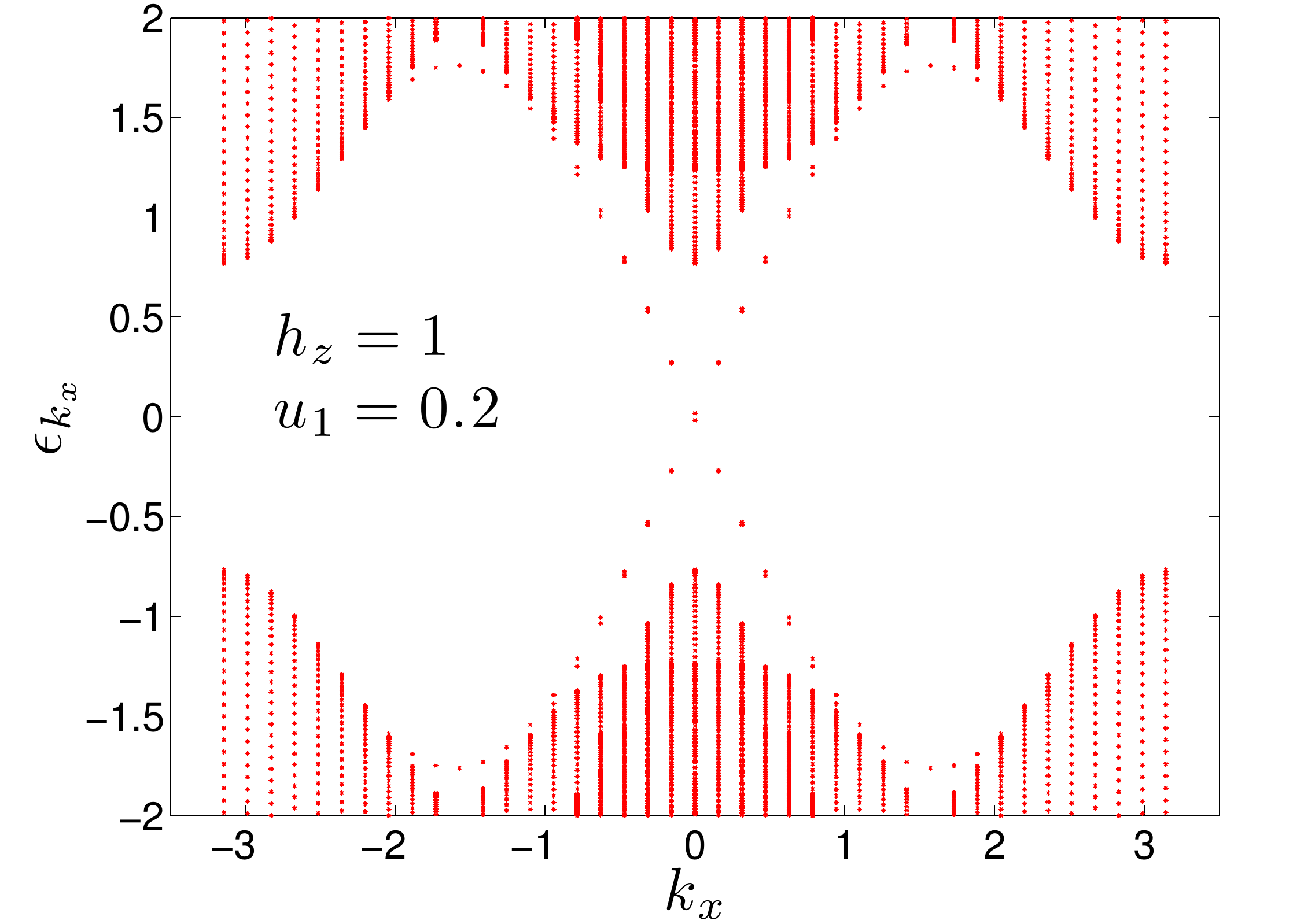}
\vspace*{-2mm}\caption{\label{backs} (Color online) Excitation
spectrum of the 2D Hamiltonian $H_D+H_M+H_b$ on a cylinder, with
$H_M=H_M^{(z)}$ being a longitudinal Zeeman field with strength $h_z$
and $H_b$ a backscattering potential with strength $u_1=0.2$
respectively, for $\mu=-1, t=1= \lambda, u_{cd}=4,\Delta=2$.  Top
panel: TR-invariant case, $h_z=0$. The two Majorana modes on each edge
form a Kramers' pair and remain gapless under $H_b$.  Bottom panel:
TR-broken case, $h_z=1$.  The two Majorana modes on each edge form a
quasi-TR pair, note the appearence of a gap in the edge spectrum.
System size: $(N_x, N_y)=(40,100)$. }
\end{figure}

Interestingly, if the magnetic field is turned off while keeping all
the other control parameters unchanged in the trivial phase corresponding
to the top panel, the same partial Chern sums $C_{+,+}=1,C_{+,-}=-1$
still hold in the limit $h_z\rightarrow 0$.  These values, however,
correspond now to a {\em non}-trivial TR-invariant TS since the
$\mathbb{Z}_2$ invariant is characterized by the (odd) parity of the
{\em partial} Chern sum when TR symmetry is preserved, rather than the
parity of the Chern sum for {\em all} the occupied bands. This
illustrates how the same topological number may in fact correspond to
completely different phases. The key point here is that {\em the
topological invariant changes when the basic symmetry class 
changes}, as a consequence of TR symmetry being broken. 
Although, as already remarked, no singular behavior develops at 
$h_z=0$, a topological QPT still takes place in this sense, solely signaled 
by the change of the underlying topological invariant.

The observation that the two situations for $h_z=0$ and $h_z \ne 0$ in
the above discussion correspond to non-trivial and trivial topological
phases, respectively, may be explicitly confirmed by investigating the
response of the gapless edge states under the effect of a
TR-preserving back-scattering interaction.  For instance, let us
consider the following boundary perturbation:
\begin{eqnarray*}
H_b=i \sum_{k_x,j} u_1^{(j)} (c_{k_x,j,\uparrow}^\dag
d_{k_x,j,\uparrow}-c_{k_x,j,\downarrow}^\dag
d_{k_x,j,\downarrow})+{\text H.c.},
\end{eqnarray*} 
where $u_1^{(j)}=u_1\ne0$ for $j=1$ or $j=N_y$, and $u_1^{(j)}=0$ otherwise.
Let us also introduce the terminology of a ``quasi-TR-pair'' to refer
to two gapless Majorana edge modes when (i) TR symmetry is broken,
yet, (ii) the relationship $C_{+,+}=-C_{+,-}$ still holds for the
corresponding bulk. A comparison between the robustness of a Kramers'
pair and a quasi-TR-pair of Majorana modes against $H_b$ is shown in
Fig.~\ref{backs}. As one can see, the Kramers' pairs of Majorana modes
are robust (top panel), whereas the quasi-TR pairs are not, in the
sense that they become gapped, with a gap scaling linearly in $u_1$
(bottom panel). Thus, gapless TR-pairs of edge modes of the zero-field
Hamiltonian $H_D$ {\em may remain gapless in the presence of a
magnetic field.  However, their degree of robustness against
subsequent TR-preserving perturbations is, in general, different} 
as compared to the original Kramers' pairs. It is worth noting that a
similar behavior was also reported recently in the context of a
TR-invariant quantum spin Hall system \cite{Yang}, where a pair of
gapless edge states was found to remain gapless when an external
magnetic field was added. These gapless edge states would however
become gapped in the presence of backscattering, indicating a
low-dissipation (but not dissipation-less) spin transport.

Unlike in other models where a Zeeman magnetic field is required for
the very existence of Majorana modes (see, for instance, in
Refs.~\onlinecite{Sato, Kub}), we iterate that this is clearly {\em not} the
case in our multi-band system.  Rather, the Zeeman field may be viewed
as a control knob for potentially tuning the emergence/disappearance
of Majorana modes.  For instance, with reference to the phase diagram
in Fig.~\ref{phasedh}, imagine that for a fixed value of $\Delta$, say
$\Delta=3.2$, the magnetic field $h_z$ is increased from $h_z=0$ to
$h_z=8$.  Then it turns out that the number of 
Majorana edge modes changes in the following way: $2$ (one TR-pair, as
$h_z=0$) $\rightarrow 2$ (one quasi-TR-pair)$ \rightarrow 3\rightarrow
2 \rightarrow 1$, which effectively turns the original strong TS into
a weak/trivial one depending on the applied field strength. Likewise,
if $\Delta$ is instead fixed at, say, $\Delta=1$, a non-zero $h_z$ can
turn a weak TS into to a non-trivial topological phase with only one
robust Majorana edge mode on each boundary.  The usefulness of a
Zeeman field as a tuning mechanism in the presence of an additional
TR-breaking perturbation will also be further discussed in
Sec. \ref{away}.

%%%%%%%%%%%%%%%%%%%%%%%%%%%%%%%%%%%%%%%%%%%%%

\subsubsection{Effect of uniform transverse magnetic fields} 

\begin{figure}[t]
\centering \includegraphics[width=6.0cm]{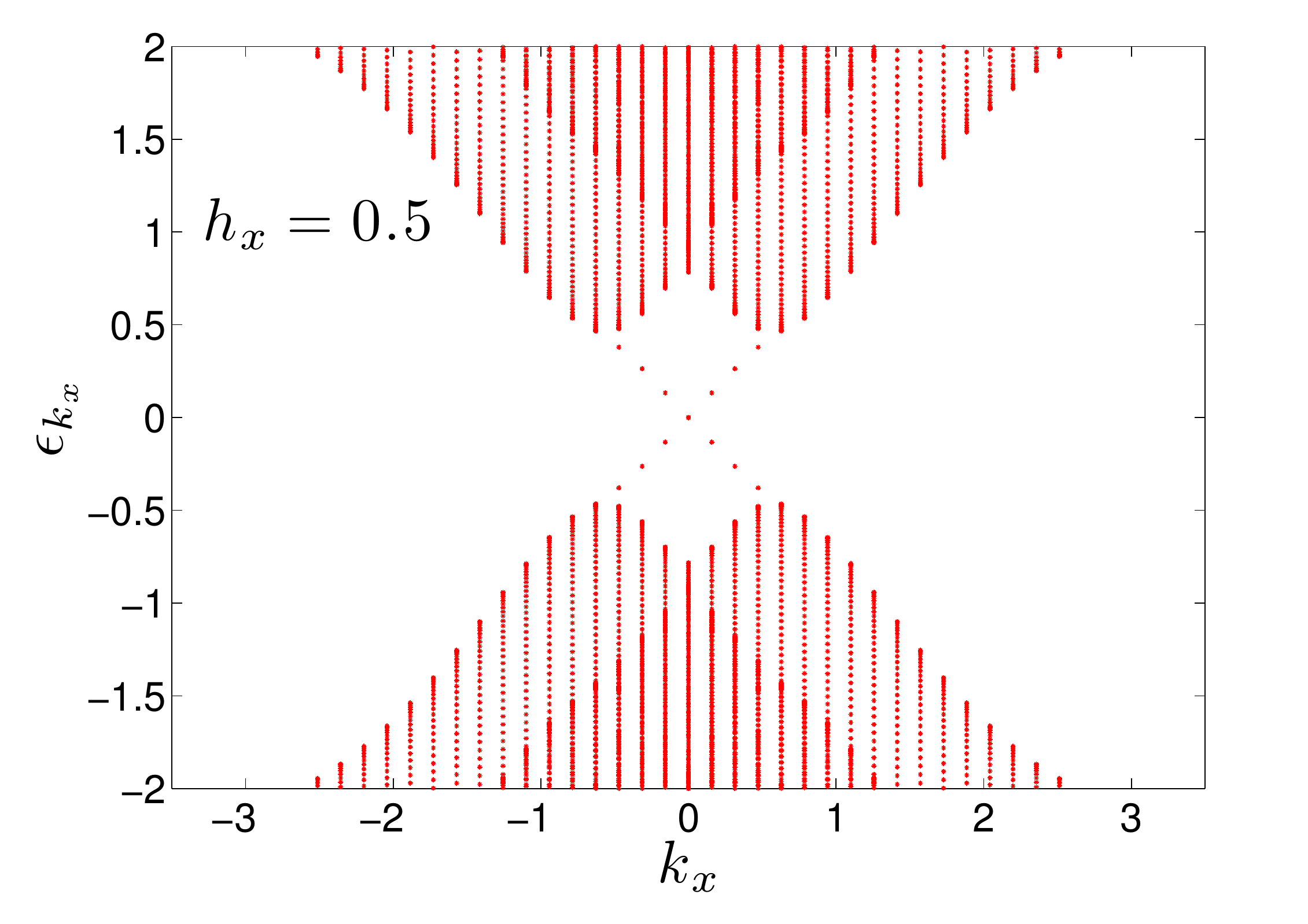}
\includegraphics[width=6.0cm]{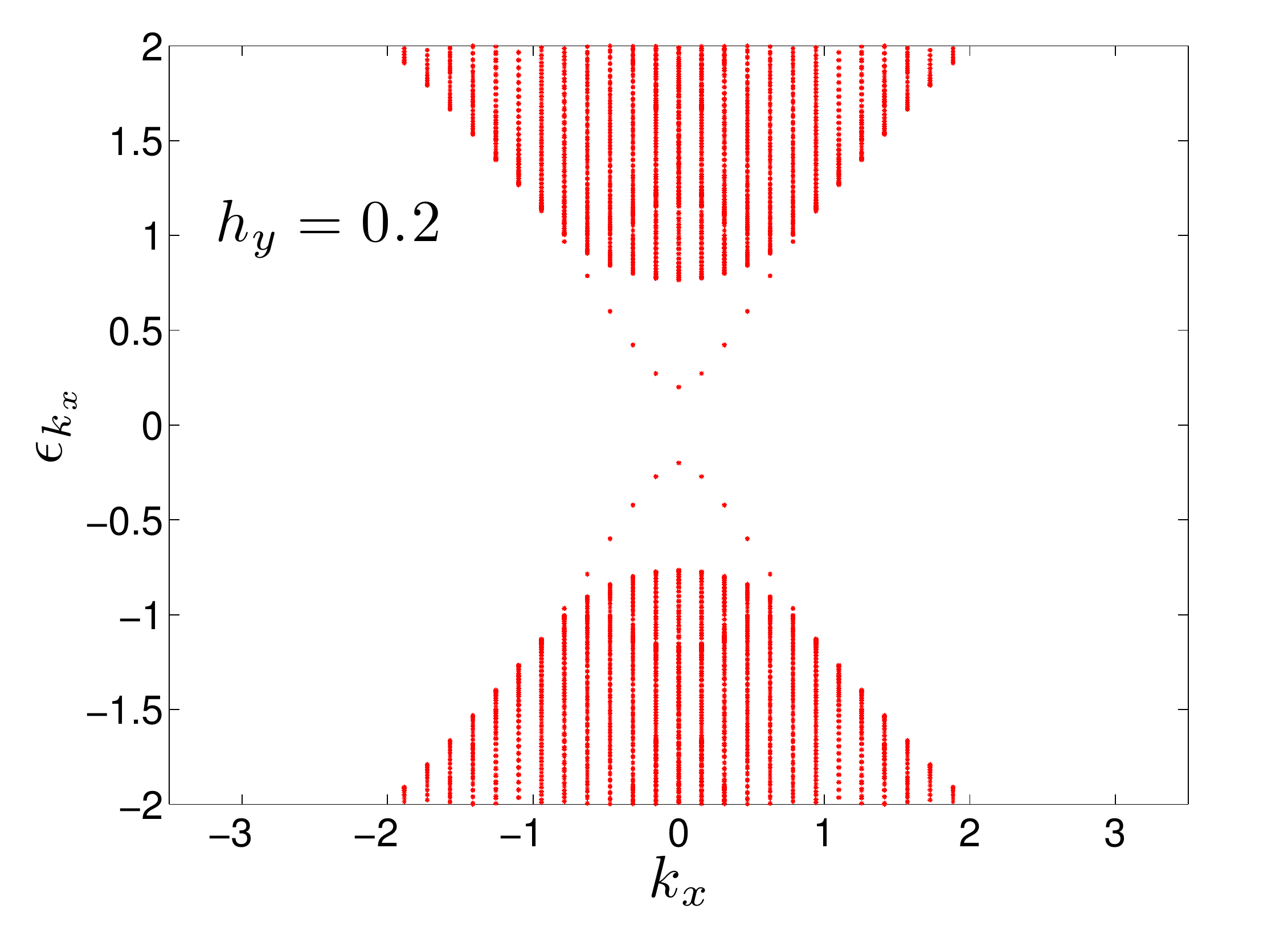}
\vspace*{-2mm}
\caption{\label{xy} (Color online) Excitation spectrum of the 2D
Hamiltonian $H_D+H_M$ on a cylinder, with $H_M=H_M^{(\nu)}$ being a uniform
transverse field, for $\mu=-1, t=1=\lambda,u_{cd}=4,\Delta=1$. Top
panel: ${x}$-field, with strength $h_x=0.5$. The edge spectrum
remains gapless.  Bottom panel: ${y}$-field, with strength
$h_y=0.2$. The edge spectrum becomes gapped.  System size: $(N_x,
N_y)=(40,100)$. }
\end{figure}

Let us now briefly consider the case where a Zeeman field is instead
applied in a transverse (${x}$ or ${y}$) direction, with focus
on the changes induced in the edge spectrum.  Specifically, imagine
first that the magnetic field acts along ${x}$,
that is, $h_\nu^{(j)}=h_x$ in Eq.~(\ref{HM}). In the presence of such
a field, the total Hamiltonian $H_D+H_M$ on a cylinder can no longer
be decoupled into two $4N_y\times4N_y$ matrices for each momentum mode
$k_x$  (see Appendix C). We can nevertheless obtain physical insight by
examining specific cases.  Imagine, in particular, that the system is
in a TS phase when $h_x=0$, say corresponding to $\mu=-1$, $u_{cd}=4,
\Delta=1$, with reference to the phase diagram in Fig. \ref{pd} [top right panel], 
and imagine that we still express the energy eigenvectors in
the basis of canonical fermion annihilation operators $a_\sigma$ and
$b_\sigma$ defined in Eq.~(\ref{canonicalf}) when $h_x=0$. Then we may
represent the two TR-invariant Majorana edge modes $(\gamma_1,
\gamma_2)$, $\gamma_2={\cal T}\gamma_1 {\cal T}^{-1}$, that exist for
$k_x=0$ on each boundary in the form
\begin{eqnarray}
\left \{ \begin{array}{c} \gamma_1 =\sum_{n=1}^{N_y} (\alpha_n
a_{n,\uparrow}^\dag+\beta_n b_{n,\downarrow}^\dag+ \text{H.c.}), \\
\gamma_{2}=\sum_{n=1}^{N_y} (\alpha_n a_{n,\downarrow}^\dag-\beta_n
b_{n,\uparrow}^\dag+ \text{H.c.}),
\end{array} \right. 
\label{gammacn3}
\end{eqnarray}
for suitable real coefficients [cf. Eq. (\ref{gammacn2})].
Direct calculation shows that the matrix
element with respect to the many-body ground state vanishes:
\begin{equation}
\langle\Psi_{\sf gs}|\gamma_2^\dag H_M^{(x)} \gamma_1 |\Psi_{\sf
gs}\rangle=0.
\label{v1}
\end{equation}
Similarly, one also finds that:
\begin{equation}
\langle\Psi_{\sf gs}|\gamma_2^\dag H_M^{(y)} \gamma_1 |\Psi_{\sf
gs}\rangle\ne0.
\label{v2}
\end{equation}
Thus, according to degenerate perturbation theory, a field in the
${x}$ direction cannot lift the degeneracy of the gapless Majorana
edge modes supported by $H_D$, whereas in general a field along the
${y}$ direction does.  These conclusions have been confirmed by
explicit numerical calculation, with illustrative results shown
in Fig. \ref{xy}. 

It is worth noting that the different roles that magnetic fields in the ${x}$ 
vs. ${y}$ direction play in our model is ultimately a consequence of the different 
boundary conditions imposed in these directions, PBC (OBC) along ${x}$ 
(${y}$) directions, respectively.  Should the latter be interchanged, then the 
effect of transverse perturbations along ${x}$ and ${y}$ would 
be as well.   Furthermore, even if the edge spectrum remains gapless
under $h_x\ne 0$, the edge modes no longer form Kramers' pairs.  Thus,
similarly to the behavior found in the presence of a $z$-field
[Fig. \ref{backs}], these modes are not expected in general to have
the same degree of robustness against disorder/backscattering as they
have in the TR-invariant case.

As a side remark, it is also interesting to observe that in the limit
$\Delta\rightarrow 0, \mu\rightarrow 0$, where the model Hamiltonian $H_D$ describes 
a TI, a perturbation in the $x$-direction does {\em not} lift the
degeneracy of the gapless edge modes either, despite the lack of Majorana
fermions.  As a result of PH symmetry, the two fermionic TR-invariant edge
modes at the Dirac point can now be written as ($h_x=0$):
\begin{eqnarray*}
\left \{ \begin{array}{c} \gamma_1=\sum_{n=1}^{N_y} \alpha _n
(a_{n,\uparrow}+ b_{n,\downarrow}),\\ \gamma_2=\sum_{n=1}^{N_y}
\alpha_n (a_{n,\downarrow}- b_{n,\uparrow}),
\end{array} \right.
\end{eqnarray*}
that is, the quasi-particles $a_{n,\uparrow} (a_{n,\downarrow})$ and
$b_{n,\downarrow} (b_{n,\uparrow})$ behave as if they were
particle-hole pairs. The possibility that edge states in a TI may remain 
robust despite TR-breaking was recently noted in a different context \cite{Dahm}.

\subsubsection{Effect of magnetic impurities} 

In reality, even in the absence of external perturbations, magnetic
fields are inevitably present due to various kinds of
impurities in the material.  Thus, it is important to get a sense
of what effect such magnetic fields will have on Majorana modes, a
main difference with respect to the uniform-field case being that
translational symmetry is now explicitly broken along one or more
spatial directions.  While more complex scenarios can be envisioned,
we limit ourselves here to impurity fields acting along
a single direction.  In particular, we consider a longitudinal
(${z}$) impurity field in a 2D geometry, and still assume 
PBC (OBC) in the ${x}$ (${y}$) direction, respectively. 

Suppose that the system is in a TS phase, with a pair of Majorana edge
states on each boundary. Two scenarios may be physically interesting:

(i) a {\em single} magnetic impurity on each boundary, in which case
we may let, for instance, $h_\nu^{(j)}=h_z \ne 0$, for $j\in \{(1,1),
(N_x,N_y)\}$, and $h_\nu^{(j)}=0$ otherwise in Eq.~(\ref{HM}); 

(ii) {\em random} magnetic impurities on each boundary, in which case
we may let, for instance, $h_z^{(j)}$ in Eq.~(\ref{HM}) to be
uniformly distributed random numbers in $[0,1]$.

\begin{figure}[t]
\centering
\includegraphics[width=7.0cm]{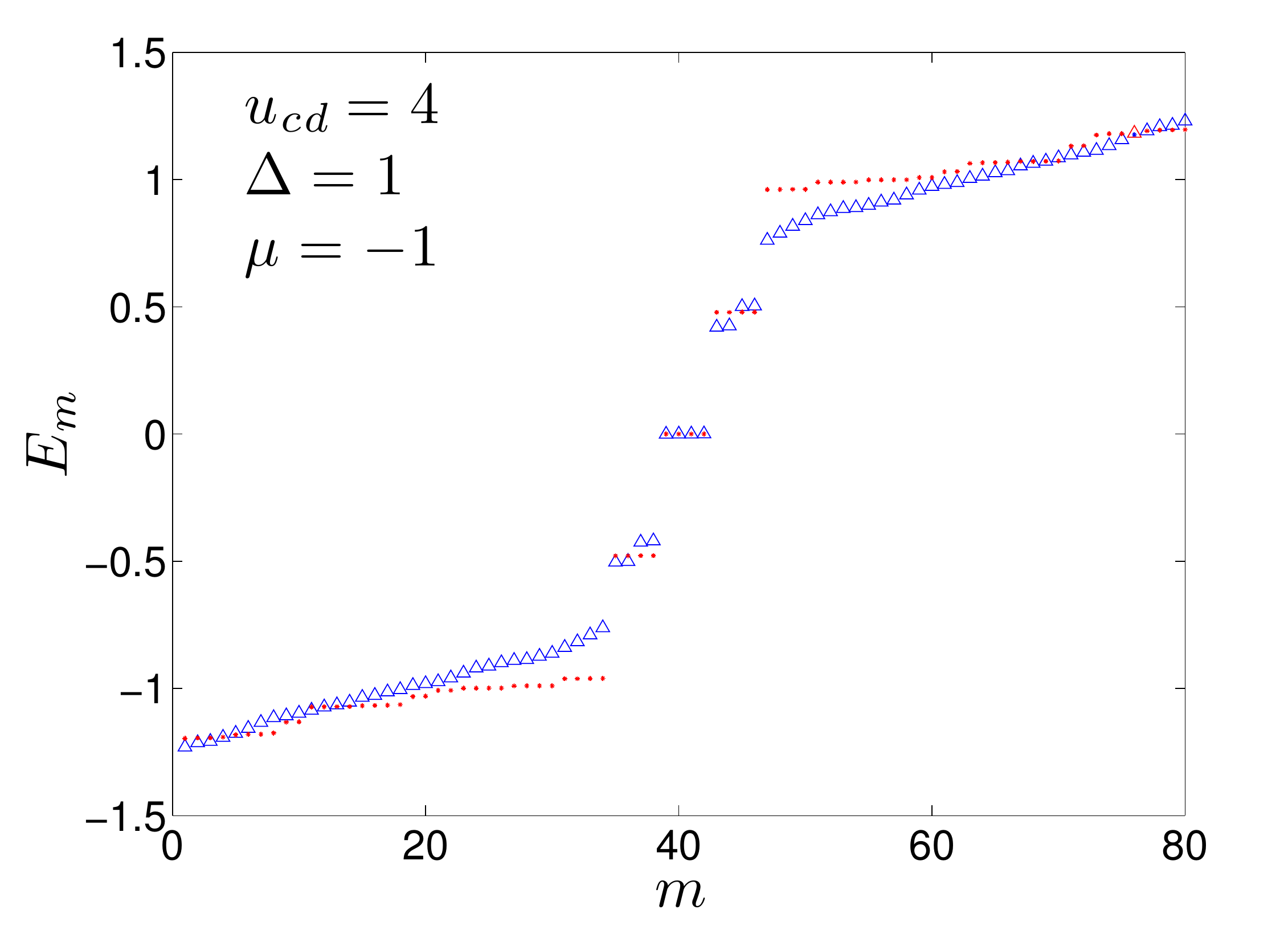}
\vspace*{-2mm}
\caption{\label{localh} (Color online) Single-particle excitation spectrum of the 2D
Hamiltonian $H_D+H_M$ on a cylinder, with $H_M =H_M^{(z)}$ representing
a longitudinal impurity magnetic field, for $\mu=-1,
t=1=\lambda,u_{cd}=4,\Delta=2$.  The blue triangle correspond to case
(i), with a single magnetic impurity with strength $h_z=0.5$ on each
boundary, whereas the red dots correspond to uniformly distributed random magnetic
impurities on each boundary (averaged over $10$ different realizations).  System
size: $(N_x,N_y)=(16,16)$. }
\end{figure}

As a result of explicitly breaking translational symmetry, 
the total Hamiltonian is now be written directly in real space in 
the form  
\begin{eqnarray*}
H_D+H_M=\sum_{i,j} \left (
\psi^\dag_i,\psi^T_i \right) \hat{H}_{i,j} 
\left ( \hspace*{-1mm} \begin{array}{c} \psi_j \\  
(\psi^\dag_j)^T
\end{array} \hspace*{-1mm} \right ), 
\label{real}
\end{eqnarray*}
for a suitable matrix $\hat{H}_{i,j}$.
Numerical results obtained by diagonalizing $H_D+H_M$
in the single-particle sector $\hat{H}_{i,j} $
in the two cases are shown in Fig.~\ref{localh}, where the label 
$E_m$ corresponds to the $m$th 
single-particle eigenvalue  and only the energy eigenvalues 
near zero are displayed. Despite TR symmetry being broken, 
gapless edge modes may still be inferred to persist in both
cases in the thermodynamic limit: in case (i), the minimum gap in the
edge spectrum is about $10^{-8}$ ($10^{-12}$) for system size $16
\times 16$ ($24 \times 24$), and similarly in case (ii) such a minimum
gap is $10^{-8}$ ($10^{-11}$) for system size $16\times16$
($24\times24$), respectively.  Hence, the Majorana edge modes are
stable against the effect of either single or random boundary
perturbation along the ${z}$ direction.

Since, from our analysis for a uniform field [Eqs. (\ref{v1})-(\ref{v2})], we 
know that the edge spectrum of our Hamiltonian remains gapless under $h_{x}$
perturbations, we may additionally conclude that Majorana modes remain
gapless in the presence of magnetic impurities (single or random)
along ${x}$, whereas magnetic impurities along the ${y}$
direction will generally result in gapped edge modes.

Although, as anticipated, we have focused above on the 2D case, a similar approach 
may be employed to determine the magnetic-field response for the 1D and 3D models.  
The situation is straightforward in 1D: the exact phase diagram under a ${z}$-field can 
be determined as in Sec. \ref{uniformz} and, if either ${z}$ or ${y}$ impurity 
fields are present, edge modes remain gapless -- unlike for a ${x}$-field due to the 
OBC imposed along ${x}$.  In 3D, although ${z}$ is no longer a special direction 
for the zero-field Hamiltonian, we may still obtain the exact phase diagram under a longitudinal 
magnetic field if we fix $k_z = k_{z,c}$ as we also did in Sec.~\ref{model}. If we still impose OBC 
along ${y}$, then similar to the 2D case, gapless edge modes remain gapless for 
magnetic fields along the ${x}$ or ${z}$ directions, 
whereas they become gapped for a field along ${y}$.

\subsection{Majorana modes away from $\pi$-shifted gaps} 
\label{away}

Throughout our discussion so far, we have assumed that 
the superconducting pairing gaps are exactly $\pi$-shifted,
$\Delta_c=-\Delta_d$.  Both because this condition need not be 
(exactly) satisfied in practice, and in order to gain additional
insight on the role it plays, it is interesting to ask what happens if
it is relaxed, while still treating the gaps as tunable parameters.
For simplicity, let us examine separately the two main mechanisms by
which the equality $\Delta_c=-\Delta_d$ may break:

(i) the two gaps may be mismatched in \emph{amplitude}, that is,
$|\Delta_c| \ne |\Delta_d|$ but the corresponding phases still obey
$\theta_c-\theta_d=\pi$;

(ii) the two gaps may be mismatched in \emph{phase}, that is,
$|\Delta_c| =|\Delta_d|$ but $\theta_c-\theta_d \ne \pi$, say,
$\theta_c-\theta_d=\pi + \epsilon$.

\begin{figure}[b]
\includegraphics[width=4.2cm]{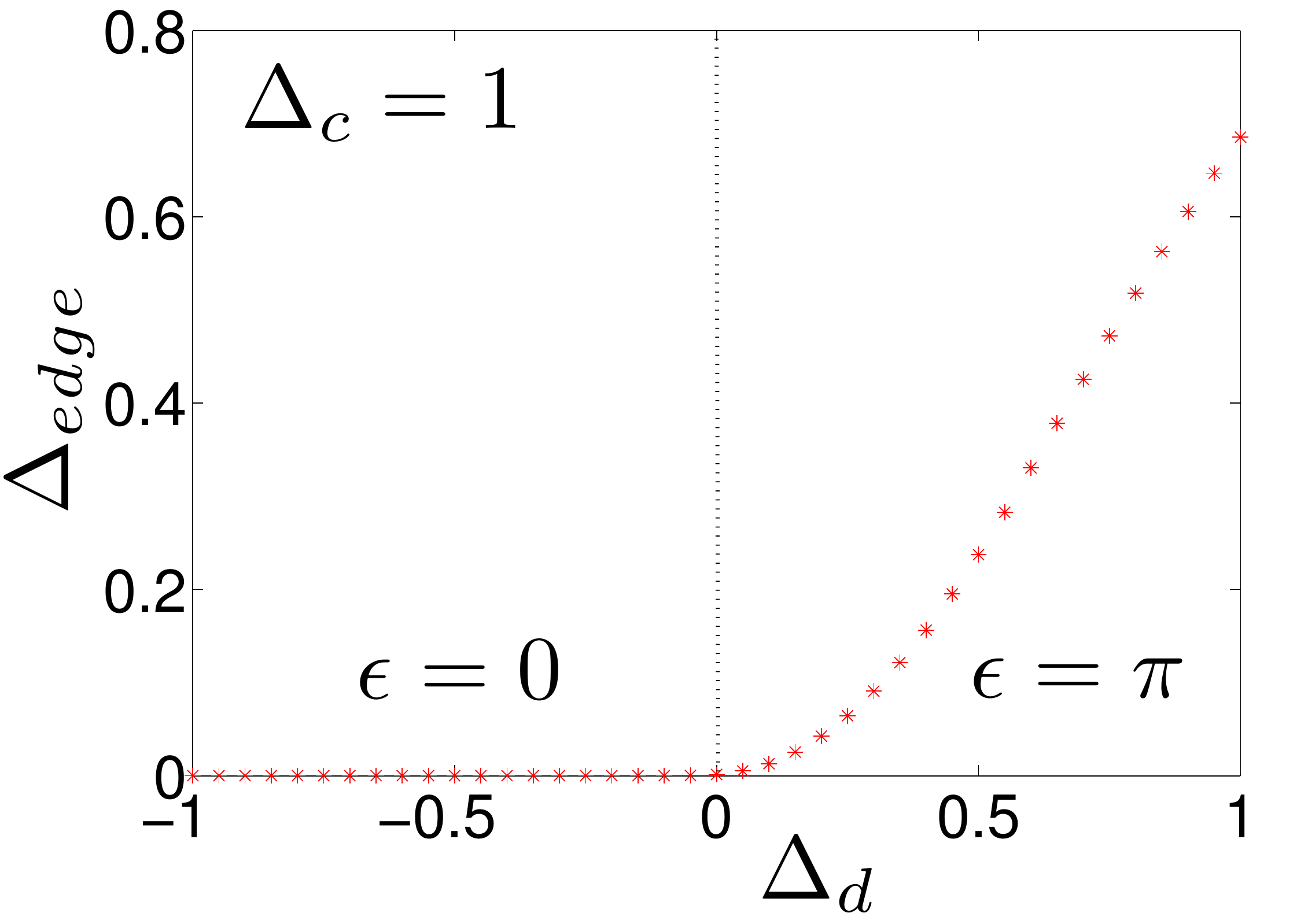}
\includegraphics[width=4.1cm]{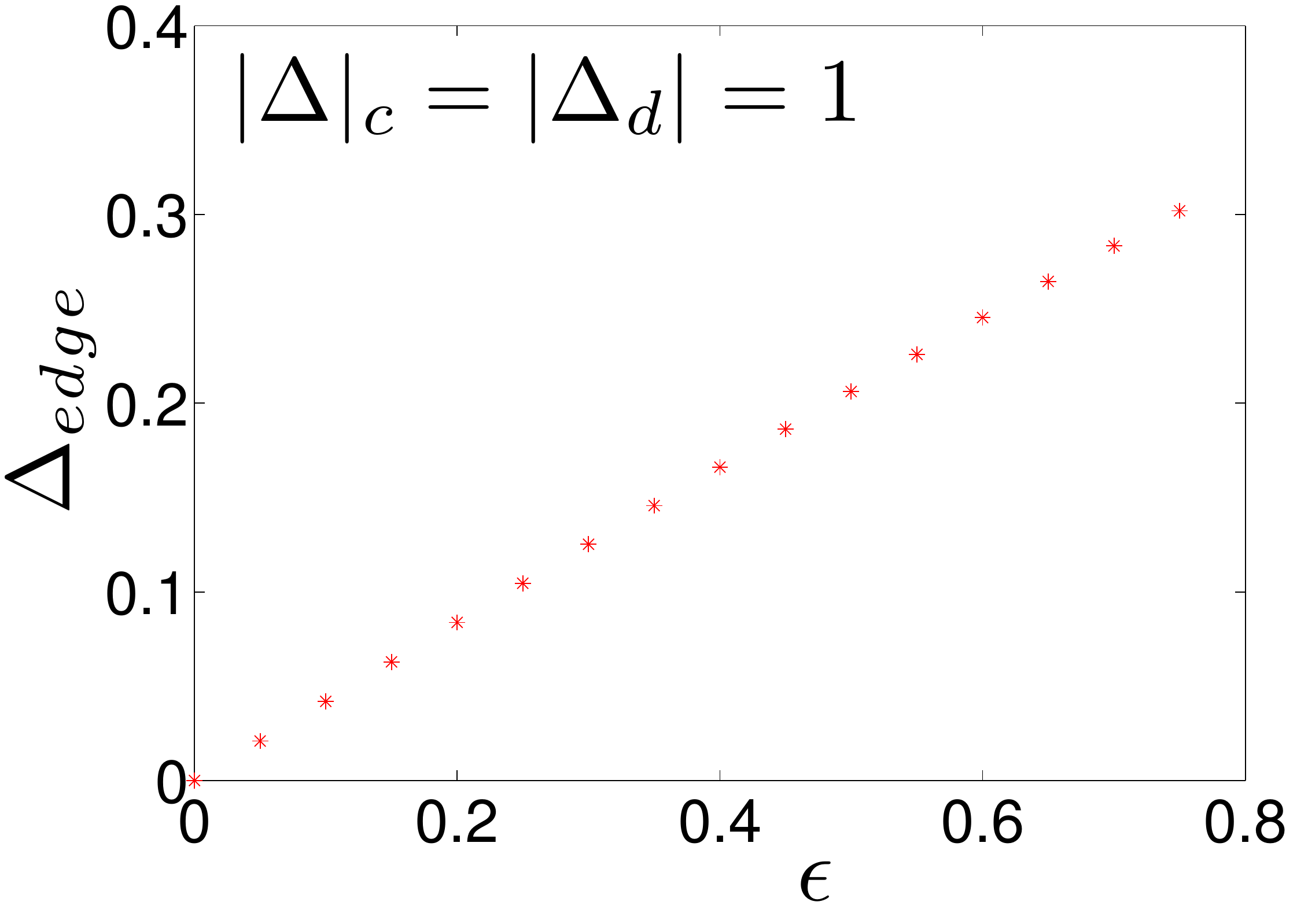}
\vspace*{-2mm}\caption{\label{mismatch1} (Color online) Minimum gap 
$\Delta_{edge}$ of
the edge spectrum of the 2D Hamiltonian $H_D$ on a cylinder, away from
the symmetry point $\Delta_c=-\Delta_d$.  Top panel: effect of a
TR-preserving amplitude mismatch in the absence ($\epsilon=0$) or in
the presence ($\epsilon =\pi$) of a concomitant phase mismatch.
Bottom panel: effect of a TR-breaking phase mismatch perturbation.
System size: $(N_x, N_y)=(40,100)$.  }
\end{figure}
\begin{figure}[t]
\centering
\includegraphics[width=6.2cm]{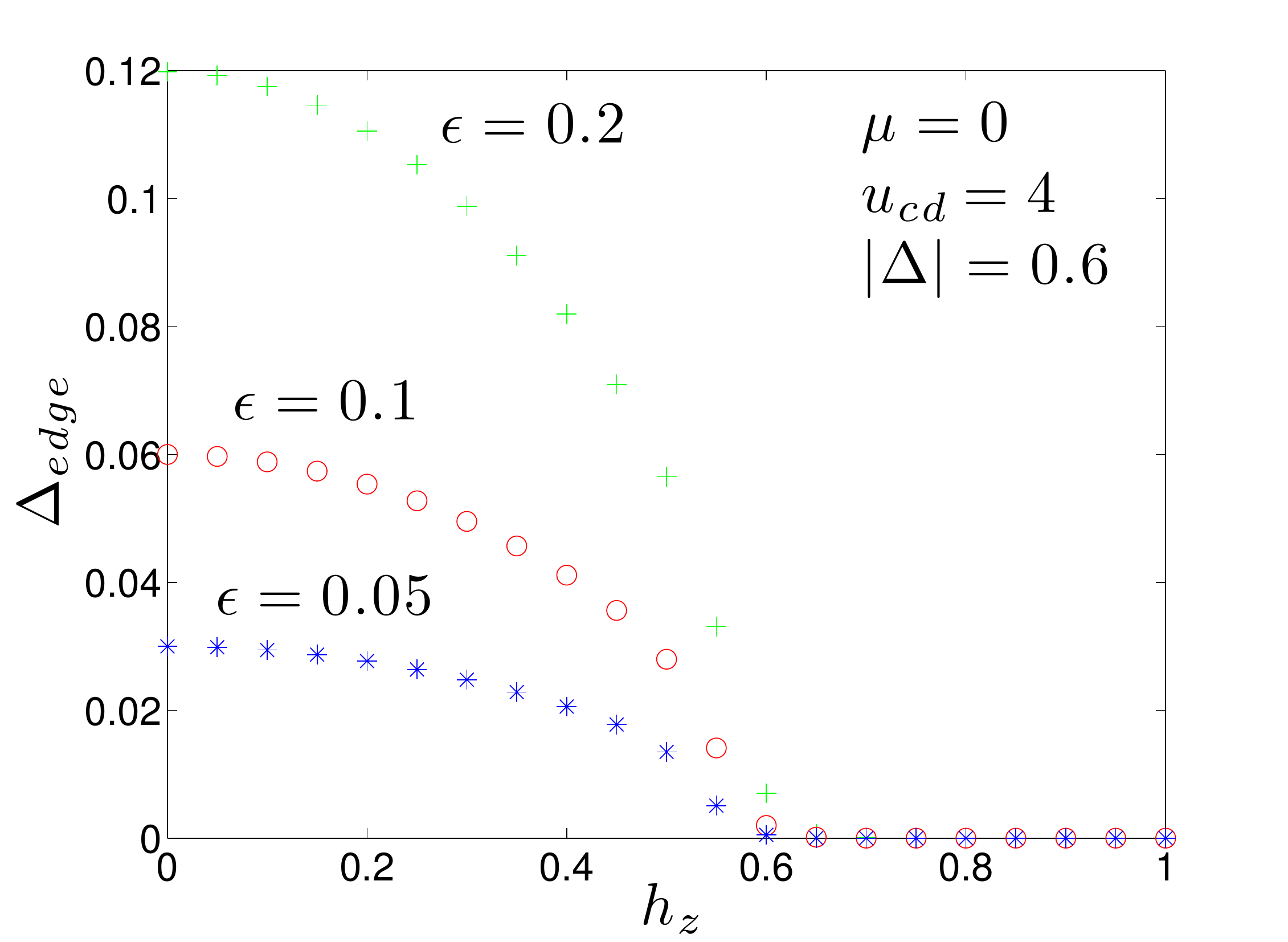}
\includegraphics[width=6.3cm]{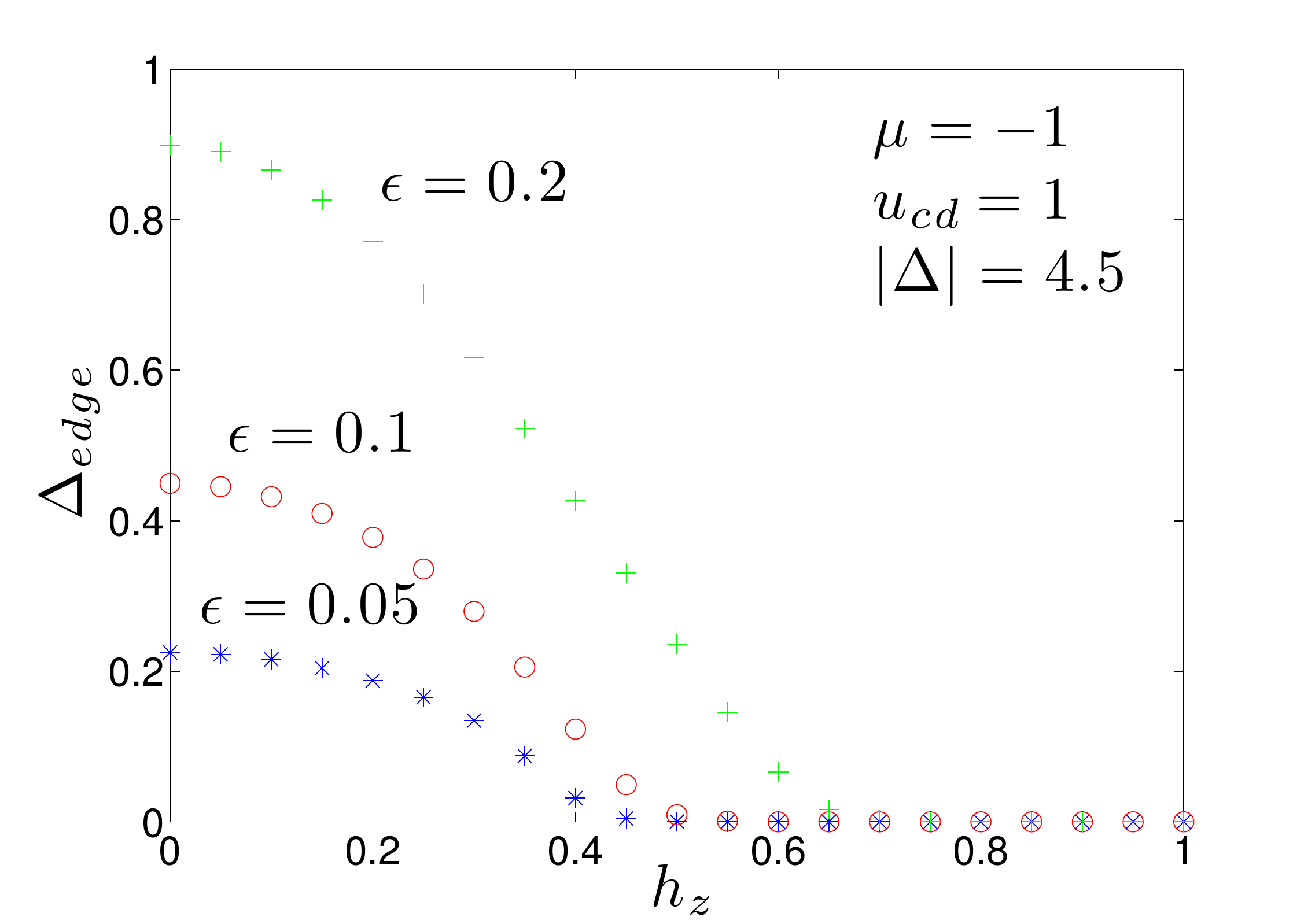}
\vspace*{-2mm}\caption{\label{mismatch2} (Color online) Effect of a
longitudinal Zeeman field in restoring gapless Majorana excitations in
the presence of a phase mismatch $\epsilon \ne 0$, for two different
sets of parameters in 2D. In the bottom panel, the minimum
magnetic-field strength required for restoring gapless Majorana modes
depends more strongly upon $\epsilon$ as a result of the relatively
large amplitude of $|\Delta|$.  System size: $(N_x, N_y)=(40,100)$. }
\end{figure}

While TR symmetry is respected in case (i), this is no
longer the case unless $\epsilon = m \pi$, $m \in {\mathbb Z}$, in case
(ii).  Representative numerical results illustrating the effect of
these two perturbations in 2D
are shown in Fig.~\ref{mismatch1}, where the value $\Delta_{edge}$ 
is twice the energy difference between the edge and zero
energy states.  As the data clearly show, Majorana edge states remain stable
(gapless) against an amplitude mismatch as in (i) (see also Ref. \onlinecite{Babak}), 
whereas they become
gapped for a phase mismatch as in (ii), with a minimum gap that scales
linearly with $\epsilon$.

\begin{figure*}[!t]
\includegraphics[width=4.4cm]{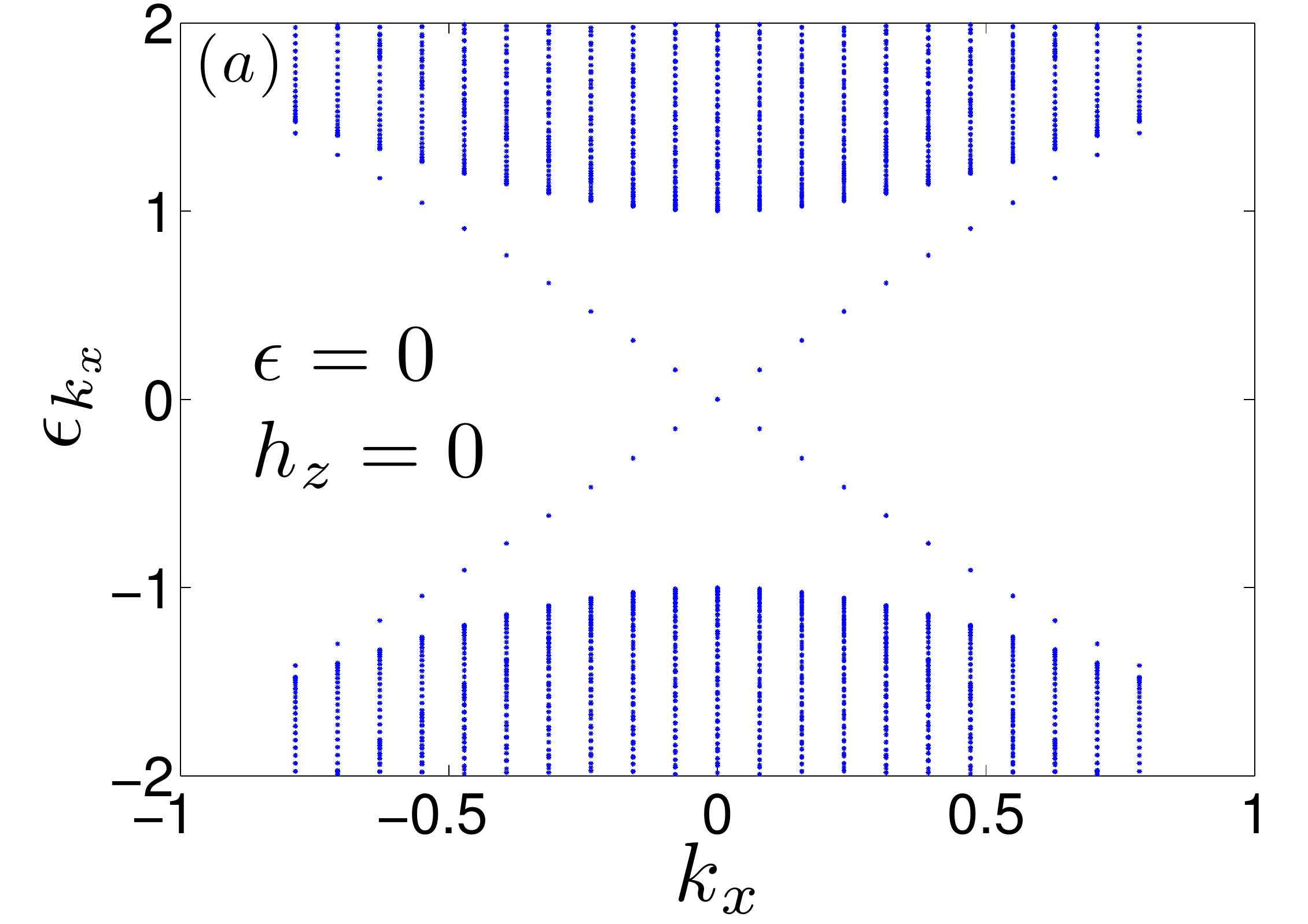}
\includegraphics[width=4.4cm]{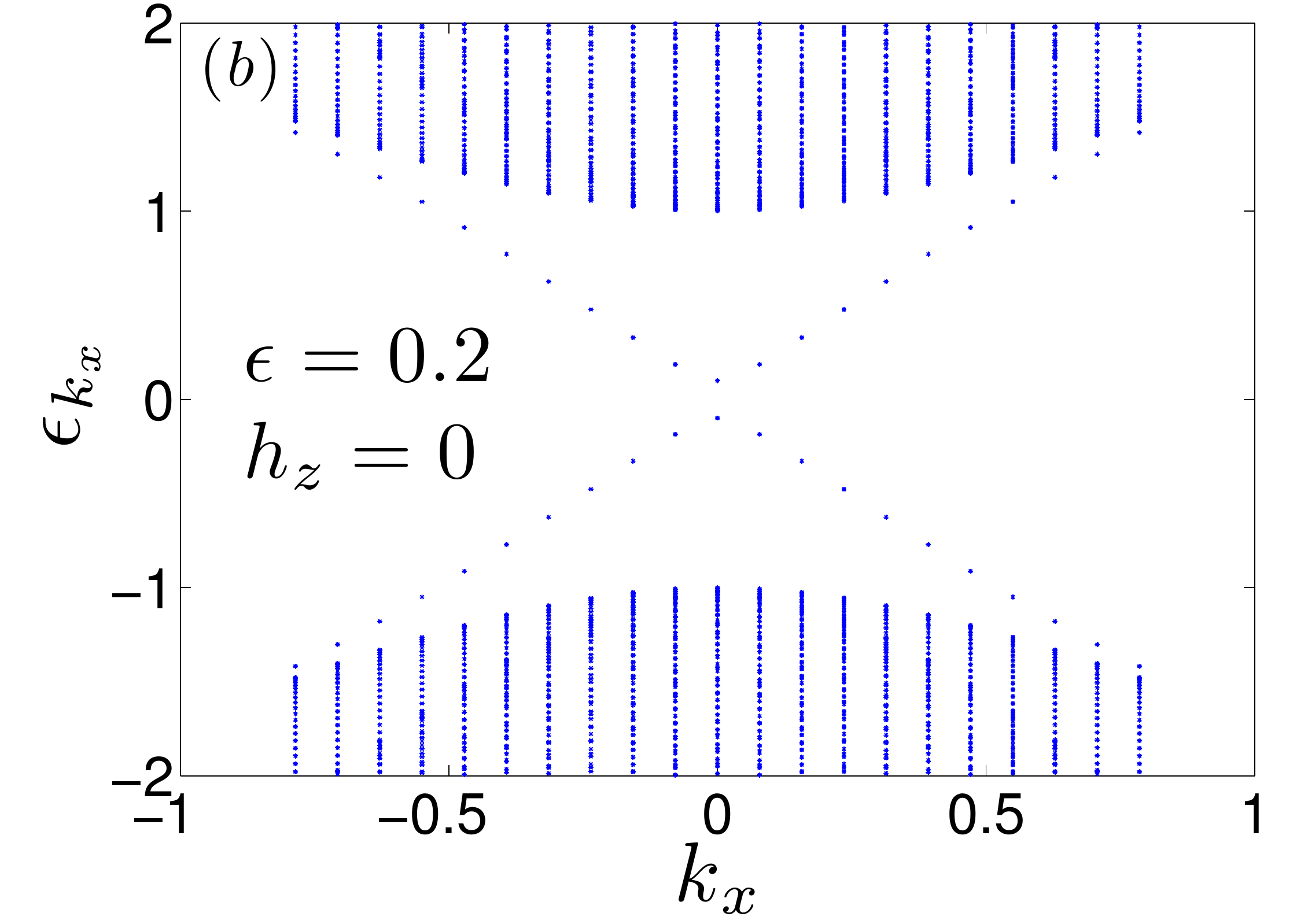}
\includegraphics[width=4.4cm]{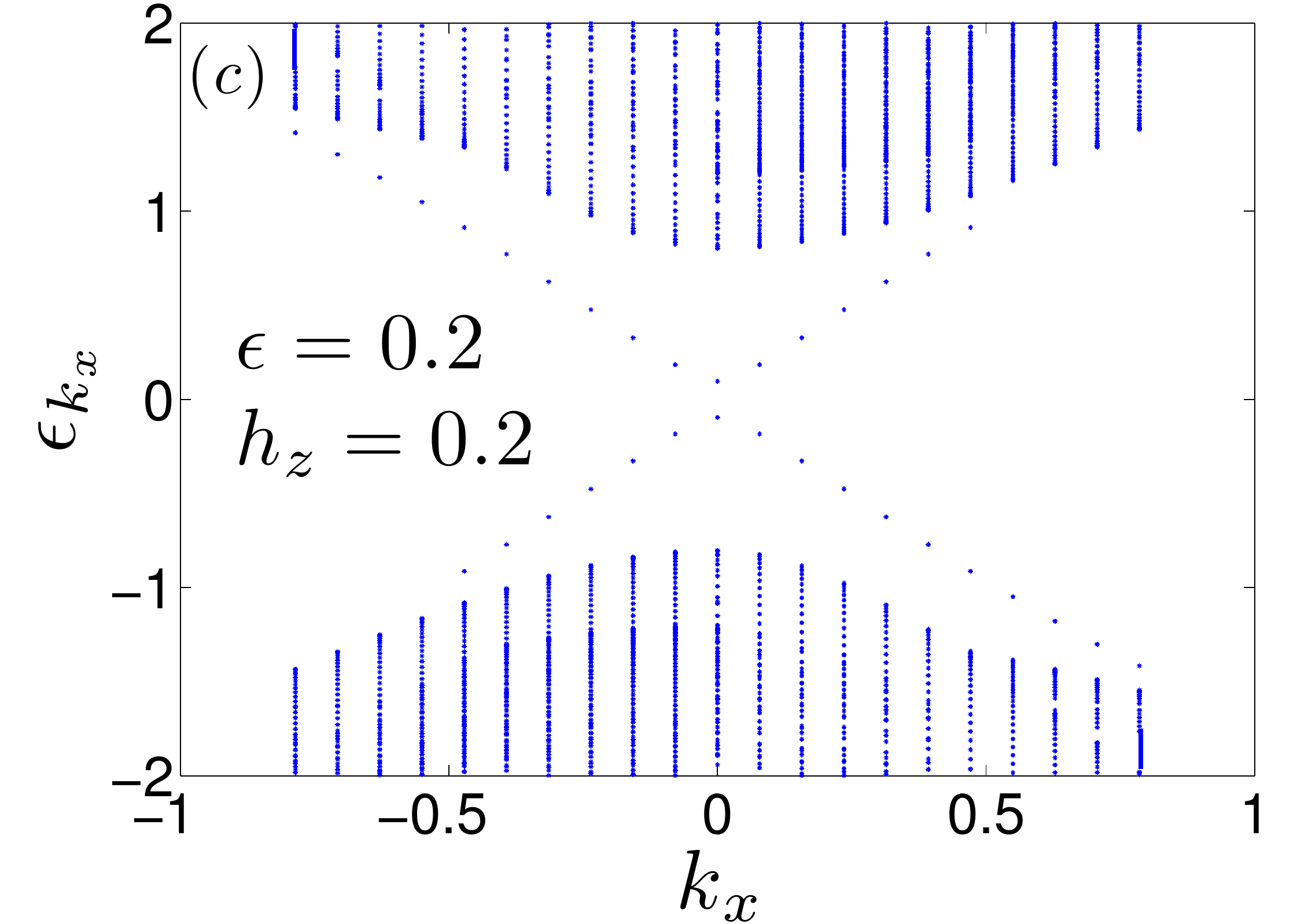}
\includegraphics[width=4.4cm]{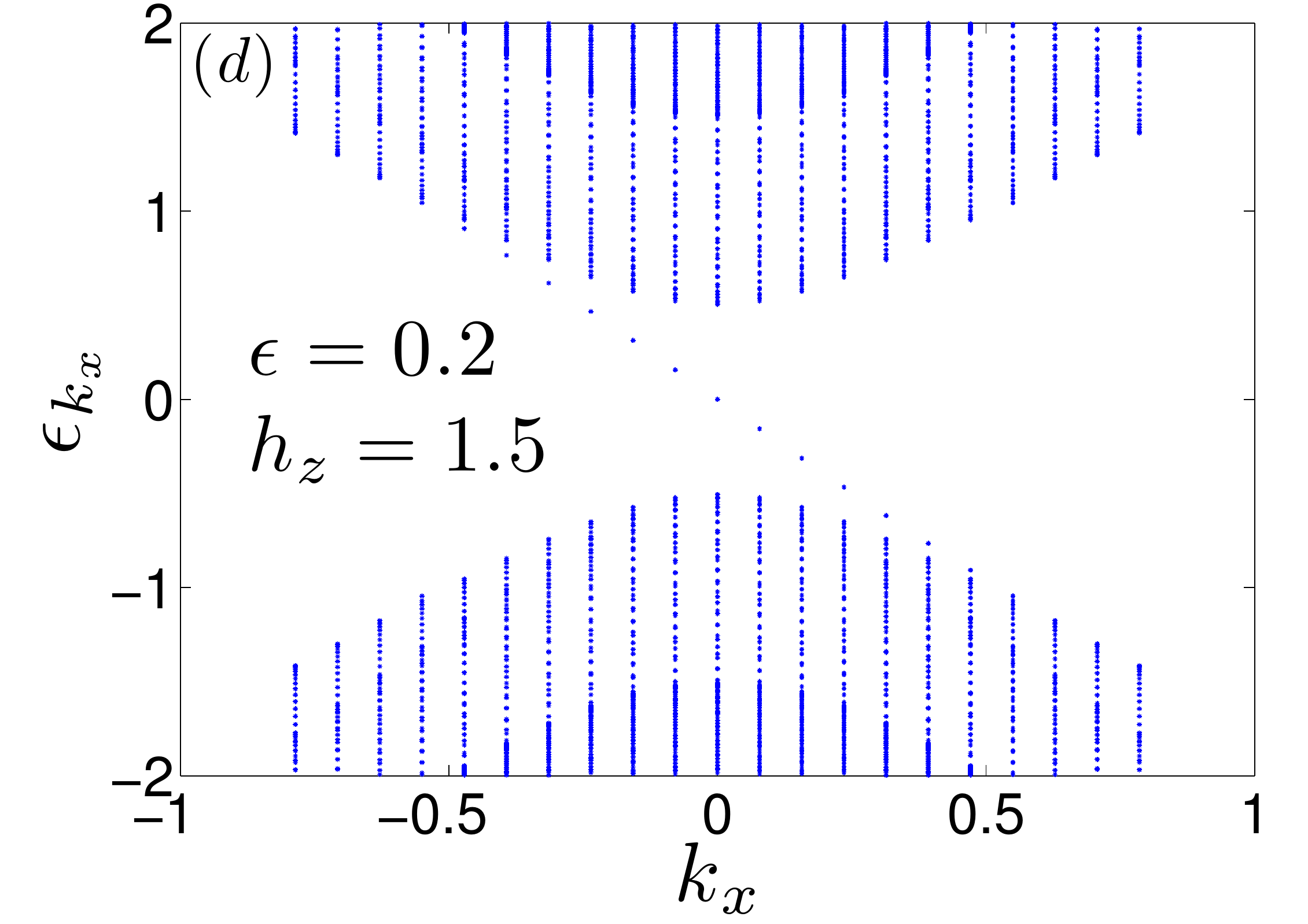}
\vspace*{-2mm}
\caption{(Color online) Excitation spectrum for the 2D Hamiltonian
$H_D+ H^{(z)}_M$ on the cylinder, for  $\mu=0$,
$u_{cd}=4$, $|\Delta_c|=|\Delta_d|=1$.  Only the excitation spectrum
of the edge modes located on {\em one} boundary is plotted for clarity
in all cases. Panel (a): $h_z=0,\epsilon=0$. A pair of gapless
Majorana modes exist, with TR-partners counterpropagating along the
boundary. Panel (b): $h_z=0, \epsilon=0.2$.  Due to the phase
mismatch, the helical edge modes becomes gapped [compare to
Fig. \ref{mismatch2}(top)]. Panel (c) $h_z=0.2,\epsilon=0.2$.  Gapped
helical edge modes still exist with asymmetric bulk excitation
spectrum for the counter-propagating modes.  The bulk gap closes 
at $\epsilon=0.2 , h_z\approx 1.0$. 
Panel (d): $h_z=1.5, \epsilon=0.2$. A pair of gapless Majorana modes
is restored, with both members propagating along the same direction
along the boundary.  System size: $(N_x, N_y)=(80,100)$.  }
\label{evolve}
\end{figure*}

Interestingly, it is possible to counter the effect of a non-zero
$\epsilon$ by applying a Zeeman longitudinal field in the bulk.  In
particular, recalling the analysis of Sec. \ref{uniformz}, we find
that in order to restore gapless Majorana excitations, it is necessary to 
have a sufficiently strong magnetic field to allow the {\em unperturbed}
Hamiltonians $\hat{H}'_{+,k}$ (with $u'_{cd}\equiv u_{cd}+h_z$) and
$\hat{H}'_{-,k}$ (with $u'_{cd}\equiv u_{cd}-h_z$) to belong to
different phases in the corresponding phase diagram (Fig. \ref{pd}).
Explicit numerical results are shown in Fig.~\ref{mismatch2}: in the
top panel, the original Hamiltonian with $\epsilon=0,h_z=0$ is in the
TR-invariant TS phase with CNs $C_{+,+}=1, C_{+,-}=-1$, 
supporting a TR Kramers' pair of Majorana modes on each edge. 
If $\epsilon \ne 0$ and the Zeeman field strength is sufficiently
large, $h_z \gtrsim 0.6$, the CNs become $C_{+,+}=0, C_{+,-}=-2$,
and the edge states become correspondingly gapless, with two {\em
chiral} (co-propagating)
Majorana modes on each boundary (note that the latter do {\em not}
form a quasi-TR-pair according to our definition).  Similarly, in the
bottom panel, the original Hamiltonian with $\epsilon=0, h_z=0$ is in the
TR-invariant TS phase with CNs $C_{+,+}=-1, C_{+,-}=1$.  If, again,
the Zeeman field strength is sufficiently large, the CNs become
$C_{+,+}=-1, C_{+,-}=0$, and gapless Majorana excitations are
restored.  In fact, since one of the original Majorana modes fuses
with the bulk and thus only {\em one} edge mode exists on each
boundary, this mode retains robustness against the effect of
TR-preserving backscattering perturbation, despite TR being explicitly
broken.

In order to verify the physical relevance of using relatively strong
magnetic fields to restore Majorana excitations, we have again also 
performed self-consistent calculations of the superconducting order
parameter $\Delta$ for representative situations, in particular for
the situation just discussed, corresponding to the bottom panel of
Fig.~\ref{mismatch2}.  Our results show that values of the pairing gap
$|\Delta|$ in the given range can still be achieved self-consistently
even in the presence of the required magnetic field. Lower magnetic-field 
strengths can also in principle be obtained by suitably modifying the parameter 
$u_{cd}$. 

Mathematically, the fact that a magnetic field can restore gapless
Majorana excitations may be understood through degenerate perturbation
theory. Suppose that $h_z = 0$ and $\epsilon=0$, the TR-pair of Majorana edge 
modes ($\gamma_{1,2}$) in the phase with $C_{+,+}=-C_{+,-}=1$
may be expressed as Eq. (\ref{gammacn3}). Now let 
$\Delta_d=-\Delta_c e^{i \epsilon}=-\Delta_c-i \epsilon
\Delta_c +O(\epsilon^2)$. Then direct calculation shows that the
effect of the perturbing term in $H_D$ scales as 
\begin{eqnarray}
\langle\Psi_{\sf gs} \big| \gamma^\dag_2 \Big(-i \epsilon
\Delta_c \sum_j d^\dag_{j,\uparrow} d^\dag_{j,\downarrow}+ \text{H.c.}
\Big)  \gamma_1\big|\Psi_{\sf gs} \rangle \sim \epsilon,
\label{phasep}
\end{eqnarray}
consistent with the behavior reported in Fig. \ref{mismatch1}. When a weak magnetic 
field is applied in the above case, the TR-pair of Majorana edge modes become quasi-TR-pair 
(since $C_{+,+}=-C_{+,-}=1$ remains true), and maybe expressed in the form 
\begin{eqnarray}
\left \{ \begin{array}{c} \tilde{\gamma}_{1} = \sum_{n=1}^{N_y}
\big(\tilde{\alpha}_n a_{n,\uparrow}^\dag + \tilde{\beta}_n
b_{n,\downarrow}^\dag + \text{H.c.}
\big), \\ \tilde{\gamma}_{2} = \sum_{n=1}^{N_y} \big(
\tilde{\alpha}'_n a_{n,\downarrow}^\dag - \tilde{\beta}'_n
b_{n,\uparrow}^\dag + \text{H.c.} \big),
\end{array} \right.
\label{quasitrpair}
\end{eqnarray}
for real expansion coefficients. [Notice that, in comparison to Eq. (\ref{gammacn3}) for 
a TR-pair of Majorana modes, the form of Eq.~(\ref{quasitrpair}) is still the same, but 
the coefficients $\tilde{\alpha}'_n (\tilde{\beta}'_n)$ are in general different 
from $\tilde{\alpha}_n (\tilde{\beta}_n)$.] 
One may then show that Eq.~(\ref{phasep}) remains valid with $\gamma_{1,2}$ 
replaced by $\tilde{\gamma}_{1,2}$,  leaving the edge spectrum
gapped, as seen in Fig. \ref{mismatch2}. 
On the other hand, if $\epsilon=0$ and $h_z \ne 0$ is large, as in the
above case where $C_{+,+}=0$, $C_{+,-}=-2$, the two chiral Majorana
edge modes on one boundary can no longer be regarded as a quasi-TR
pair but should rather be expressed as in Eq.~(\ref{gammacn2}).  By
introducing now the effect of the phase-mismatch perturbing term as
done above ($-i\epsilon \Delta_c$), a similar calculation shows that the 
degeneracy between the two Majorana modes cannot be lifted, despite the fact 
that they are {\em not} a quasi-TR pair. 

This is interesting as it demonstrates that {\em ``unpaired''
Majorana modes need not  be, a priori, less robust than modes
forming a (quasi-)TR pair}: A TR Majorana pair is guaranteed to behave
robustly against perturbations that preserve TR symmetry, however,
once the latter is broken (via $h_z$ in our example), it may happen that
unpaired Majorana modes are more robust against additional TR-breaking
(such as a phase mismatch, and possibly even backscattering if only
one mode is present on each edge).  Thus, the robustness of edge modes
ultimately depends on the specific form of the perturbations.

As implied by the above discussion, the application of a TR-breaking
Zeeman magnetic field allows for effectively changing the helical
nature of the original TR-invariant edge spectrum of $H_D$ into a
chiral one [Figs. \ref{tso} and \ref{mismatch2}].  In fact, the
interplay between a phase mismatch in the pairing gaps and an applied
Zeeman field may be exploited to steer the system across a topological
QPT between helical and chiral phases.  
This is explicitly demonstrated in Fig. \ref{evolve}, where for
clarity the edge modes propagating {\em only on one} boundary are plotted.
Starting from a TS phase at $\epsilon=0=h_z$ supporting a Kramers' pair
of helical Majorana modes [panel (a)], a non-zero phase mismatch causes
these two modes to become gapped [panel (b)].  As a Zeeman field is
turned on, both the edge and bulk spectrum are modified [panel (c)],
until for a sufficiently strong field gapless Majorana modes are
restored [panel (d)].  As in Fig. \ref{mismatch2}(top), there are
still two Majorana modes, which travel in the same direction along
each edge, thus forming a chiral pair.
This helical-to-chiral transformation is accompanied by a closing of
the bulk gap, which we have verified happens for parameter values
intermediate between (c) and (d) [data not shown].  

As noted in
discussing Fig. \ref{mismatch2}(bottom), it is also possible that upon
increasing the Zeeman magnetic field, one of the original Majorana
modes dissolves into the bulk, leaving the system in a topologically
non-trivial chiral phase.  Interestingly, a topological QPT between a
helical quantum spin Hall phase and a chiral spin-imbalanced quantum
Hall state was also predicted in Ref.~\onlinecite{Goldman} for a 2D
honeycomb fermionic lattice.  While such a QPT is induced by tuning a
Rashba SO coupling rather than a Zeeman field, our results point to
suggestive similarities between the underlying physics and additional
routes for topological phases manipulation.

%%% LV: Several changes and new fig in this section

\section{Candidate material realizations}
\label{material}

Identifying superconducting materials for which the Hamiltonian $H_D$
in Eq. (\ref{Ham}), or its spin-singlet variant $\tilde{H}_D$ in
Eq. (\ref{Hamtilde}), may provide an adequate physical model requires
an in-depth dedicated study which is beyond our current scope.
In this section, we nevertheless provide some perspective that may be
useful to guide further exploration, also in the context of ongoing 
experiments.  

Recently discovered materials such as 
Cu$_x$Bi$_2$Se$_3$ \cite{Sasaki11,Stroscio} and Sn$_{1-x}$In$_x$Te
\cite{Sasaki12} are attracting significant attention as possible 
TR-invariant (centro-symmetric) TSs in 3D.  In general, strongly SO-coupled 
doped semiconductors are natural candidates to search 
for topological superconductivity.  
The presence of a zero-bias conductance peak (ZBCP) in the measured 
point-contact spectrum has been interpreted, in particular, as a signature 
of the surface helical Majorana fermions associated with a non-trivial topological 
behavior.  Experimentally, it is observed that such a ZBCP has a
distinctive magnetic-field dependence, its amplitude being strongly
suppressed by a relatively weak Zeeman field applied perpendicularly
to the cleaved surface, consistent with delocalization of Majorana
modes \cite{Sau}.  While a theoretical description of Cu$_x$Bi$_2$Se$_3$ 
has been proposed based on an unconventional {\em spin-triplet pairing with 
odd parity} \cite{Fu10,Fu2012}, scanning tunneling spectroscopy 
measurements of the superconducting gap appear to be consistent, at least for 
current Cu concentration, with a fully gapped $s$-wave spectrum and no 
mid-gap energy state. 

\begin{figure}[t]
\includegraphics[width=7cm]{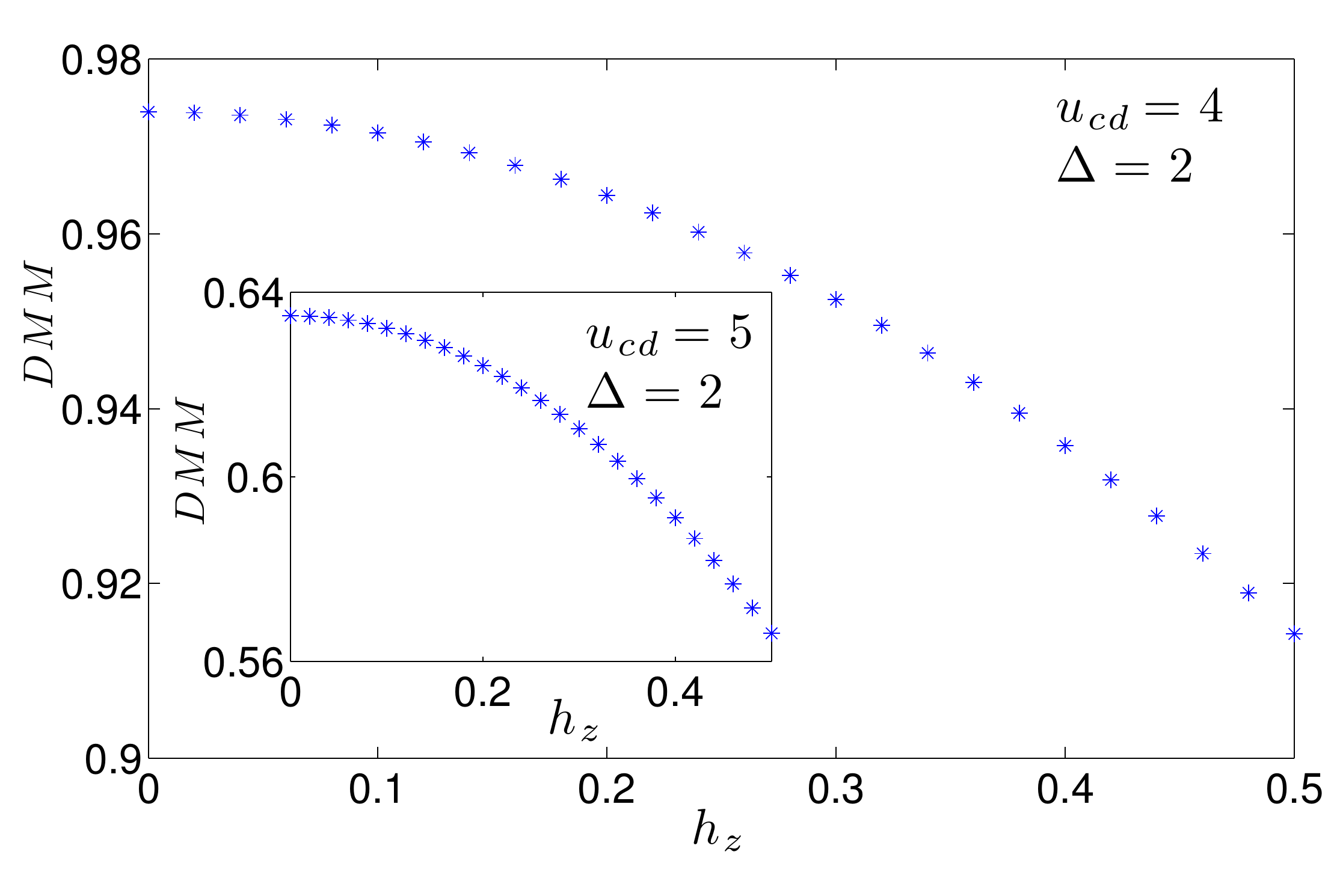}
\vspace*{-2mm}\caption{\label{ldosm} (Color online) Density of
Majorana modes on the boundary [Eq. (\ref{dmm})] for the 2D (main panel) 
and 3D (inset) Hamiltonian $H_D+H_M^{(z)}$ as a function of the applied magnetic
field strength $h_z$, and $\mu=-1$. System size: $(N_x, N_y)=(40,100)$ for 2D, 
and $(N_x, N_y,N_z)=(40,100,40)$ for 3D.}
\end{figure}

With reference to our two-band model, there are two aspects we would like to 
highlight in regard to the above discussion.   First, we have 
investigated the density of Majorana modes ({\em DMM}) on the boundary as a
function of the magnetic field strength in both 2D and 3D. Suppose,
specifically, that the system is originally in a TS phase 
characterized by CNs $(C_+^0,C_+^\pi)=(1,0)$ in 3D
and  $C_{+}=1$ in 2D, in which cases a TR-pair of Majorana 
modes at the Dirac cones exist on each edge. 
When a weak magnetic field is applied, such that TR is broken but the
CNs remain unchanged, the two Majorana modes become a quasi-TR-pair and 
may thus be expressed as in Eq. (\ref{quasitrpair}).  Therefore, the 
{\em DMM} on a given boundary (say, $j=1$)
%(say, $N_y=1$)   LV:?? 
may be computed as
\begin{eqnarray}
\text{\em DMM}=\Big(\big{|}\tilde{\alpha}_1 \big{|}^2
+\big{|}\tilde{\beta}_1 \big{|}^2+\big{|}\tilde{\alpha}'_1\big{|}^2+
\big{|}\tilde{\beta}'_1\big{|}^2 \Big),
\label{dmm}
\end{eqnarray} 
with an equivalent definition holding for the other boundary.
The numerically computed {\em DMM} in 2D (3D) is shown in the main (inset) 
panels of Fig.~\ref{ldosm}, respectively. Clearly, in our two-band TS, a weak 
applied magnetic field delocalizes the Majorana fermions in both cases, 
qualitatively similar to the observed ZBCP dependence.   

\begin{figure}[t]
\includegraphics[width=7.5cm]{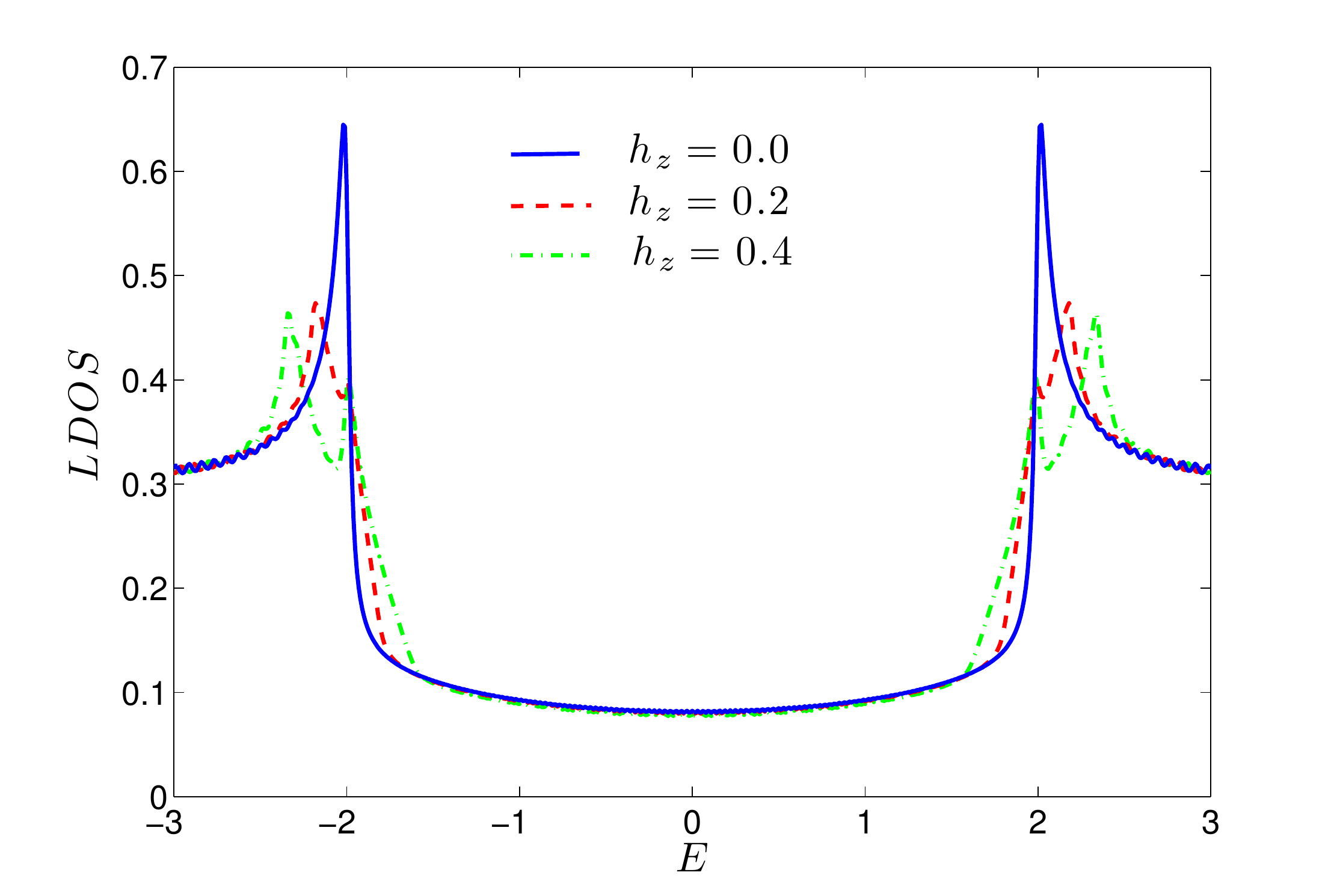}
\vspace*{-2mm}\caption{\label{ldos2D} (Color online) Local density of states 
on the boundary [Eq. (\ref{eqldos})]
for the 2D Hamiltonian $H_D+H_M^{(z)}$ for different values of the applied magnetic
field strength $h_z$.  The remaining parameters are the same as in Fig. \ref{ldosm}, except
the system size is now $(N_x,N_y)=(400,80)$.
}
\end{figure}

Second, it is interesting to contrast the {\em DMM} behavior to the one of 
a quantity which is more directly related to the measured 
scanning-tunneling spectrum, namely, the local density of states (LDOS), 
computed as 
\begin{eqnarray}
\text{\em LDOS}(j_y,\hspace{-0.5mm}E)\hspace{-1mm}&=&\hspace{-1mm}\frac{1}{N_x}\hspace{-0.5mm}\sum_{n,\hspace{-0.2mm}k_x}\hspace{-0.5mm} \sum_{i=1,4}\hspace{-0.8mm}\Big[\big|u_i(n,\hspace{-0.5mm}k_x,\hspace{-0.5mm}j_y)\big|^2 \delta(E\hspace{-0.5mm}-\hspace{-0.5mm}\epsilon_{\hspace{-0.2mm}n,\hspace{-0.2mm}k_x})   \nonumber \\
&+&\hspace{-0.2mm}\big|v_i(n,\hspace{-0.5mm}k_x,\hspace{-0.5mm}j_y)\big|^2\hspace{-0.5mm}\delta(E\hspace{-0.5mm}+\hspace{-0.5mm}\epsilon_{n,k_x})\hspace{-0.5mm}\Big],
\label{eqldos}
\end{eqnarray}
where $(u_1 ,\ldots,u_4,v_1,\ldots,v_4)^\dag$
%$(u_1(n,k_x,j_y),\ldots,u_4(n,k_x,j_y),v_1(n,k_x,j_y),\ldots,v_4(n,k_x,j_y))^\dag$
is the single-particle eigenvector corresponding to 
energy $\epsilon_{n,k_x}$ for $H_D+H_M^{(z)}$ with PBC in the $\hat{x}$ 
direction, and OBC in the $\hat{y}$ direction.
While a more in-depth analysis will be presented elsewhere 
\cite{DengFlat}, numerical results for the \emph{LDOS} profile of the 
same  2D TS considered in Fig. \ref{ldosm} are shown in Fig. \ref{ldos2D}, 
for selected values of the applied Zeeman field.  Despite the existence of 
zero-energy Majorana modes in the phase under consideration, no peak 
is manifest at zero energy.  Although the absence of such a peak is consistent
with the measurements in Ref. \onlinecite{Stroscio}, our {\em DMM} and {\em LDOS} 
results taken together suggest that care may be needed in diagnosing 
the presence or absence of Majorana modes from such quantities.  
While a conclusive determination will require additional cross-checks and 
spectroscopic measurements on higher doped materials, experimental 
signatures of {\em spin-singlet pairing and/or of a two-gap behavior} would
provide evidence in favor of our TS model in describing these or possibly 
%Cu$_x$Bi$_2$Se$_3$/Sn$_{1-x}$In$_x$Te, or possibly 
similar doped-semiconducting materials.

A different intriguing possibility is provided by 
$s$-wave pairing with a sign reversal of the superconducting order
parameter between different Fermi surface sheets, resulting in
so-called $s_{\pm}$ {\em symmetry} \cite{Ishida,Mazin}.  While an unambiguous
experimental characterization has yet to be established, $s_{\pm}$
symmetry is widely believed to be realized and play an important role in
iron-based superconductors of both the iron-pnictide and the
chalcogen-based family \cite{MazinNat}.  Interestingly, for such
superconductors, it has been experimentally established that the Cooper 
pairs consist of spin singlets, and their $s_{\pm}$ symmetry would automatically 
realize the $\pi$-symmetry condition that is required for our TR-invariant
Hamiltonian to support non-trivial topological phases and Majorana
modes to emerge. Although additional quantitative study is certainly
needed before any solid conclusion can be reached (see also Ref.~\onlinecite{Zhang12}
for a recent proposal on topological superconductivity via proximity effects  
between $s_{\pm}$ wave iron-based superconductor and semiconductors), 
we conjecture that iron-based materials may provide a
natural candidate for realizing our model in Nature, as first suggested in Ref. 
\onlinecite{Deng}.

\section{Conclusions}

We have continued our investigation of {\it time-reversal 
invariant multiband $s$-wave 
topological superconductors} as introduced in Ref. \onlinecite{Deng}. 
This class of superconductors involves at least two spin-1/2 
fermionic bands,  spin-orbit coupling, and bulk $s$-wave 
superconducting fluctuations. In the two-band case, provided that a 
{\em sign reversal} ($\pi$ shift) between the superconducting pairing 
gaps is realized, the  model can be analytically  solved in 1D, 2D, and 3D upon 
restricting to the excitation spectrum of time-reversal invariant momentum modes.  
Our model may then be interpreted as either an intra-band or an inter-band 
{\em spin-singlet bulk topological superconductor}, since there exists an exact unitary 
mapping connecting both representations.  From the standpoint of the 
classification introduced in Ref. \onlinecite{Altland},  our model belongs to 
the DIII symmetry class, as shown in full mathematical detail in Appendix 
\ref{appA}.  In the absence of superconductivity 
our Hamiltonian describes a topological insulator, and thus one may think 
of our superconducting state as emerging from a doped topological 
insulator. 

An important part of our work is the topological characterization, 
into trivial as opposed to non-trivial topological states,  
of the various possible phases generated as a result of changes in the 
parameters of our model Hamiltonian.  To this end, we have introduced
(and explained in Appendix 
\ref{appC} how to numerically compute) corresponding bulk
$\mathbb{Z}_2$ topological invariants that are,  in principle, generalizable to 
interacting systems. In the time-reversal invariant case, only one Kramers' sector 
is included in the computation of the topological invariants, while when TR 
symmetry is explicitly broken we have showed how to modify these bulk 
topological invariants to include all the states. 
Physically, these topological invariants are key to 
understanding the bulk-boundary correspondence between 
the (trivial and non-trivial) topological nature of the bulk state and the existence
of robust Majorana edge modes.  Topologically non-trivial superconducting 
phases support an odd number of Kramers' pairs (in the time-reversal-invariant 
case) or an odd number (in the broken time-reversal case) of Majorana 
modes on each boundary, respectively. 
Topologically non-trivial  phases are associated to robust Majorana edge states:
in the time-reversal-invariant case they are robust against perturbations that preserve 
such symmetry, while in the broken time-reversal symmetry case they are robust 
against different kinds of perturbations. 

Throughout this work, emphasis has been put into exploring: (1) the relation between 
the spatial dimensionality of our model and the emergence of non-trivial topological phases; 
and (2) the stability/robustness of gapless Majorana edge states under different 
mechanisms for breaking time-reversal symmetry, such as applied or 
impurity magnetic fields or broken time-reversal invariance due 
to a deviation from $\pi-$shifted superconducting gaps. While a number of 
interesting problems remain open for further investigation as indicated in the text, 
our main findings may be itemized as follows:

\noindent
(i) Nontrivial topological phases that exists in our 2D
two-band time-reversal invariant model \cite{Deng} can be extended to 1D and 3D. 
Topologically trivial and non-trivial  $s$-wave spin-singlet superconducting phases exist 
in all the spatial dimensions  studied; \\ 
(ii) Lower-dimensional $\mathbb{Z}_2$ invariant can be extended
to characterize higher dimensional topological phases; \\
%% LV: Changed 
(iii) Gapless pairs of Majorana modes that exist in a topologically 
non-trivial phase in the absence of an applied or impurity magnetic field
may remain gapless under such a perturbation.  
However, since these Majorana modes are no longer protected by 
time-reversal symmetry, their degree of robustness against
subsequent perturbations is, in general, different 
as compared to the original Kramers' pairs, and dependent upon 
the perturbation details; \\
(iv) Suitable static magnetic fields may be used to restore Majorana
edge modes when the phase of the gaps of the superconductor is not
$\pi$-shifted; \\
(v) Under the effect of a Zeeman magnetic field, helical Majorana
excitations in our time-reversal invariant model may be transformed into chiral
Majorana excitations, providing an interesting example of a topological 
quantum phase transition; \\
(vi) Candidate materials that may potentially realize 
our proposal for time-reversal invariant topological superconductivity 
and Majorana edge modes may be provided by strongly spin-orbit coupled 
doped semiconductors or iron-based superconductors with $s_\pm$ pairing 
symmetry. \\
(vii) The existence of Majorana modes in 2D (and 3D) fully gapped TS need not imply 
a zero-energy peak in the local density of states on the surface. Thus, care is needed in 
interpreting point-contact and scanning-tunneling spectroscopy measurements and in 
identifying unambiguous experimental signatures.

\vspace*{1mm}

\section*{Acknowledgements}

It is a pleasure to thank B. Seradjeh, A. W. W. Ludwig and S. Ryu for
useful input and correspondence.  Support from the NSF through
Grant No. PHY-0903727 (to LV) is gratefully acknowledged.

{}

\begin{appendix}

\section{Symmetry class}
\label{appA}

Here, we provide an explicit representation of the discrete symmetry
 properties of our basic Hamiltonian,
Eq. (\ref{Ham}), as relevant to the topological classification of
Refs. \onlinecite{Altland,Ryu}.  In addition, we describe the 
hidden discrete symmetry mentioned in Sec. \ref{BBC}.

In second-quantized language, we define a anti-unitary TR operator
$\mathcal{T}$, with $\mathcal{T}^2=-I$, through its action on the
fermion creation and annihilation operators.  Specifically, let
$\mathcal{T} \equiv U_T K$, where $U_T$ and $K$ are a unitary operator
and complex conjugation, respectively, such that 
\begin{eqnarray}
\mathcal{T} (A_{\bf{k}})_j \mathcal{T}^{-1} \equiv \sum_l (U_T)_{j
l}(A_{-\bf{k}})_l
\label{UT},
\end{eqnarray}
where $(A_{\bf{k}})_j$ is the $j$th component of the vector
$A_{\bf{k}}$ given after Eq. (\ref{fourier}) in the main text. Thus, if we define
$$U_T=I_{4\times4} \otimes i\sigma_y,$$
\noindent 
then the above condition yields 
$$\mathcal{T} c(d)_{{\bf k},\uparrow} \mathcal{T}^{-1}=c(d)_{{-\bf
k},\downarrow}, \;\; \mathcal{T} c^\dag(d^\dag)_{{\bf k},\downarrow}
\mathcal{T}^{-1}=-c^\dag(d^\dag)_{{-\bf k},\uparrow},$$
$$\mathcal{T} c(d)_{{\bf k},\downarrow} \mathcal{T}^{-1}=-c(d)_{{-\bf
k},\uparrow}, \;\; \mathcal{T} c^\dag(d^\dag)_{{\bf k},\uparrow}
\mathcal{T}^{-1}=c^\dag(d^\dag)_{{-\bf k},\downarrow}.$$ 
\noindent 
Recall that $H_D=(1/2)\sum_{\bf{k}} (\hat{A}_{\bf{k}}^{\dag}
\hat{H}_{\bf{k}}\hat{A}_{\bf{k}}^{\;}-4\mu)$, where the
single-particle Hamiltonian $\hat{H}_{\bf k}$ is given in
Eq.~(\ref{8by8}).  Direct calculation then yields 
\begin{eqnarray}
U^\dag_T \hat{H}^\ast_{\bf k} U_T=\hat{H}_{-{\bf k}}.
\label{UThk}
\end{eqnarray}
Accordingly,  satisfying the requirement of TR symmetry [cf. Eq. (3) in
Ref. \onlinecite{Ryu} in momentum space].

Similarly, let us define an anti-unitary PH operator $\mathcal{C}$,
with $\mathcal{C} \equiv U_C K$ and $\mathcal{C}^2 =+I$, through its action on the
fermion creation and annihilation operators, where $U_C$ is a unitary
operator, such that
\begin{eqnarray}
\mathcal{C} (A_{\bf{k}})_j \mathcal{C}^{-1}=\sum_l (U_C)_{j
l}(A_{-\bf{k}})_l 
\label{UC}.
\end{eqnarray}
Thus, if we define $U_C$ as 
$$U_C= \sigma_x\otimes I_{4\times4},$$
\noindent 
it follows that 
$$\mathcal{C} d(c)_{{\bf k},\sigma}
\mathcal{C}^{-1}=d^\dag(c^\dag)_{{\bf k},\sigma}, \;\;\; \mathcal{C}
d^\dag(c^\dag)_{{\bf k},\sigma} \mathcal{C}^{-1}=d (c)_{{\bf
k},\sigma}.$$ 
\noindent 
Direct calculation then yields
\begin{eqnarray}
U^\dag_C 
\hat{H}^\ast_{\bf k}  U_C=-\hat{H}_{-{\bf k}}
\label{UChk}
\end{eqnarray}
satisfying the requirement of PH symmetry in the single-particle
representation [cf. Eq. (4) in Ref. \onlinecite{Ryu} in momentum
space].  
In addition, it is also straightforward to verify that $H_D$ also possesses a 
unitary inversion symmetry [cf. Eq. (C.29) in Ref. \onlinecite{Ryu}], 
\begin{eqnarray}
U^\dag_I \hat{H}_{\bf k}  U_I = \hat{H}_{-{\bf k}}, 
\label{invs}
\end{eqnarray}
where the inversion operator may be expressed as 
$$ U_I= \sigma_z \otimes \sigma_x \otimes I_{2 \times 2}, $$
independently of the system's dimension $D$. 

Interestingly, besides exhibiting the above manifest discrete symmetries, our Hamiltonian may 
also preserve additional accidental ``hidden'' symmetries.  While a characterization is far from 
trivial, we have explicitly identified one such 
${\mathbb Z}_2 \otimes {\mathbb Z}_2 \otimes {\mathbb Z}_2 \otimes {\mathbb Z}_2$ symmetry 
in the limit where $\mu=0, \lambda=t$.  In order to describe this symmetry, let us now introduce 
the following new canonical Dirac fermion operators: 
\begin{eqnarray*}
\left \{ \begin{array}{c} \tilde{a}_{j,\uparrow} = \frac{1}{2}(c_{j,\uparrow}+d_{j,\uparrow}+c^\dag_{j,\downarrow}-d^\dag_{j,\downarrow}), \\
\tilde{b}_{j,\uparrow} = \frac{1}{2}(c_{j,\uparrow}+d_{j,\uparrow}-c^\dag_{j,\downarrow}+d^\dag_{j,\downarrow}), \\
 \tilde{a}_{j,\downarrow} = \frac{1}{2}(c_{j,\downarrow}+d_{j,\downarrow}+c^\dag_{j,\uparrow}-d^\dag_{j,\uparrow}), \\
\tilde{b}_{j,\downarrow} = \frac{1}{2}(c_{j,\downarrow}+d_{j,\downarrow}-c^\dag_{j,\uparrow}+d^\dag_{j,\uparrow}) , 
\end{array} \right . 
\end{eqnarray*}
which actually take a simpler form once expressed in terms of the $a$ and $b$ fermion operators defined in Eq.~(\ref{canonicalf}):
\begin{eqnarray}
\left \{ \begin{array}{c} \tilde{a}_{j,\uparrow} = \frac{1}{\sqrt{2}}(a_{j,\uparrow}+b^\dag_{j,\downarrow}), \\
\tilde{b}_{j,\uparrow} = \frac{1}{\sqrt{2}}(a_{j,\uparrow}-b^\dag_{j,\downarrow}), \\
 \tilde{a}_{j,\downarrow} = \frac{1}{\sqrt{2}}(a_{j,\downarrow}+b^\dag_{j,\uparrow}), \\
\tilde{b}_{j,\downarrow} = \frac{1}{\sqrt{2}}(a_{j,\downarrow}-b^\dag_{j,\uparrow}) .
\end{array} \right.
\label{newab}
\end{eqnarray}
Upon expressing the Hamiltonian $H_D$, with $\mu=0$ and $\lambda=t$,  
in terms of the newly defined fermion operators $\tilde{a}_{j,\sigma}$, $\tilde{b}_{j,\sigma}$, 
the following commutation relationships are found to hold:
\begin{eqnarray*}
&\big[H_{D=1}, \bigotimes_{\sigma, \tilde{c}} \big(e^{i \pi \sum_{j} \tilde{c}^\dag_{j,\sigma} \tilde{c}_{j,\sigma}} \big)\big] = 0, \\ \nonumber
&\big[H_{D=2}, \bigotimes_{\sigma, \tilde{c}} \big(e^{i \pi \sum_{j_y} \tilde{c}^\dag_{k_{x,c},j_y,\sigma} \tilde{c}_{k_{x,c},j_y,\sigma}} \big)\big] = 0, \\
&\big[H_{D=3}, \bigotimes_{\sigma, \tilde{c}} \big(e^{i \pi \sum_{j_y} \tilde{c}^\dag_{k_{x,c},j_y,k_{z,c},\sigma} \tilde{c}_{k_{x,c},j_y,k_{z,c}\sigma}} \big)\big] \hspace*{-1mm}= 0,
\end{eqnarray*}
where the products run over $\sigma=\uparrow, \downarrow$ and $\tilde{c}=\tilde{a},\tilde{b}$. 

Recall that in Sec.~\ref{BBC} we have considered the fate
of the Majorana edge modes under a boundary perturbation $H_p$ in 3D,
Eq. (\ref{pert0}).  Thus, it is important to determine whether the 
total Hamiltonian $H_D+H_p$ still belongs to the same symmetry class.
To this purpose, it is convenient to imagine that $H_p$ acts {\em both} on 
the surface and in the bulk, in which case we can equivalently work under PBC.  
Thus, in place of Eq. (\ref{pert0}), we may consider
\begin{eqnarray*}
H'_p &=& \sum_{\bf{k},\sigma} u_p ( c^\dag_{k_x,k_y,k_z,\sigma}
c_{-k_x,k_y,-k_z,\sigma} \\ \nonumber&+& d^\dag_{k_x,k_y,k_z,\sigma} d_{-k_x,k_y,-k_z,\sigma}) +
\text{H.c.} \\ \nonumber 
&=& \sum_{\bf{k},\sigma} u_p ( \tilde{a}^\dag_{k_x,k_y,k_z,\sigma}
\tilde{b}_{-k_x,k_y,-k_z,\sigma} \\ \nonumber&+& \tilde{b}^\dag_{k_x,k_y,k_z,\sigma} 
\tilde{a}_{-k_x,k_y,-k_z,\sigma} )+\text{H.c.} +\text{const}, 
\end{eqnarray*}
where in the last two lines we have rewritten the perturbation using 
the new canonical operators $\tilde{a}_{k,\sigma}, \tilde{b}_{k,\sigma}$, and const is a c-number. Due to $H'_p$, 
the dimension of the single-particle Hamiltonian matrix $\hat{H}_{\bf k}$ is now 
doubled (to $16 \times 16$), and the
matrices $U_T$ and $U_C$ for the TR and PH symmetries need to be
changed correspondingly, that is, 
$$U'_T=I_{8\times8} \otimes i\sigma_y, \;\;\; U'_C= I_{2 \times 2} \otimes \sigma_x\otimes
I_{4\times4},$$
\noindent 
respectively, and similarly for $U'_I$. It can then be verified that the new
perturbed Hamiltonian {\em still} exhibits both TR and PH symmetry, that is,
Eq.~(\ref{UThk}), Eq.~(\ref{UChk}), and Eq. (\ref{invs}) still hold. 
However, the perturbation $H_p$ ($H'_p$) {\em does break} the hidden symmetry shown 
above, for $D=3$. 
Notice that when the perturbation acts on the surface only, the dimensions of $U_T$ and
$U_C$ are changed accordingly, nevertheless the conclusion that the
perturbation still conserves the basic manifest discrete symmetries remains true.

\section{The role of self-consistency}
\label{self}

Throughout most of the discussion in the main text, and in particular in obtaining 
the phase diagrams of Fig.~\ref{1Dpd} and Fig.~\ref{pd}, the
superconducting pairing gap $\Delta$ has been treated as a \emph{free}
control parameter, tunable at will.  In real systems, however,
$\Delta$ can only be obtained by minimizing self-consistently the
total free energy.  Let $V_{cc}=V_{dd} \equiv V >0$ in
Eq. (\ref{gaps}) denote the effective attraction strength in each
band, and assume that the $\pi$-symmetry condition is obeyed,
$\Delta_c =-\Delta_d$.  Then, in 2D and at zero temperature, this
amounts to minimizing the many-body ground-state energy,
\begin{eqnarray*}
E_{\sf gs}= 2 N_x N_y \frac{\Delta^2}{V} + \sum_{\bf k}
(\epsilon_{1,{\bf k}} + \epsilon_{2,{\bf k}} -2 \mu), 
\label{selfc}
\end{eqnarray*}
where the first term represents the condensation energy, and 
similar expressions hold in 1D and 3D. 

As shown in Ref.~\onlinecite{Deng}, for $D=2$ all the topological 
phases identified in the non-self-consistent regime are found to be 
{\em stable} for suitable choices of the control parameters in the self-consistent 
phase diagram, although new features may also emerge [see Fig.~3 therein 
and related discussion].  It would be interesting to obtain a
full self-consistent description without imposing that 
$V_{cc}=V_{dd}$ and $\Delta_c =-\Delta_d$, that is, by leaving
the two pairing gaps as independent parameters to be determined
separately for generic intra-band parameters, and also allowing for
an inter-band scattering term $V_{cd} \ne0$.  While
such a complete study is beyond our current scope, we have
verified that, as long as Eq. (\ref{pi}) is obeyed, all
the trivial and the non-trivial topological phases in both 1D and 3D
remain physically accessible for suitable parameters after imposing
the self-consistency constraint.  In a similar spirit, self-consistent calculations 
have also been performed for representative parameter values in the presence 
of a magnetic field, as discussed in Sec. \ref{break}.

\section{Explicit form of Hamiltonian matrices}
\label{appB}

Consider the most general case of 3D geometry, under the condition of 
$\pi$-shifted gaps,  $\Delta_c =-\Delta_d=\Delta$.  Recall that for
general parameter values and PBC, the Hamiltonian matrix with respect
to the operator basis $\{ A_{\bf k}\}$ takes the form given in
Eq. (\ref{8by8}).  Upon transforming to the fermionic operators 
$\{ a_{{\bf k}, \sigma}, b_{{\bf k}, \sigma} \}$ defined in
Eq. (\ref{canonicalf}) and, for convenience, moving to the slightly different 
operator basis 
$$\hat{B}_{\bf{k}}^{' \dag} \equiv (a_{{\bf{k}},\uparrow}^\dag, b_{{\bf{k}},
\downarrow}^\dag,a_{{-\bf{k}},\uparrow}^{\;},b_{{-\bf{k}},\downarrow}^{\;},
a_{{\bf{k}},\downarrow}^\dag,
b_{{\bf{k}},\uparrow}^\dag,a_{{-\bf{k}},
\downarrow}^{\;},b_{{-\bf{k}},\uparrow}^{\;}),$$  
\noindent 
the new Hamiltonian matrix $\hat{H}'_{\bf k}$ becomes:
\begin{widetext}
\begin{eqnarray*}
\hat{H}'_{\bf{k}}\hspace{-0.8mm}=
\hspace{-0.8mm}\left (\hspace{-1mm}\begin{array}{cccc}
-\hspace{-0.7mm}\mu+\hspace{-0.8mm}\lambda_{k_x}\sigma_x\hspace{-0.8mm}+\hspace{-0.8mm}\lambda_{k_y}\sigma_y\hspace{-0.8mm}+m_{\bf k}\sigma_z \hspace{-0.8mm}  & \hspace{-0.8mm}i\Delta\sigma_y\hspace{-0.8mm} & \hspace{-0.8mm}i\lambda_{k_z} \sigma_y \hspace{-0.8mm} & \hspace{0.8mm}0\hspace{-0.8mm} \\
\hspace{-0.8mm}-i\Delta\sigma_y \hspace{-0.8mm} &\hspace{-0.8mm}\mu+\hspace{-0.8mm}\lambda_{k_x}\sigma_x\hspace{-0.8mm}-\hspace{-0.8mm}\lambda_{k_y}\sigma_y\hspace{-0.8mm}-\hspace{-0.8mm}m_{\bf k}\sigma_z \hspace{-0.8mm}   & \hspace{-0.8mm}0\hspace{-0.8mm} & \hspace{-0.8mm}i\lambda_{k_z} \sigma_y \hspace{-0.8mm}\\
\hspace{-0.8mm}-i\lambda_{k_z} \sigma_y \hspace{-0.8mm} & \hspace{-0.8mm}0\hspace{-0.8mm} & -\hspace{-0.8mm}\mu+\hspace{-0.8mm}\lambda_{k_x}\sigma_x\hspace{-0.8mm}-\hspace{-0.8mm}\lambda_{k_y}\sigma_y\hspace{-0.8mm}+\hspace{-0.8mm}m_{\bf k}\sigma_z \hspace{-0.8mm} & -i\Delta\sigma_y \\ 
\hspace{-0.8mm}0\hspace{-0.8mm} & \hspace{-0.8mm}-i\lambda_{k_z}\sigma_y\hspace{-0.8mm}  &  \hspace{-0.8mm}i\Delta\sigma_y \hspace{-0.8mm} &  \hspace{-0.8mm}\mu+\hspace{-0.8mm}\lambda_{k_x}\sigma_x\hspace{-0.8mm}+\hspace{-0.8mm}\lambda_{k_y}\sigma_y\hspace{-0.8mm}-m_{\bf k}\sigma_z \hspace{-0.8mm}
\end{array}  \hspace{-0.8mm}\right)\hspace{-0.8mm},
\label{8by8td}
\end{eqnarray*}
\end{widetext}
where as before we define ${\vec{\lambda}_{\bf k}} = -2\lambda
\sum_{\nu \in u_D} \sin{k_\nu} \hat{e}_{\nu} \equiv ( \lambda_{k_x},
\lambda_{k_y}, \lambda_{k_z})$, and $m_{\bf{k}} =
u_{cd}-2t\sum_{\nu \in u_D} \cos{k_\nu}$. The above expression makes it clear
why, due to the SO component $\lambda_{k_z}$, a decoupled
structure no longer arises in 3D for arbitrary momentum values.

In the presence of an applied magnetic field with components $(h_x,
h_y, h_z)$ as in Eq. (\ref{HM}), the general expression for the
resulting Hamiltonian matrix of $H_D+H_M$ becomes:
\begin{widetext}
\begin{eqnarray*}
\hat{H}''_{\bf{k}}=
\hspace{-1mm}\left (\hspace{-1mm}\begin{array}{cccc}
-\hspace{-0.8mm}\mu\hspace{-0.8mm}+\hspace{-0.8mm}\lambda_{k_x}\sigma_x\hspace{-0.8mm}+\hspace{-0.8mm}\lambda_{k_y}\sigma_y\hspace{-0.8mm}+\hspace{-0.8mm}m_{+,{\bf k}}\sigma_z \hspace{-0.8mm}  & \hspace{-0.8mm}i\Delta\sigma_y\hspace{-0.8mm} & \hspace{-0.8mm}i\lambda_{k_z} \sigma_y \hspace{-0.8mm}+\hspace{-0.8mm}h_x \hspace{-0.8mm}-\hspace{-0.8mm}ih_y\sigma_z\hspace{-0.8mm} & \hspace{0.8mm}0\hspace{-0.8mm} \\
\hspace{-0.8mm}-i\Delta\sigma_y \hspace{-0.8mm} &\hspace{-0.8mm}\mu\hspace{-0.8mm}+\hspace{-0.8mm}\lambda_{k_x}\sigma_x\hspace{-0.8mm}-\hspace{-0.8mm}\lambda_{k_y}\hspace{-0.8mm}\sigma_y\hspace{-0.8mm}-\hspace{-0.8mm}
m_{+,{\bf k}} \sigma_z \hspace{-0.8mm}   & \hspace{-0.8mm}0\hspace{-0.8mm} & \hspace{-0.8mm}i\lambda_{k_z}\hspace{-0.8mm} \sigma_y\hspace{-0.8mm}-\hspace{-0.8mm}h_x \hspace{-0.8mm}-\hspace{-0.8mm}ih_y\sigma_z \hspace{-0.8mm}\\
\hspace{-0.8mm}-\hspace{-0.6mm}i\lambda_{k_z}\hspace{-0.8mm} \sigma_y \hspace{-0.8mm}+\hspace{-0.8mm}h_x \hspace{-0.8mm}+\hspace{-0.8mm}ih_y\sigma_z \hspace{-0.8mm}& \hspace{-0.8mm}0\hspace{-0.8mm} & -\hspace{-0.8mm}\mu+\hspace{-0.8mm}\lambda_{k_x}\hspace{-0.8mm}\sigma_x\hspace{-0.8mm}-\hspace{-0.8mm}\lambda_{k_y}\hspace{-0.8mm}\sigma_y\hspace{-0.8mm}+\hspace{-0.8mm}m_{-, {\bf k}} \sigma_z \hspace{-0.8mm} & \hspace{-0.8mm}-\hspace{-0.6mm}i\Delta\sigma_y\hspace{-0.8mm} \\ 
\hspace{-0.8mm}0\hspace{-0.8mm} & \hspace{-0.8mm}-\hspace{-0.7mm}i\lambda_{k_z}\hspace{-0.8mm}\sigma_y\hspace{-0.8mm}-\hspace{-0.8mm}h_x \hspace{-0.8mm}+\hspace{-0.8mm}ih_y\hspace{-0.8mm}\sigma_z\hspace{-0.8mm}  &  \hspace{-0.8mm}i\Delta\sigma_y \hspace{-0.8mm} &  \hspace{-0.8mm}\mu\hspace{-0.8mm}+\hspace{-0.8mm}\lambda_{k_x}\hspace{-0.8mm}\sigma_x\hspace{-0.8mm}+\hspace{-0.8mm}\lambda_{k_y}\hspace{-0.8mm}\sigma_y\hspace{-0.8mm}-\hspace{-0.8mm}m_{-, {\bf k}} \sigma_z \hspace{-0.8mm}
\end{array}  \hspace{-0.8mm}\right)\hspace{-1mm}, 
\label{8by8r}
\end{eqnarray*}
\end{widetext}
where $m_{\pm, {\bf k}} \equiv m_{\bf k} \pm h_z$, as also defined in the main text. 
The above expression makes it clear that, when $k_z= k_{z,c} \in \{ 0, \pi\}$ ($\lambda_{k_{z}}=0$), 
we may still obtain an analytical solution 
of the excitation spectrum for a magnetic field in the ${z}$ direction, with respect to the 
above operator basis $\hat{B}_{\bf{k}}^{'}$. Similarly, it is possible in principle to find a suitable 
basis (not shown) such that an analytical solution of the excitation spectrum exists 
for $k_\nu = k_{\nu,c}$,  $\nu={x}$ or ${y}$, in the presence of 
a magnetic field along the $x$ or $y$ direction, respectively.

\section{Numerical evaluation of topological invariants}
\label{appC}

In numerical computations of topological invariants, it is crucial to 
guarantee that the results be {\em numerically gauge-invariant}\cite{Ortiz94,Ortiz96},
otherwise one gets non-sensical results because of the random phases 
generated from numerical diagonalization of $H_D$. 
We briefly describe here the procedure we followed to ensure numerical 
gauge-invariance. 

Recall the definition of the Berry phase [Eq.~(\ref{bn})]:
\begin{eqnarray*}
B_n=i \int_{-\pi}^\pi dk \, \langle \psi_{n,k}|\partial_{k}
\psi_{n,k}\rangle ,
\label{berryi}
\end{eqnarray*} 
in terms of normalized states, $\langle \psi_{n,k}| \psi_{n,k}\rangle=1$.
The phase $\varphi(n,k,k')$ of the matrix element $\langle \psi_{n,k}| 
\psi_{n,k'}\rangle$  satisfies the following relation:
\begin{eqnarray}
i \langle \psi_{n,k}|\partial_{k}
\psi_{n,k}\rangle &=& -\partial_{k'}\varphi(n,k,k')|_{k'=k} \\ \nonumber
&=& -\partial_{k'}\big( \text{Im} \ln \langle \psi_{n,k}|
\psi_{n,k'}\rangle \big)|_{k'=k} , 
\label{berryp}
\end{eqnarray} 
where in the last line the definition of the phase $\varphi(n,k,k')$ was used. Then, 
the Berry phase can be rewritten as 
\begin{eqnarray*}
B_n=- \int_{-\pi}^\pi dk \, \partial_{k'}\varphi(n,k,k')|_{k'=k},
\label{berryi1}
\end{eqnarray*} 
admitting a simple discretized approximation \cite{Ortiz94,Ortiz96} 
\begin{eqnarray}
B_n&=&\lim_{N \rightarrow \infty} \sum_{i=0}^{N-1} \Delta\varphi(n,i,i+1), \nonumber \\
&=&-\lim_{N \rightarrow \infty} \text{Im} \ln \prod_{i=0}^{N-1} \langle
\psi_{n,k_i}|\psi_{n,k_{i+1}}\rangle ,
\label{bnn}
\end{eqnarray} 
with the identification of the states ($k_0=-\pi, k_N=\pi$)
\begin{eqnarray*}
|\psi_{n,k_N}\rangle \equiv |\psi_{n,k_0}\rangle .
\label{bnn1}
\end{eqnarray*} 
In practice one needs only a few (some tenths) points in the above product
for a stable result to be found. 

Similarly, the CN $C_n$ is given by [Eq. (\ref{cn})]:
\begin{eqnarray*}
C_n=\frac{1}{\pi} \int_{-\pi}^\pi dk_x\int_{-\pi}^\pi dk_y \,
\text{Im} \,\langle \partial_{k_x}\psi_{n,{\bf k}}|\partial_{k_y}
\psi_{n,{\bf k}}\rangle.
\label{cni}
\end{eqnarray*}
In numerical computations of $C_n$, we approximate the
integrand in Eq.~(\ref{cn}) as
\begin{eqnarray*}
\text{Im} \,\langle \partial_{k_x}\psi_{n,{\bf k}}|\partial_{k_y}
\psi_{n,{\bf k}}\rangle&\approx& \\ 
&&\hspace*{-3cm} \frac{1}{\epsilon^2} 
\text{Im}\,[ \ln ( \langle \psi_{n,{\bf k}}^{\;} | \psi_{n,{\bf
k}_x} \rangle \langle \psi_{n,{\bf k}_x} | \psi_{n,{\bf k}_y}
\rangle\langle \psi_{n,{\bf k}_y} | \psi_{n,{\bf k}}^{\;} \rangle)]  \nonumber ,
\end{eqnarray*}
where ${\bf k}_{\nu}\equiv {\bf k}+\epsilon \hat{k}_{\nu}$,
$\hat{k}_{\nu}$ are unit vectors in momentum space, and $\epsilon \ll
1$. Finally, we compute $C_n$ as 
\begin{eqnarray}
C_n=\frac{1}{\pi} 
\text{Im}\,\prod_{{\bf k}} \ln ( \langle \psi_{n,{\bf k}}^{\;} | \psi_{n,{\bf
k}_x} \rangle \langle \psi_{n,{\bf k}_x} | \psi_{n,{\bf k}_y}
\rangle\langle \psi_{n,{\bf k}_y} | \psi_{n,{\bf k}}^{\;} \rangle)] \nonumber .
\end{eqnarray}

\end{appendix} 


\begin{thebibliography}{100}

\bibitem{GBook} H. Nishimori and G. Ortiz, {\em Elements of Phase
Transitions and Critical Phenomena} (Oxford University Press, Oxford,
2011).

\bibitem{Kitaev03} A. Y. Kitaev, Ann. Phys. {\bf 321}, 2 (2003).

\bibitem{Nayak} C. Nayak {\em et al.}, Rev. Mod. Phys. {\bf 80}, 1083
(2008).

\bibitem{Kane05} C. L. Kane and E. J. Mele, Phys. Rev. Lett. {\bf 95},
146802 (2005).

\bibitem{Kanereview} M. Z. Hasan and C. L. Kane, Rev. Mod. Phys. {\bf
82}, 3045 (2010). 

\bibitem{Green} N.  Read and D. Green, Phys. Rev. B {\bf 61}, 10267
(2000).

\bibitem{Ivanov} D. A. Ivanov, Phys. Rev. Lett. {\bf 86}, 268 (2001).

\bibitem{Kitaev01} A. Y. Kitaev, Phys. Uspekhi {\bf 44}, 131 (2001).

\bibitem{Qi} X. L. Qi, and S. C. Zhang, Rev. Mod. Phys. {\bf 83}, 1057
  (2011).

\bibitem{Ettore} E. Majorana, Nuovo Cimento {\bf 14}, 171 (1937).

\bibitem{Moore} G. Moore and N. Read, Nucl. Phys. B {\bf 360}, 362 (1991).

\bibitem{Frank} F. Wilczek, Nature Phys. {\bf 5}, 614 (2009).

\bibitem{Lutchyn} R. M. Lutchyn {\em et al.}, Phys. Rev. Lett. {\bf
105}, 077001 (2010); {\em ibid.} {\bf 106}, 127001 (2011); R. M. Lutchyn 
and M. A. Fisher, Phys. Rev. B {\bf 84},  214528 (2011).

\bibitem{Oreg} Y. Oreg, G. Refael, and F. von Oppen,
Phys. Rev. Lett. {\bf 105}, 177002 (2010).

\bibitem{Potter10} A. C. Potter and P. A. Lee, Phys. Rev. Lett. {\bf
105}, 227003 (2010); A. C. Potter and P. A. Lee, Phys. Rev. B {\bf
83}, 094525 (2011).

\bibitem{2dhetero} J. D. Sau, R. M. Lutchyn, S.  Tewari, and S. Das
Sarma, Phys. Rev. Lett. {\bf 104}, 040502 (2010); J. Alicea,
Phys. Rev. B {\bf 81}, 125318 (2010).

\bibitem{Shen11} B. Zhou and S.-Q. Shen, Phys. Rev. B {\bf 84}, 054532 
(2011).

\bibitem{Brouwer} P. W. Brouwer, M. Duckheim, A. Romito, and F. V. Oppen, 
Phys. Rev. B {\bf 84}, 144526 (2011).

\bibitem{Chung} S. B. Chung, H. J. Zhang, X. L. Qi, and S. C. Zhang, Phys. 
Rev. B {\bf 84}, 060510 (R) (2011). 

\bibitem{Kane} L. Fu and C. L. Kane, Phys. Rev. Lett. { \bf 100},
096407 (2008).

\bibitem{Sarma} T. D. Stanescu, J. D. Sau, R. M. Lutchyn, and
S. D. Sarma, Phys. Rev. B {\bf 81}, 241310(R) (2010).

\bibitem{Volkov} B. A. Volkov and O. A. Pankratov, JETP Lett. {\bf
42}, 178 (1985); O. A. Pankratov, S. V. Pakhomov, and B. A. Volkov,
Solid State Commun. {\bf 61}, 93 (1987).

\bibitem{Qi09} X.-L. Qi {\em et al.}, Phys. Rev. Lett. {\bf 102},
187001 (2009).

\bibitem{SatoOdd} M. Sato, Phys. Rev. B {\bf 81}, 220504(R) (2010).

\bibitem{Fu10} L. Fu, and E. Berg, Phys. Rev. Lett. {\bf 105}, 097001
(2010).

\bibitem{SMW} H. Suhl, B. T. Matthias, and L. R. Walker,
Phys. Rev. Lett. {\bf 3}, 552 (1959).

\bibitem{Naga} J. Nagamatsu {\em et al.}, Nature {\bf 410}, 63 (2001).

\bibitem{materials} R. Khasanov {\em et al.}, Phys. Rev. Lett. {\bf
  98}, 057007 (2007); M. Jourdan, A. Zakharov, M. Foerster, and H. Adrian, {\em ibid.} {\bf 93},
  097001 (2004); Y. Kamihara,T. Watanabe, M. Hirano, and H. Hosono, J. Am.  Chem. Soc. {\bf
    130}, 3296 (2008); A. P. Petrovi\'c {\em et al.}, Phys. Rev. Lett.
  {\bf 106}, 017003 (2011); T. Hanaguri, S. Niitaka, K. Kuroki, H. Takagi, Science {\bf
    328}, 474 (2010).

\bibitem{MazinNat} I. I. Mazin, Nature {\bf 464}, 183 (2010).

\bibitem{Deng} S. Deng, L. Viola, and G. Ortiz, Phys. Rev. Lett. {\bf
108}, 036803 (2012).

\bibitem{Nakosai} S. Nakosai, Y. Tanaka, and N. Nagaosa, Phys. Rev. Lett. {\bf 108},
147003 (2012).

\bibitem{Babak} B. Seradjeh, Phys. Rev. B {\bf 86}, 121101(R) (2012).

\bibitem{FuKaneNote} As noted in Ref. \onlinecite{Deng}, interpreting the band index as a layer 
index also allows to establish a \emph{formal} similarity with Fu \& Kane's proposal 
\cite{Kane}.  While the physics underlying the two models differ significantly, a phase 
difference of $\pi$ between the pairing gaps of neighboring $s$-wave superconductors 
is also required in order for the corresponding junction to support Majorana fermions 
(in 1D) or more generally Majorana bound states.

\bibitem{Sasaki11} S. Sasaki {\em et al.}, Phys. Rev. Lett. {\bf 107},
217001 (2011).

\bibitem{Sasaki12} S. Sasaki {\em et al.}, arXiv:1208.0059.

\bibitem{Stroscio} N. Levy {\em et al.}, arXiv:1211.0267. 

\bibitem{Franz} G. Rosenberg and M. Franz, Phys. Rev. B {\bf 82},
035105 (2010).

\bibitem{Isaev} L. Isaev, Y. H. Moon, and G. Ortiz, Phys. Rev. B {\bf
84}, 075444 (2011).

\bibitem{Altland} A. Altland, and M. R. Zirnbauer, Phys. Rev. B {\bf
55}, 1142 (1997).

\bibitem{Roy} R. Roy, New J. Phys. {\bf 12} 065009 (2010).

\bibitem{Kane06} L. Fu and C. L. Kane, Phys. Rev. B {\bf 74}, 195312
  (2006).

\bibitem{Roy09} R. Roy, Phys. Rev. B {\bf 79}, 195322 (2009).

\bibitem{Ortiz94} G. Ortiz, and R. M. Martin, Phys. Rev. B {\bf 49},
14202 (1994).

\bibitem{Ortiz96} G. Ortiz, P. Ordej{\' o}n, R. M. Martin, and
G. Chiappe, Phys. Rev. B {\bf 54}, 13515 (1996).

\bibitem{Berry} M. V. Berry, Proc. R. Soc. Lond., {\bf A392}, 45
(1984).

\bibitem{Zak} J. Zak, Phys. Rev. Lett. {\bf 62}, 2747 (1989).

\bibitem{LSM} E. Lieb, T. Schultz, and D. Mattis, Ann. Phys. (N.Y.)
{\bf 16}, 407 (1961).

\bibitem{Chakravarty} Y. Niu {\em et al.}, Phys. Rev. B {\bf 85}, 035110 (2012).

\bibitem{Ryu} S. Ryu, A. P. Schnyder, A. Furusaki, and
A. W. W. Ludwig, New J.  Phys. {\bf 12}, 065010 (2010).

\bibitem{Kub} A. Kubasiak, P. Massignan, and M. Lewenstein, EPL {\bf
  92}, 46004 (2010).

\bibitem{Kitaev2010} L. Fidkowski and A. Kitaev, Phys. Rev. B {\bf 81}, 134509 
(2010).

\bibitem{experiments} D. Hsieh {\em et al.}, Nature {\bf 452}, 970
  (2008); D. Hsieh {\em et.al.}, Science {\bf 323}, 919 (2009);
  P. Roushan {\em et al.}, Nature {\bf 460}, 1106 (2009).

\bibitem{Yang} Y. Yang {\em et al.}, Phys. Rev. Lett. {\bf 107}, 066602 (2011).

\bibitem{Liu} Q. Liu {\em et al.}, Phys. Rev. Lett. {\bf 102}, 156603 (2009).

\bibitem{Goldman} N. Goldman, W. Beugeling, and C. Morais Smith,
Europhys. Lett. {\bf 97}, 23003 (2012).

\bibitem{Dahm} T. Paananen and T. Dahm, arXiv:1210.4422.

\bibitem{Sato} M. Sato, Y. Takahashi, and S. Fujimoto,
Phys. Rev. Lett. {\bf 103}, 020401 (2009).

\bibitem{Sau} J. D. Sau, S. Tewari, R. M. Lutchyn, T. D. Stanescu, and S. 
D. Sarma, Phys. Rev. B {\bf 82}, 214509 (2010).

\bibitem{Fu2012} T. H. Hsieh and L. Fu, Phys. Rev. Lett. {\bf 108},
  107005 (2012).

\bibitem{DengFlat} S. Deng, G. Ortiz, and L. Viola, ``Majorana flat bands in 
$s$-wave gapless topological superconductors,'' forthcoming. 

\bibitem{Mazin} I. I. Mazin, D. J. Singh, M. J. Johannes, and
M. D. Hu, Phys. Rev. Lett. {\bf 101}, 057003 (2008).

\bibitem{Ishida} K. Ishida, Y. Nakai, and H. Hosono, J. Phys. Soc. Jpn.
{\bf 78}, 062001 (2009). 

\bibitem{Zhang12} F. Zhang, C. L. Kane, and E. J. Mele, 
arXiv:1212.4232.

\end{thebibliography}
\end{document}